\documentclass[12pt,a4paper]{article}
 \pdfoutput=1
 \usepackage[left=2.00cm, right=2.00cm, top=2.00cm, bottom=2.00cm]{geometry}
\usepackage{float}
\usepackage{mathtools}
\usepackage{yfonts}
\usepackage{esint}
\usepackage{setspace}
\usepackage{subfigure}
\usepackage{soul}
\usepackage{dcolumn}
\usepackage{color}
\usepackage{xcolor}
\usepackage{ulem}
\usepackage{graphicx}
\usepackage{bm}
\usepackage{amsmath,amsthm,amssymb}
\usepackage{authblk}
\usepackage{hyperref}

\newtheorem{remark}{Remark}
\newcommand{\bxi}{\bm{\xi}}

\begin{document}
 \title{ \textbf{Non-Hermitian acoustic waveguides with periodic electroacoustic feedback}}
\author[1,*]{{\small Danilo Braghini}}
\author[1]{{\small Vinicius D. de Lima}}
\author[2]{{\small Danilo Beli}}
\author[3]{{\small Matheus I. N. Rosa}}
\author[1]{{\small José R. de F. Arruda}}

\affil[1]{School of Mechanical Engineering, University of Campinas, Campinas, São Paulo 13083-970, Brazil }
\affil[2]{Sao Carlos School of Engineering, University of Sao Paulo, Sao Carlos/SP 13563-120, Brazil}
\affil[3]{Department of Mechanical Engineering, University of Colorado Boulder, Boulder CO 80309, USA}
\affil[*]{ Corresponding author: d166353@dac.unicamp.br}

\maketitle

\begin{abstract}
In this work, we investigate non-Hermitian acoustic waveguides designed with periodically applied feedback efforts using electrodynamic actuators. One-dimensional spectral (infinite-dimensional) and finite element (finite-dimensional) models for plane acoustic waves in ducts are used. It is shown that dispersion diagrams of this family of metamaterials exhibit non-reciprocal imaginary frequency components, manifesting as wave attenuation or amplification along opposite directions for all pass bands. The effects of different feedback laws are investigated. Furthermore, the non-Hermitian skin effect manifesting as topological modes localized at the boundaries of finite domains is investigated and successfully predicted by the topology of the reciprocal space. This work extends previous numerical results obtained for a piezoelectric rod system and contributes to recent efforts in designing active metamaterials with novel properties associated with the physics of non-Hermitian systems, which may find fruitful technological applications related to noise control, wave localization, filtering and multiplexing.
\end{abstract}
\newpage

\section{Introduction} \label{sec: intro}

In Physics and Engineering, at the first steps of the investigation of a system, it is usual to assume that no energy is lost when losses are small when compared with the total energy for the investigated periods. This is the basic condition for classifying systems as Hermitian. However, the exchange of energy between a system and its surrounding environment in amounts that cannot be assumed small is ubiquitous. When these effects are considered, the system is defined as non-Hermitian (NH).

In recent years, the investigations regarding the odd bulk boundary correspondence of NH systems \cite{xiong2018does,koch2020bulk,kunst2018biorthogonal,yao2018edge,yao2018non} have led to a deeper understanding of the effects of symmetry, such as the topological skin modes. Differently from topological modes previously observed in Hermitian systems, these modes are extremely sensitive to boundary conditions \cite{kawabata2019symmetry,okuma2020topological,bergholtz2021exceptional}. Particularly, in tight-biding models it was shown that the sensitivity is exponential \cite{edvardsson2022sensitivity}. As shall become clearer in this work, the appropriate boundary conditions must be chosen to observe skin modes on NH systems, and yet different boundary conditions if the topological aspects are to be analyzed.

Non-Hermiticity has also been used to design metamaterials. Topological mechanics has been applied for the investigation of the dynamics of such metamaterials \cite{wang2018topological,ma2019topological,huber2016topological,delplace2020geometry}. Developments associated with geometrical phases in metamaterial research have allowed the prediction of novel topological matter exhibiting the NH skin effect (NHSE) on both reciprocal and non-reciprocal systems, both in one-dimensional and in higher-dimensional systems \cite{zhong2021nontrivial,okugawa2020second,kawabata2020higher,hofmann2020reciprocal}. 

One of such topological modes happens in non-reciprocal platforms built upon NH periodic metamaterials. Non-reciprocity has already been the object of investigations in the context of metamaterial engineering \cite{nassar2020nonreciprocity,scheibner2020odd, chen2021realization}. For such a system, the topological invariant of the bulk is defined as the winding number of the corresponding dispersion diagram of Bloch-Bands (BB) and is related to observable NHSE in the real space \cite{zhang2020correspondence}. Such properties have been explored in quantum systems, in classical electric devices \cite{hofmann2020reciprocal,helbig2020generalized}, and, more recently, in mechanical platforms~\cite{ghatak2020observation,rosa2020dynamics,braghini2021non,zhang2021acoustic}.

Recently, the investigation of non-Hermitian dispersion relations regarding boundary conditions and the NHSE were extended to distributed-parameter (infinite number of degrees of freedom) models \cite{longhi2021non,zhong2021nontrivial,braghini2021non,doi:10.1063/5.0097530} rather than tight-binding or lumped-parameter (finite number of degrees of freedom) models. Mechanical configurations able to generate arbitrary topologies have also been previously reported \cite{wang2021generating,zhang2021acoustic} on quantum systems (modulated ring resonator with a synthetic frequency dimension) and classical mechanical systems (acoustic cavities), both in the context of lumped-parameter models. In distributed-parameter models, arbitrary topologies have been investigated by using non-local feedback interactions \cite{rosa2020dynamics,braghini2021non} or by varying geometrical parameters of the waveguides \cite{doi:10.1063/5.0097530}.   

Although all topological aspects of the NHSE - in contrast with Hermitian topological modes - can be studied in simpler single-band periodic systems with one degree of freedom per unit cell, distributed-parameter models with an infinite number of bands in the reciprocal space are more realistic in representing practical applications. This work makes contributions in this direction, using acoustic one-dimensional waveguides with periodically applied electroacoustic feedback, following a design strategy that emulates both nearest-neighborhood and long-range non-reciprocal coupling by applying local and non-local feedback interactions, respectively \cite{rosa2020dynamics,braghini2021non,PhysRevApplied.18.014067,chen2021realization}. 

We explore different possible feedback laws as a way to achieve different BB topologies. The stability of the designed metastructures (finite systems) is investigated as a previous step in performing experiments on the designed electroacoustic platforms. These systems may find a myriad of applications in engineering, wherever mechanical waves need to be localized and filtered. For instance, investigations suggest that they can be used as design strategies to control filaments and membranes in biological systems \cite{chen2021realization} and may also be highly effective for broad-band energy harvesting, when compared to traditional approaches\cite{PhysRevApplied.18.014067}.

\section{The system under study}

\begin{figure}[H]
	\centering
	\includegraphics[width=\textwidth]{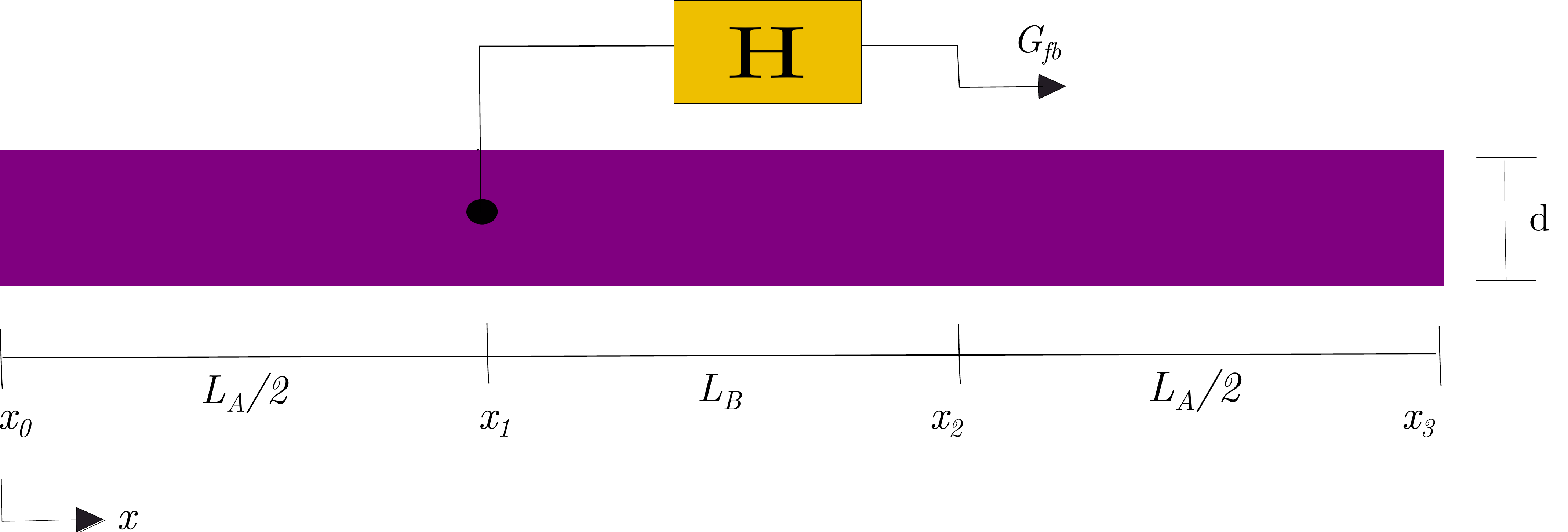}
	\caption{Acoustic NH metamaterial unit cell.}
	\label{cell}
\end{figure}
Figure \ref{cell} shows the configuration of the unit cell of the acoustic NH system. It consists of a 1D acoustic duct with circular cross-section of constant diameter $d$, equipped with an acoustic pressure sensor (e.g., a microphone) and an electrodynamic actuator (e.g., a loudspeaker). The pressure is measured at $x_1$ and a feedback volume velocity, defined by the operator $H$ (controller), is applied at $x_2$ using, for instance, a loudspeaker. In Fig.~\ref{cell}, $x_2$ is located in the same cell as $x_1$, defining a local feedback actuation. If the feedback is applied to a different cell, the feedback actuation is referred to as non-local. Three methods were used to study the effects of NH waveguides: the Finite Element Method (FEM), the Spectral Element Method (SEM), and the Plane Wave Expansion Method (PWE). Details concerning these methods and their application in this work are provided in the Appendices.

For all the simulations considered, the methods detailed in \ref{A1: methods} were applied to a 1D acoustic duct filled with air at ambient conditions ($\rho_0 \approx 1.225 kg/m^3$, $c \approx 343 m/s$), $L_c = 50 cm$, $d=4 cm$, $L_A = L_B = \frac{L_c}{2}$. 


\section{Results}

\subsection{Effects of the feedback law on the system spectrum}

\begin{figure}[H]
\centering
\subfigure[]{\includegraphics[width=0.495\textwidth]{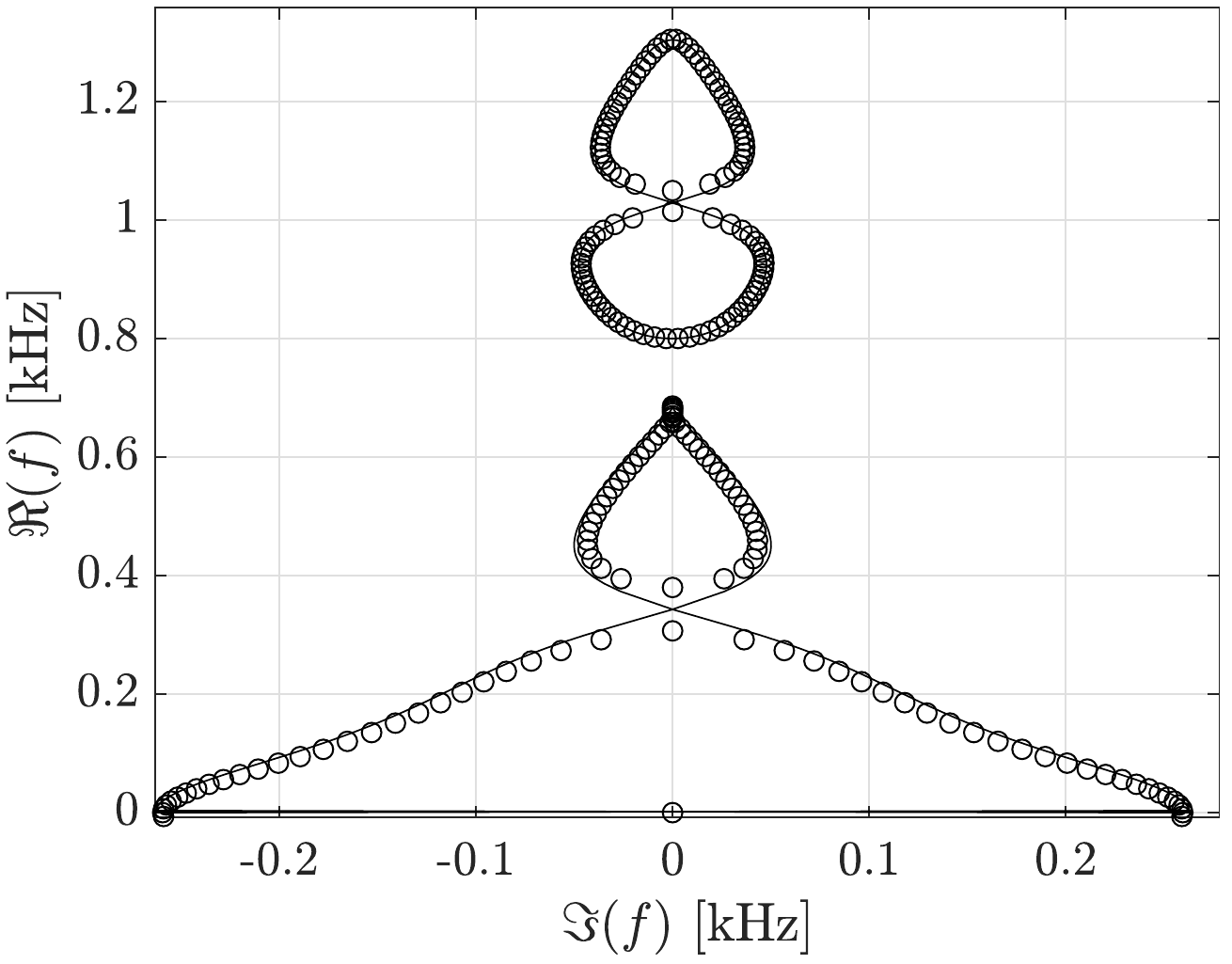}}\label{fig:PBCxOBCa}
\subfigure[]{\includegraphics[width=0.495\textwidth]{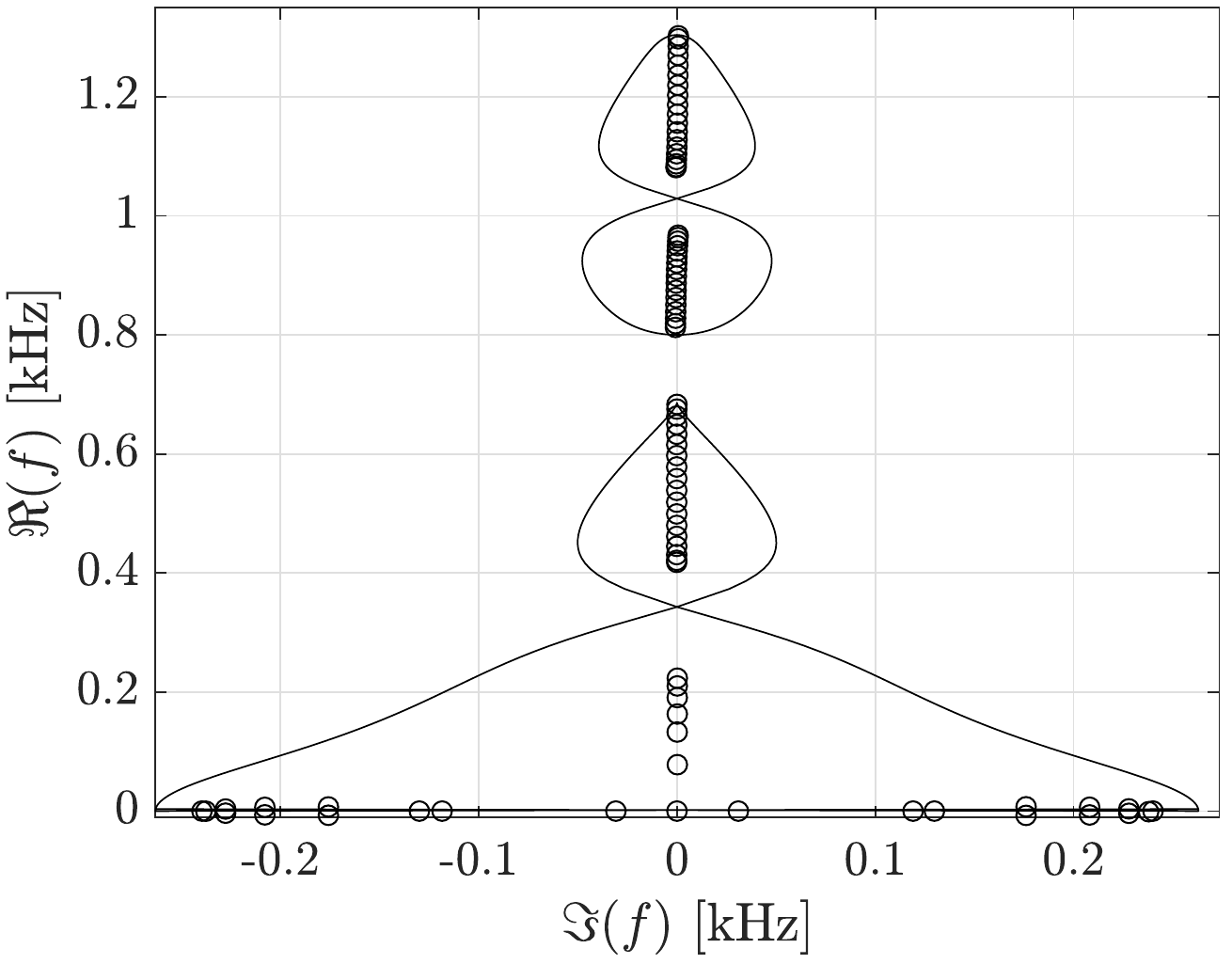}}\label{fig:PBCxOBCb}
\caption{Complex frequency plane of the system with the feedback defined in Eq.~(\ref{eq:pidfb}) for local ($a=0$) and integral action ($\gamma_D = \gamma_P = 0$), with gain $\gamma_I = 0.015$. Dispersion relations (solid curves) and eigenmodes of the structure (circles) are compared under (a) PBC and (b) OBC.} 
\label{fig:PBCxOBC}
\end{figure}

Fig.~\ref{fig:PBCxOBC} shows the complex frequency plane for two different boundary conditions: periodic boundary conditions (PBC) and open (free) boundary conditions (OBC). A system with OBC is a system composed of a finite number of cells with closed ends. Thus, it is finite in length and will be dubbed a metastructure herein. To impose PBCs on phononic crystals, the \textit{Bloch-Floquet} theorem is usually invoked. However, a novel approach is used herein. If the domain is one-dimensional, one may ``wrap around'' the system by connecting its ends as a way to impose the infinite periodic repetition of the system in a cyclic way. In the periodic NH system treated here, this was achieved by connecting (imposing continuity and equilibrium) the left boundary of the first cell with the right boundary of the last cell (metastructure ends). We use this wraparound boundary condition and name it periodic boundary condition (PBC). 


In both cases - OBC and PBC -, FEM was used to obtain the dynamic stiffness matrix of the closed-loop feedback metastructure, which leads to a finite-dimensional eigenproblem. The correspondence between the eigenspectrum and the complex plane of the dispersion relation computed via SEM is direct, by the change of variables $s$ to $f = -j \frac{s}{2 \pi}$, where $s$ is any eigenvalue, $f$ is the complex frequency and $j$ denotes the imaginary unit. Figure~\ref{fig:PBCxOBCa} shows that, with PBC, the eigenfrequencies are on the dispersion curve. This is due to the fact that the FEM mesh is a discretization (lumped model) of the system, and, thus, the eigenspectrum found is actually a discretization of the corresponding continuum spectrum of the metamaterial (infinite system \cite{hussein2006dispersive}) depicted in solid lines. In Fig.~\ref{fig:PBCxOBCb}, with OBC, the eigenmodes found by FEM are placed on the real and imaginary axes and represent a discretization of the metastructure eigenspectrum. The difference between results on Fig.~\ref{fig:PBCxOBCa}(a) and (b) expose the unique bulk boundary correspondence of NH systems, where eigenmodes are extremely sensitive to the boundary conditions.

These results are in agreement with theorem 1 and Eq.(5) of reference \cite{okuma2020topological}, which state that the eigenspectrum of the system under OBC is contained on the set formed by the union of the eigenspectrum of the metamaterial (which forms closed paths on the complex plane) with the subset of the plane divided by the paths that have a non-zero winding number (i.e., inside the closed paths). 
Also, since the matrices are real, the spectrum is symmetric relative to the imaginary frequency axis. Thus, we will herein consider only positive real frequencies, which have physical meaning from the wave propagation viewpoint.

In the sequence, the eigenspectrum of metastructures with each individual component of a typical proportional-integral-derivative (PID) controller used as the feedback interaction in each unit cell is depicted over a wide range of frequencies ($0-30 kHz$) using 21 finite elements per unit cell and applying PBC as a way to approximate the dispersion relation.

\begin{figure}[H]
\centering
\subfigure[]{\includegraphics[width=0.495\textwidth]{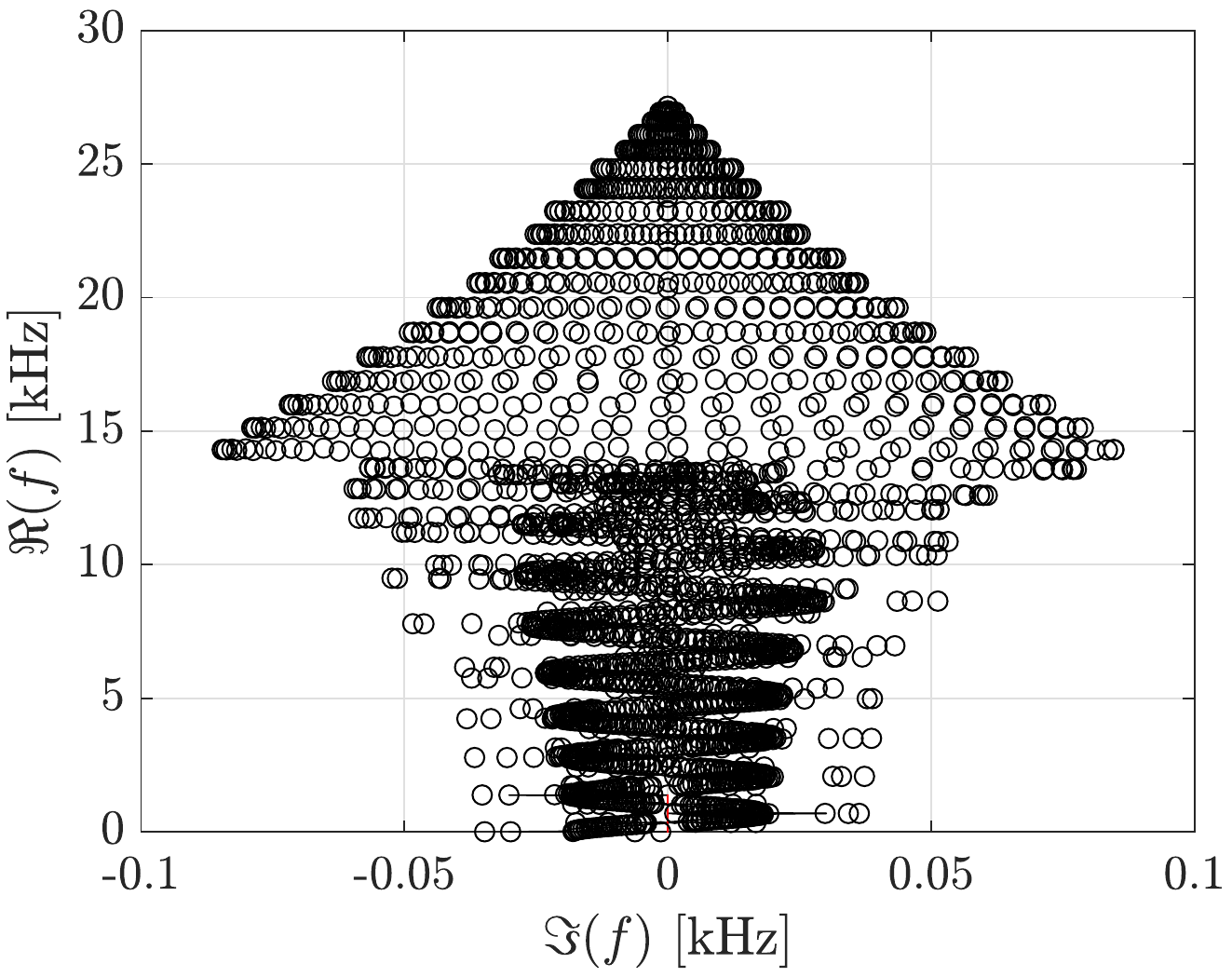}}
\subfigure[]{\includegraphics[width=0.495\textwidth]{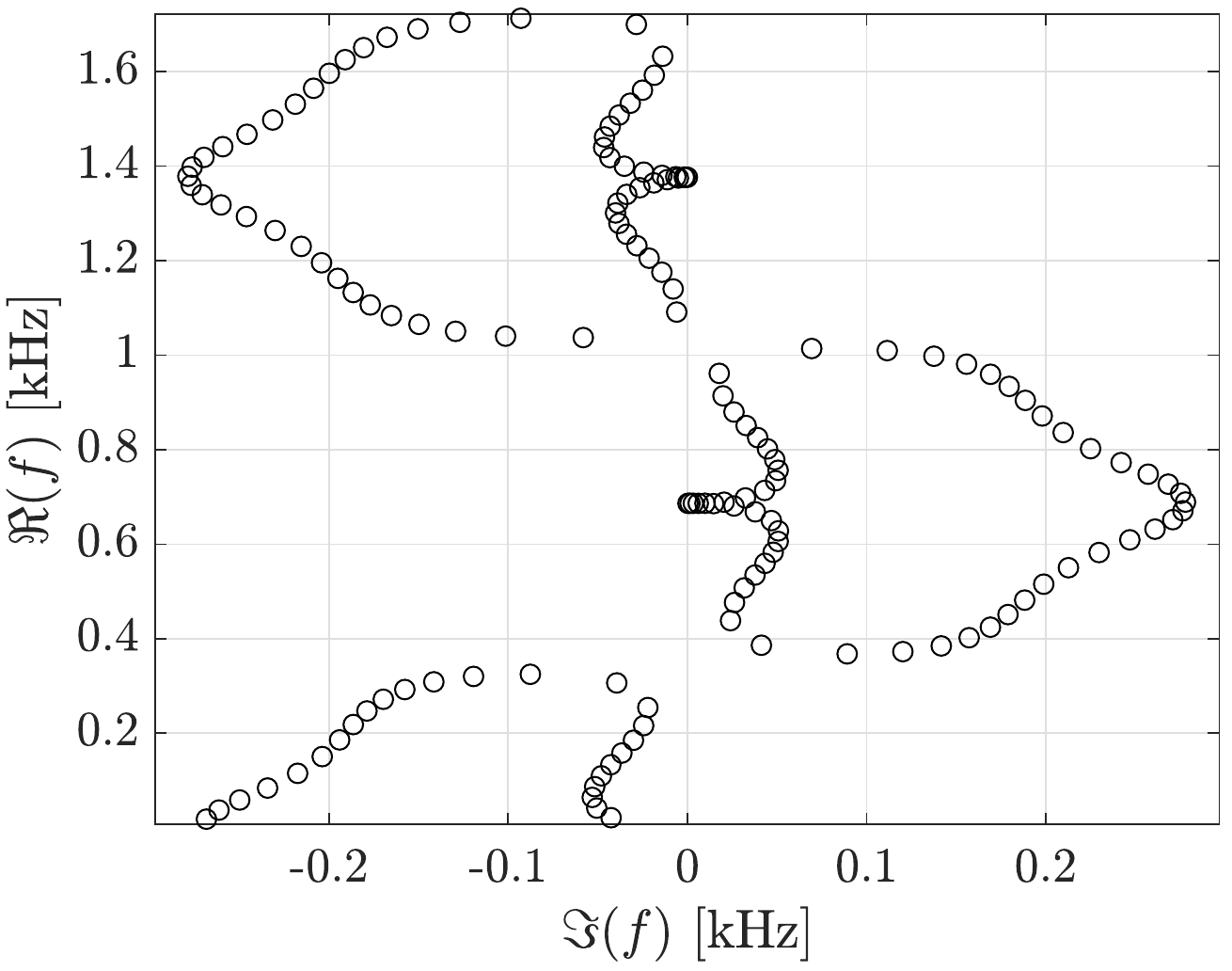}}
\caption{(a) Complex frequency plane of the system under PBC with local proportional feedback -$\gamma_P = 1e-5$. (b) zooming in the low frequencies.}
\label{fig:PBCproportional}
\end{figure}

In Fig.~\ref{fig:PBCproportional} it can be seen that a non-trivial topology was achieved with proportional feedback, as indicated by the closed paths on the complex plane. Moreover, the imaginary part of the frequency tends to increase up to an optimal frequency, and decrease for higher frequencies.

\begin{figure}[H]
\centering
\subfigure[]{\includegraphics[width=0.495\textwidth]{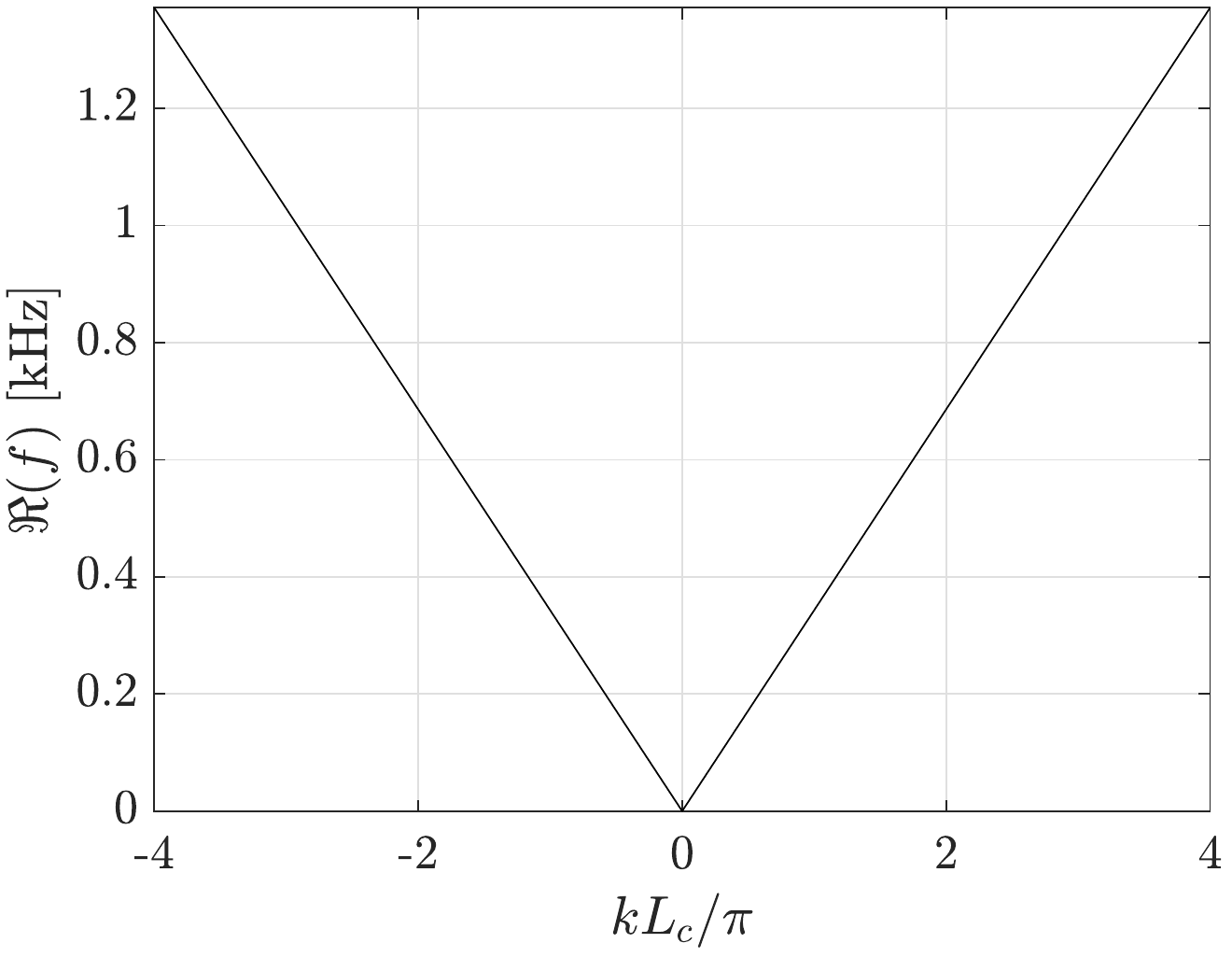}}
\subfigure[]{\includegraphics[width=0.495\textwidth]{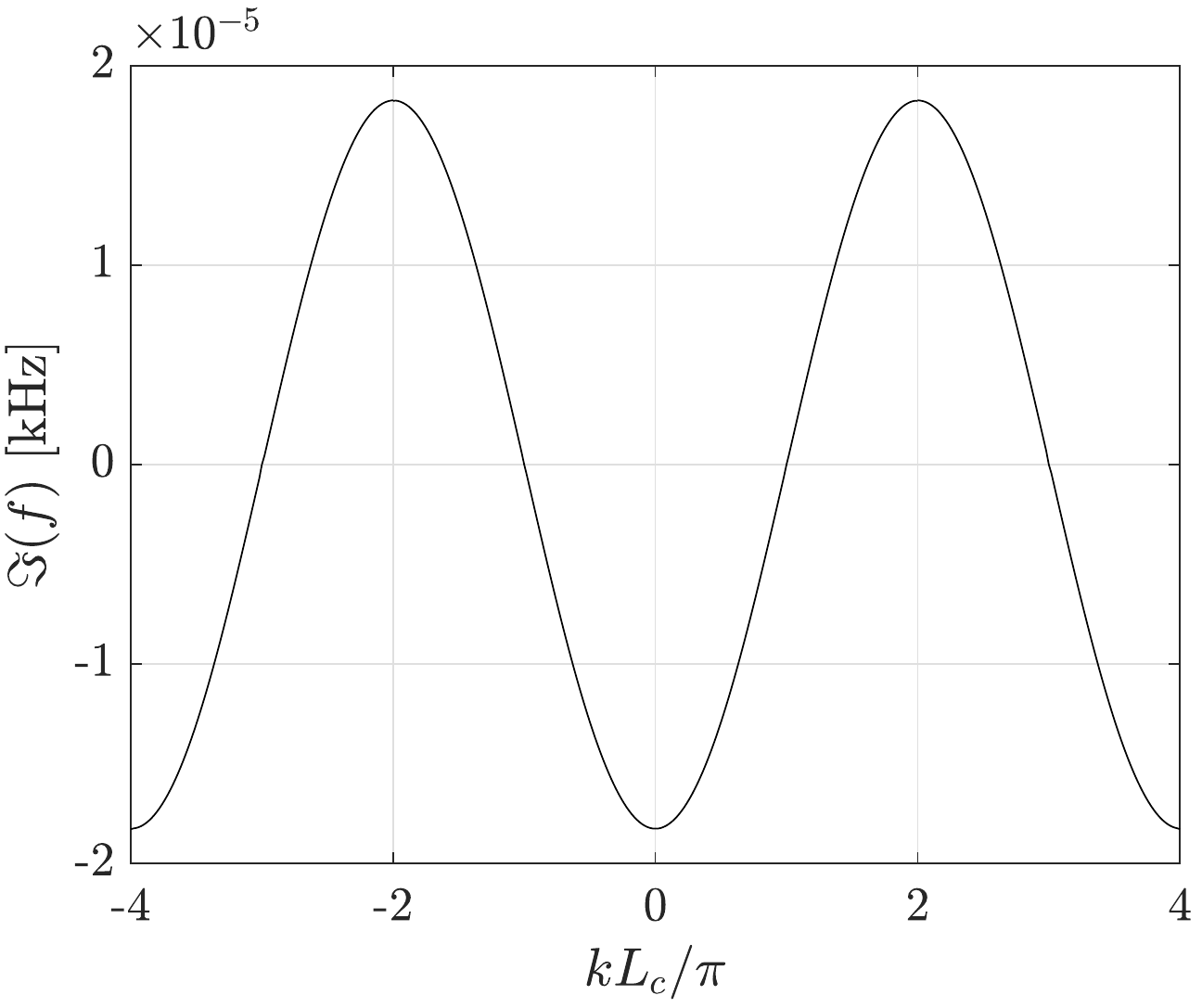}}
\caption{ Dispersion relation (a) real frequency against wavenumber (b) imaginary frequency against wavenumber.}
\label{fig:trivialproportional}
\end{figure}

It should be observed that, with lower values of feedback gain, this proportional feedback law provides a non-Hermitian trivial topology, as depicted in Fig.~\ref{fig:trivialproportional}, which is different from the Hermitian one (purely real frequency). Even though the real part of the dispersion relation shows no difference from the Hermitian counterpart of the acoustic system (passive and without damping), the imaginary parts exhibit a small but non-zero value, with ranges of wavenumbers for which the wave response is attenuated (positive imaginary frequency), as well as ranges for which it is amplified (negative imaginary frequency). Moreover, the diagram is reciprocal, which indicates the absence of the NHSE. 

\begin{figure}[H]
\centering
\subfigure[]{\includegraphics[width=0.495\textwidth]{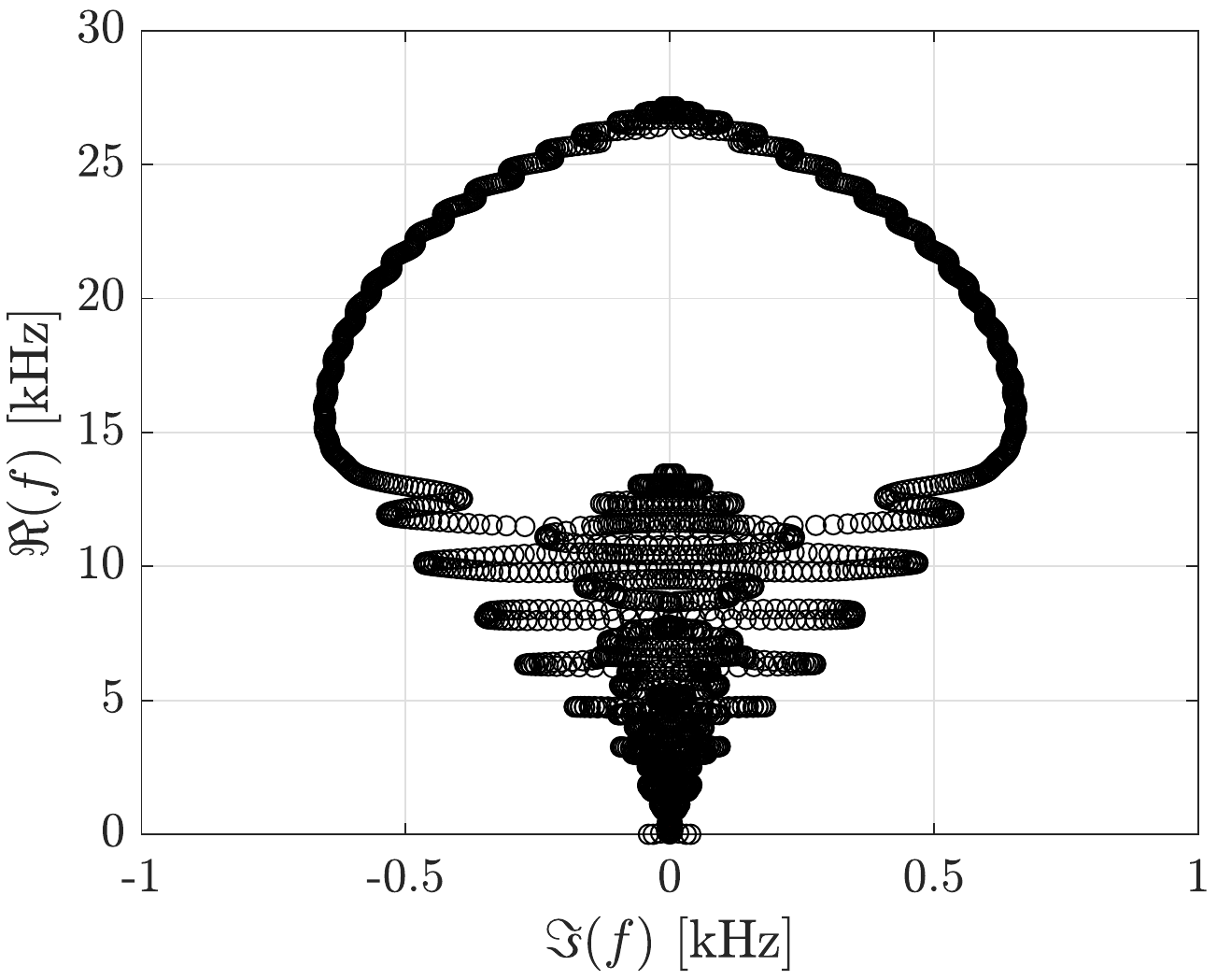}\label{PBC_derivative}}
\subfigure[]{\includegraphics[width=0.495\textwidth]{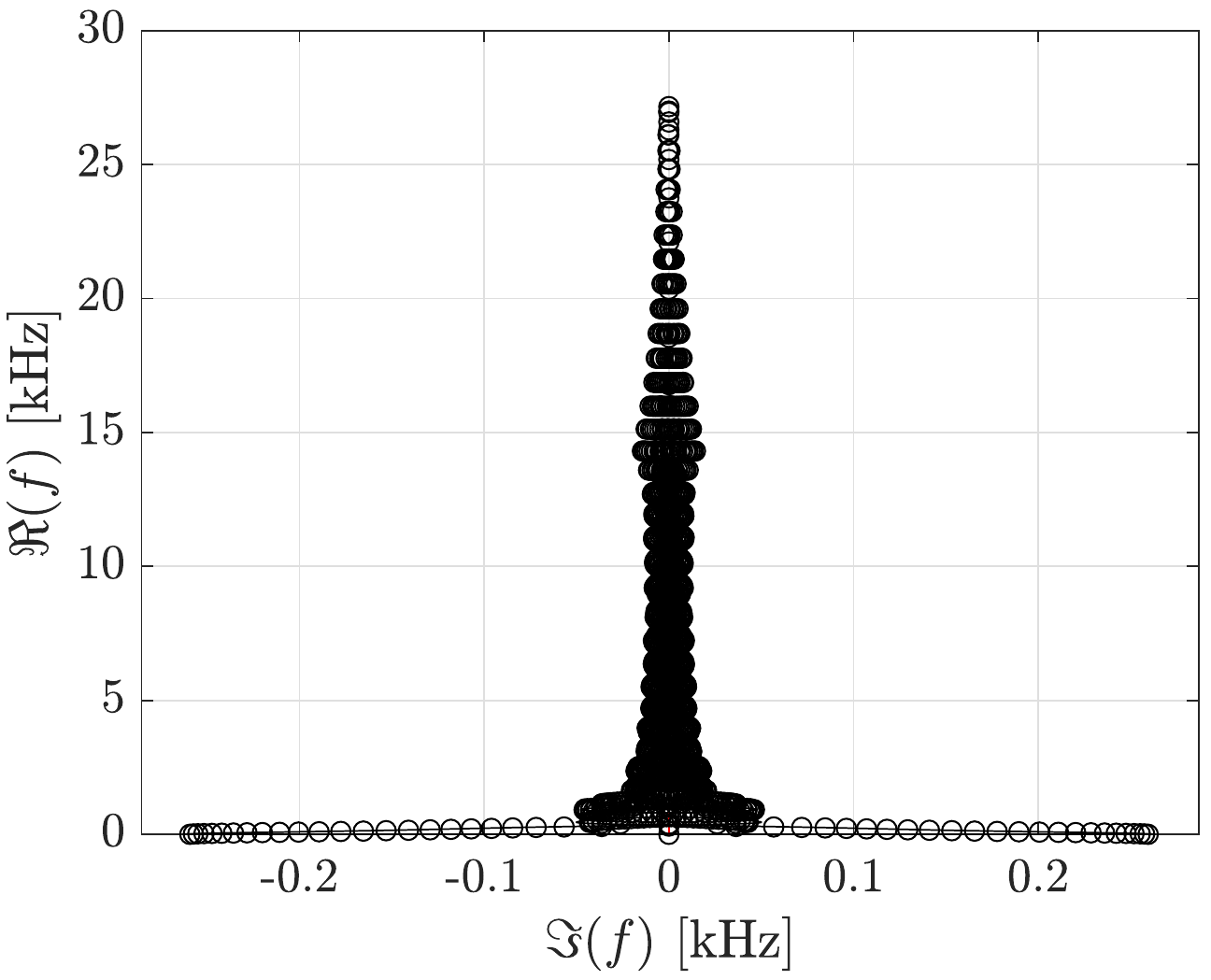}\label{PBC_integral}}
\caption{ Complex frequency plane of the system under PBC with (a) derivative and (b) integral feedback laws.}
\label{fig:PBC_otherlaws}
\end{figure}

As can be seen in Fig.~\ref{fig:PBC_otherlaws}, with the same number of finite elements, the derivative feedback concentrates the imaginary part of the frequency, related to non-reciprocal attenuation or amplification, on loops that get larger with increasing frequency. On the other hand, integral feedback causes loops that get smaller as frequencies gets higher. Thus, we can conclude that the derivative feedback effect on the BB dominates at high frequencies, whereas the integral feedback effect dominates at low frequencies. Since lead-lag feedback laws can be viewed as approximations of PI and PD feedback, the PID feedback law generalizes a wide range of possible classical control laws to be tested as ways to achieve broadband NHSE.

\subsection{NHSE and stable metastructures}
\label{stable}

\begin{figure}[H]
\centering
\subfigure[]{\includegraphics[width=0.495\textwidth]{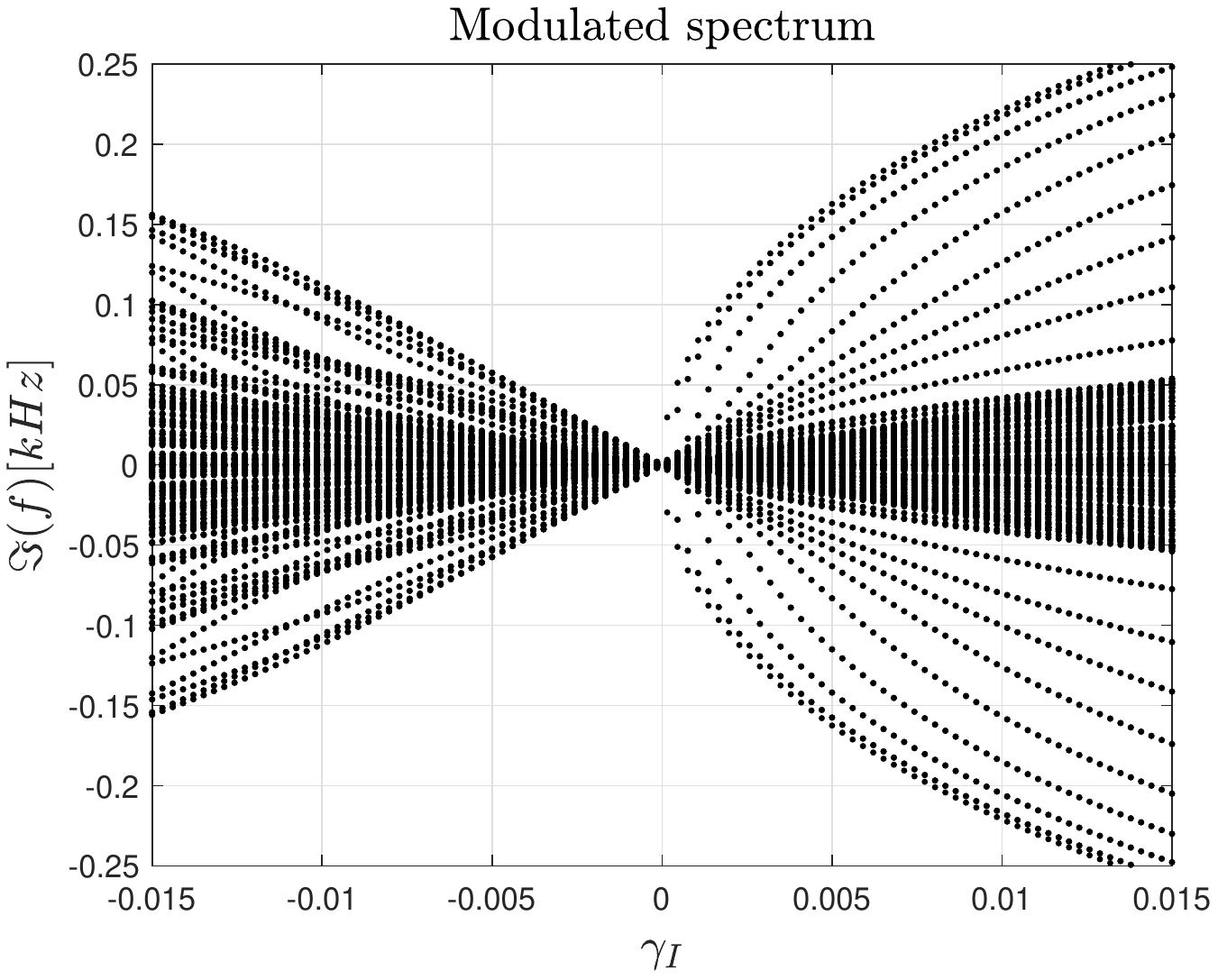}\label{fig:parametrica}}
\subfigure[]{\includegraphics[width=0.495\textwidth]{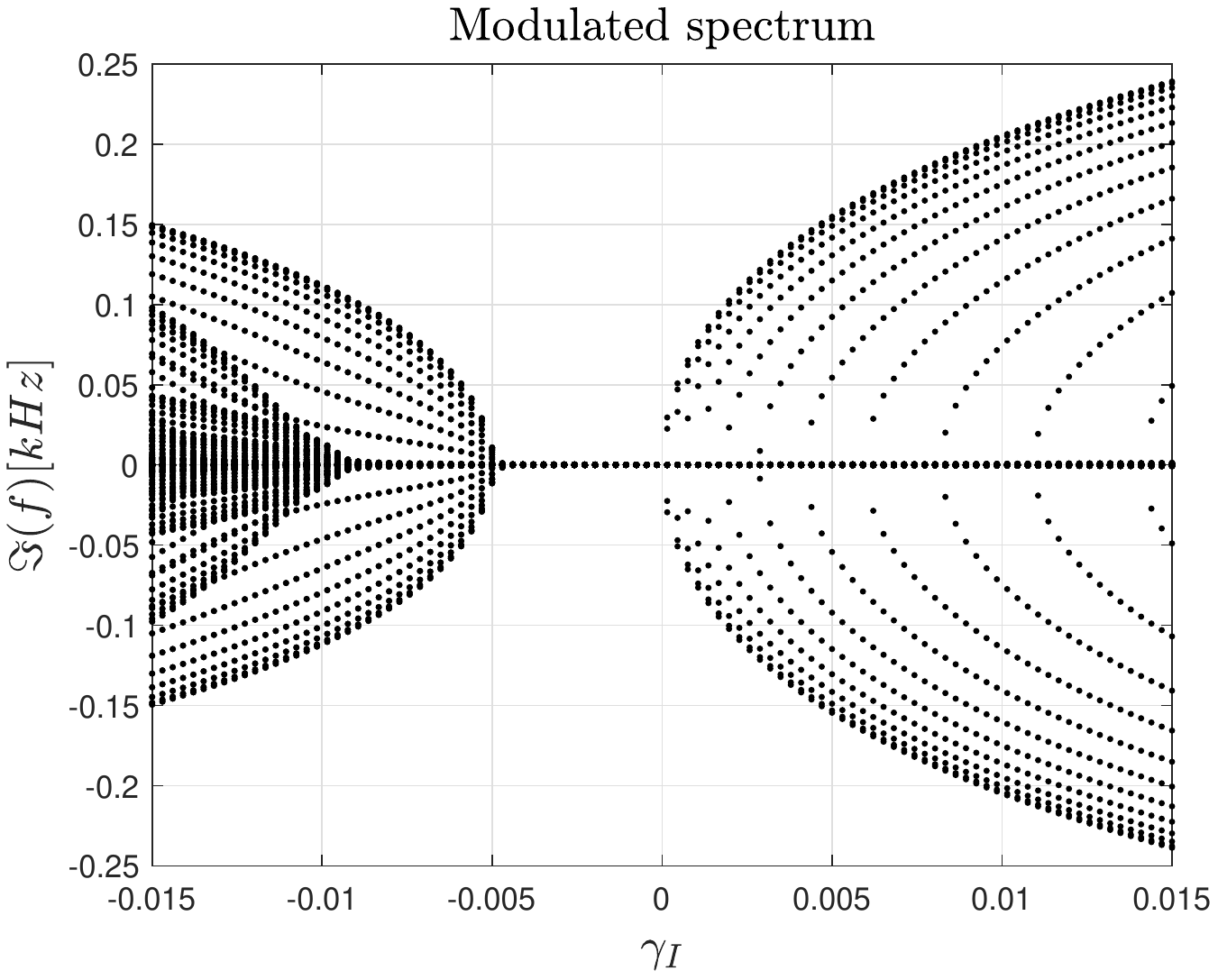}\label{fig:parametricb}}
\caption{Imaginary part of the spectrum against control gain for integral feedback under (a) PBC (b) OBC.}
\label{fig:parametric}
\end{figure}

  Figure \ref{fig:parametric} depicts the imaginary part of frequency, which is, as shall become clear, of particular interest in this analysis, for a purely integral local feedback case, with control gain varying in the interval $-0.015<\gamma_I<0.015$.
 
 From a practical point of view, it is interesting to find values of feedback gain that give rise to structures endowed with the edge states resulting from the NHSE and, at the same time, are stable when externally excited. Thus, the definition of stability that needs to be verified is the input-to-state stability, particularly the bounded-input-bounded-output (BIBO) stability \cite{khalil2015nonlinear}. Nonetheless, only a sufficient condition will be used herein. It is known that matrix $\mathbf{A_{cl}}$ of the state-space representation related to the FEM model of the system (see appendix, Eq.~(\ref{SH})) should be \textit{Hurwitz}, which means that the square matrix has all its eigenvalues $s$ with strictly negative real parts. In other words, the system in Eq.~(\ref{SH}) should be asymptotically stable. This implies that $f$ has a strictly positive imaginary part. 
 
 The NHSE will occur when the dispersion is non-trivial. A necessary condition is that the imaginary part of the dispersion relation admits non-zero values. We can verify in Fig.~\ref{fig:parametrica} that any non-zero gain of this particular feedback law satisfies this condition. On the other hand, following the stability condition described in the previous paragraph, no gain can provide strictly positive eigenfrequencies for the metastructure under OBC. At this point, one can relax the condition and look for eigenfrequencies with non-negative imaginary parts. By doing so and looking at Fig.~\ref{fig:parametricb}, it can be seen that within the range of values of $-0.005 < \gamma_I < 0$, the modes of the metastructure have almost zero imaginary part. Purely proportional and derivative cases, however, seem to show no range of gains for potentially stable structures, although further investigation is required.
 
\begin{figure}[H]
\centering
\subfigure[]{\includegraphics[width=0.495\textwidth]{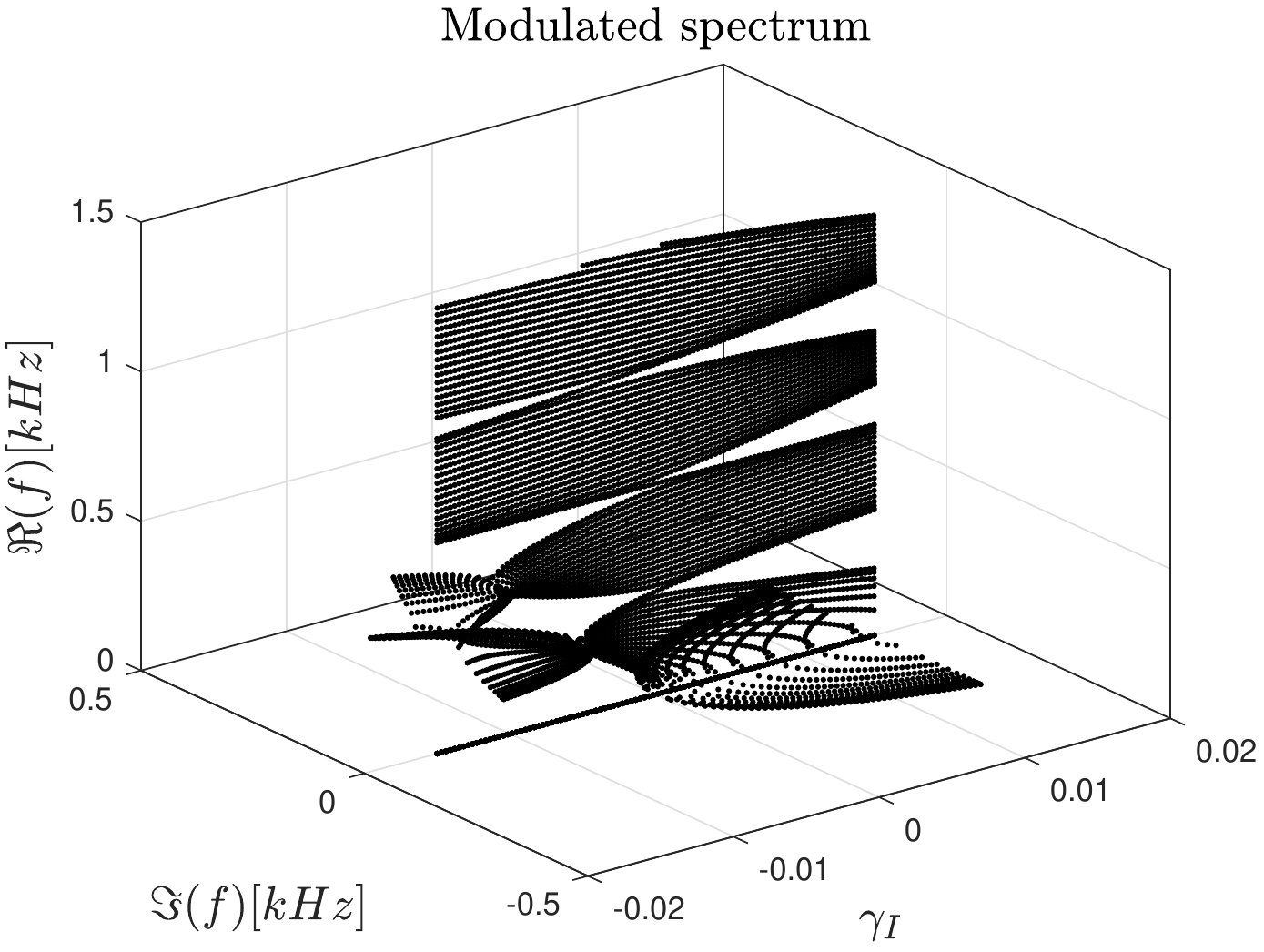}\label{OBCa}}
\subfigure[]{\includegraphics[width=0.495\textwidth]{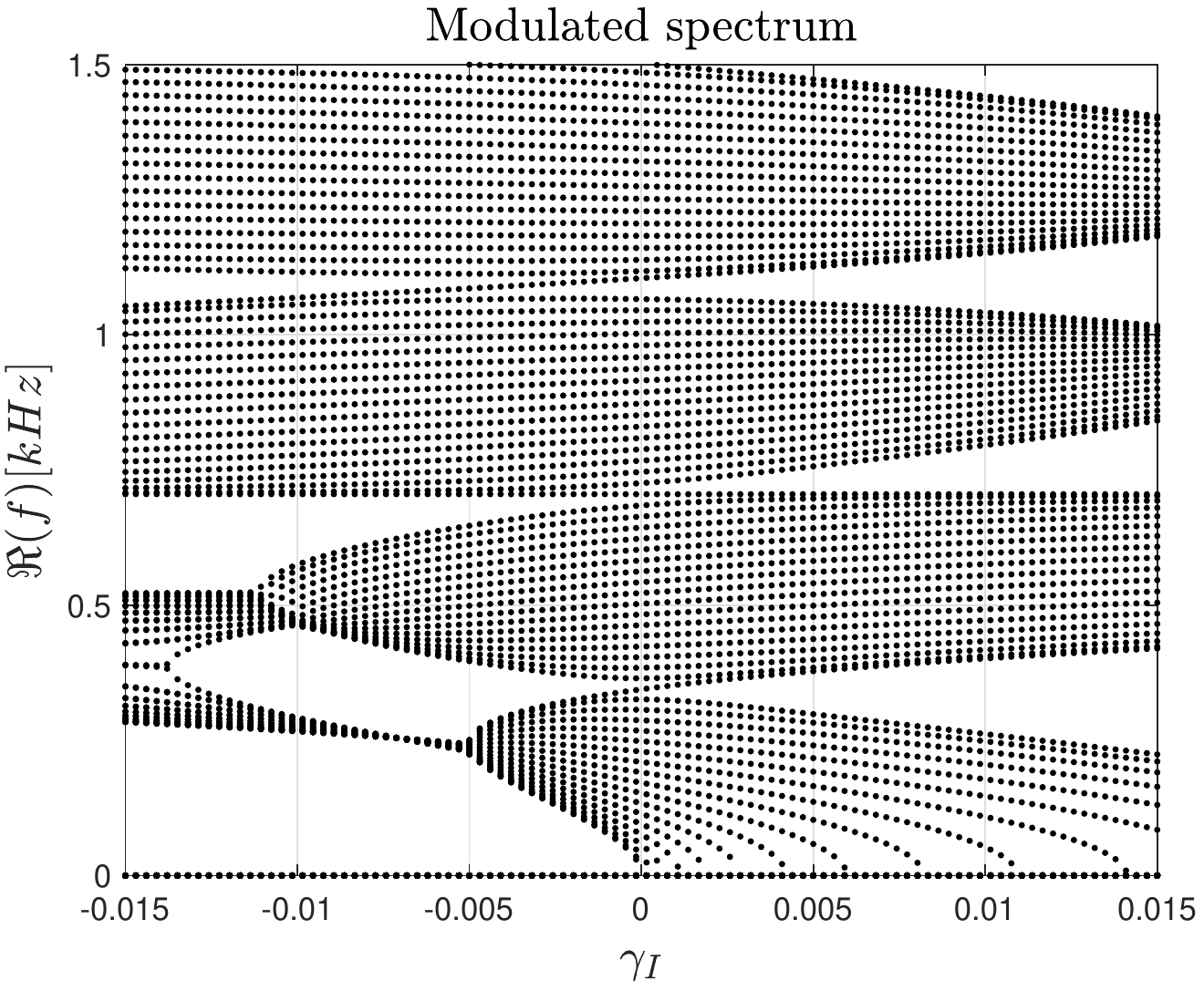}\label{OBCb}}
\caption{spectra against control gain for integral feedback under OBC (a) complex frequency against control gain (b) real frequency against control gain.}
\label{fig:OBC}
\end{figure}

Figure~\ref{fig:OBC} depicts both the complex frequency and its real part against the feedback gain for the system under OBC. The 3D plot in Fig.~\ref{OBCa} allows to observe that the unstable modes pictured in Fig.~\ref{fig:parametricb} correspond to lower frequencies (first and second BB). In Fig.~\ref{OBCb}, the BB and stop bands can be distinguished. For the chosen frequency range, four BB can be distinguished, with the first two becoming narrower as opposed to growing band gaps for negative values of feedback gain. This means that some eigenmodes coalesce.  

\begin{figure}[H]
\centering
\subfigure[]{\includegraphics[width=0.495\textwidth]{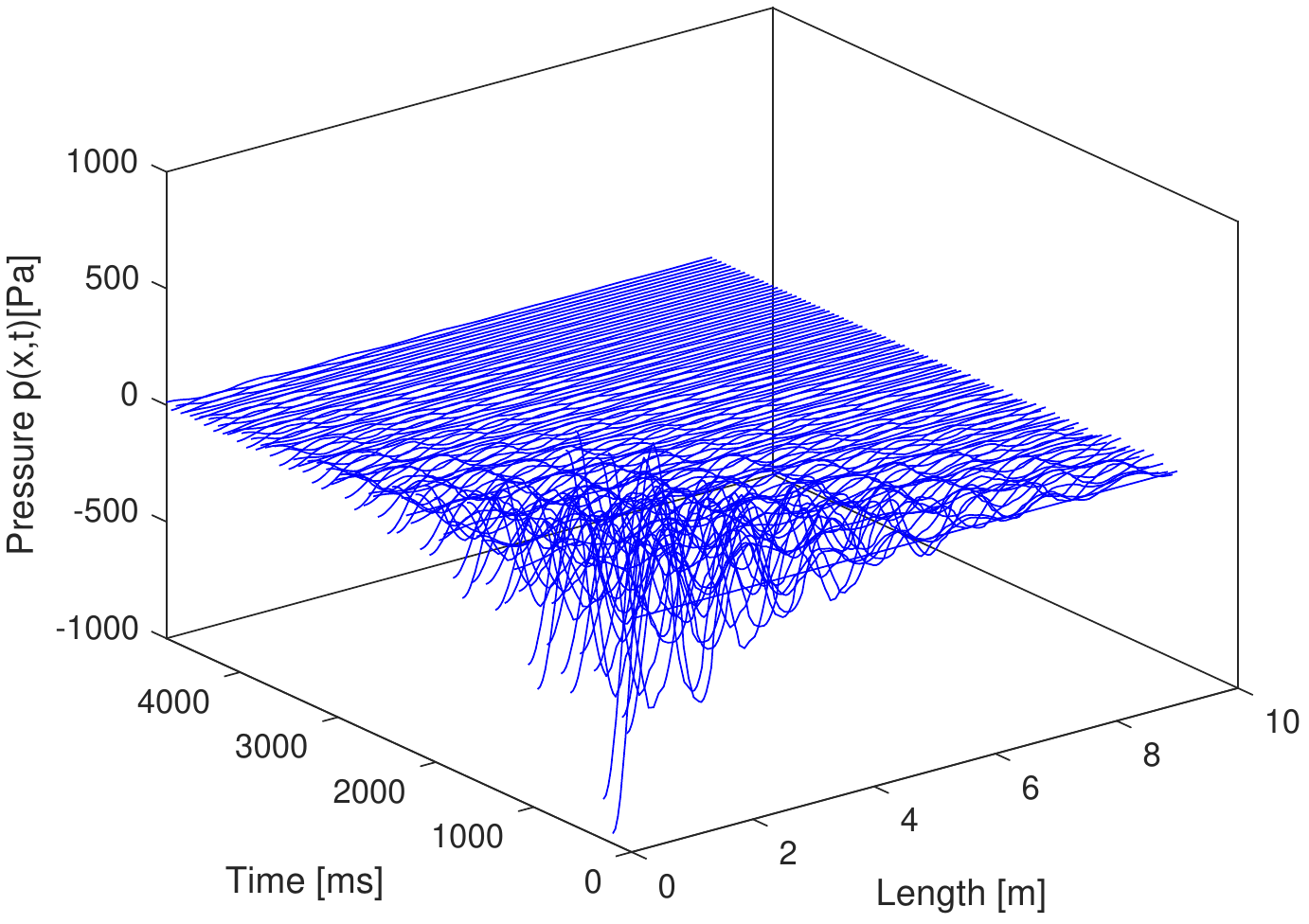}\label{transienta}}
\subfigure[]{\includegraphics[width=0.495\textwidth]{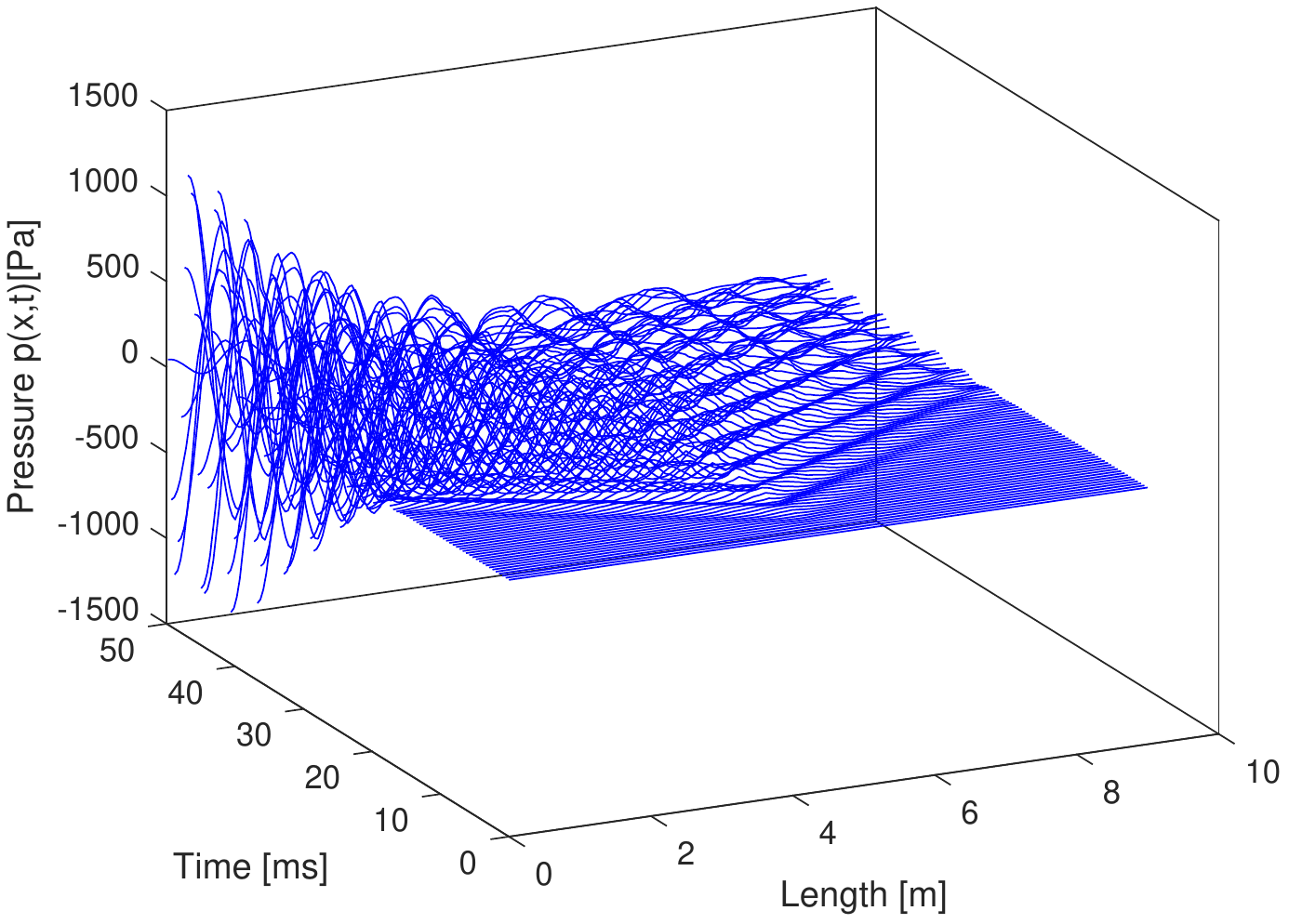}\label{transientb}}
\caption{Transient responses. (a) Stable response. (b) Zoom to highlight non-reciprocal wave propagation resulting from the NHSE.}
\label{fig:stableresponse}
\end{figure}

\begin{figure}[H]
\centering
\subfigure[]{\includegraphics[width=0.495\textwidth]{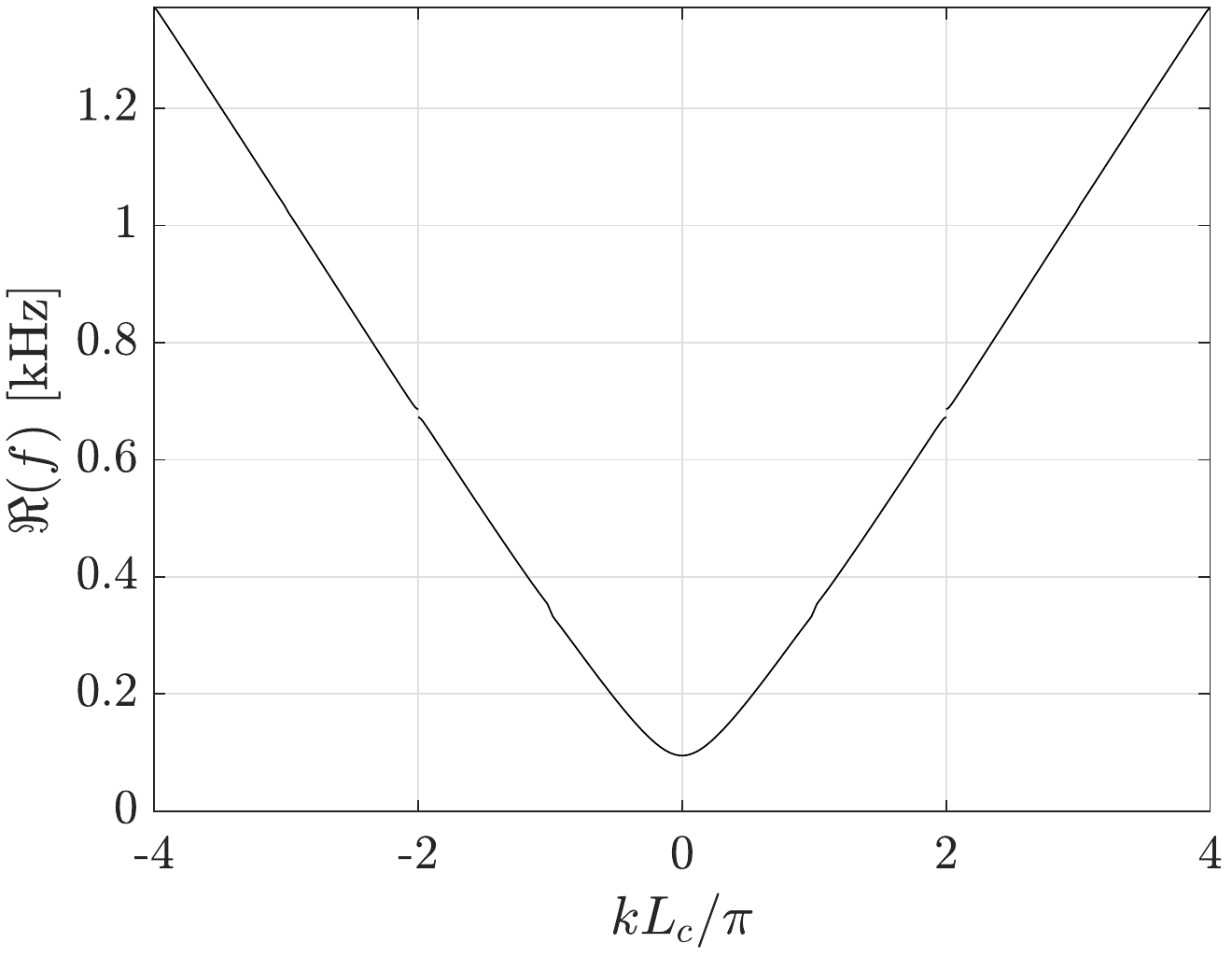}\label{dispersiona}}
\subfigure[]{\includegraphics[width=0.495\textwidth]{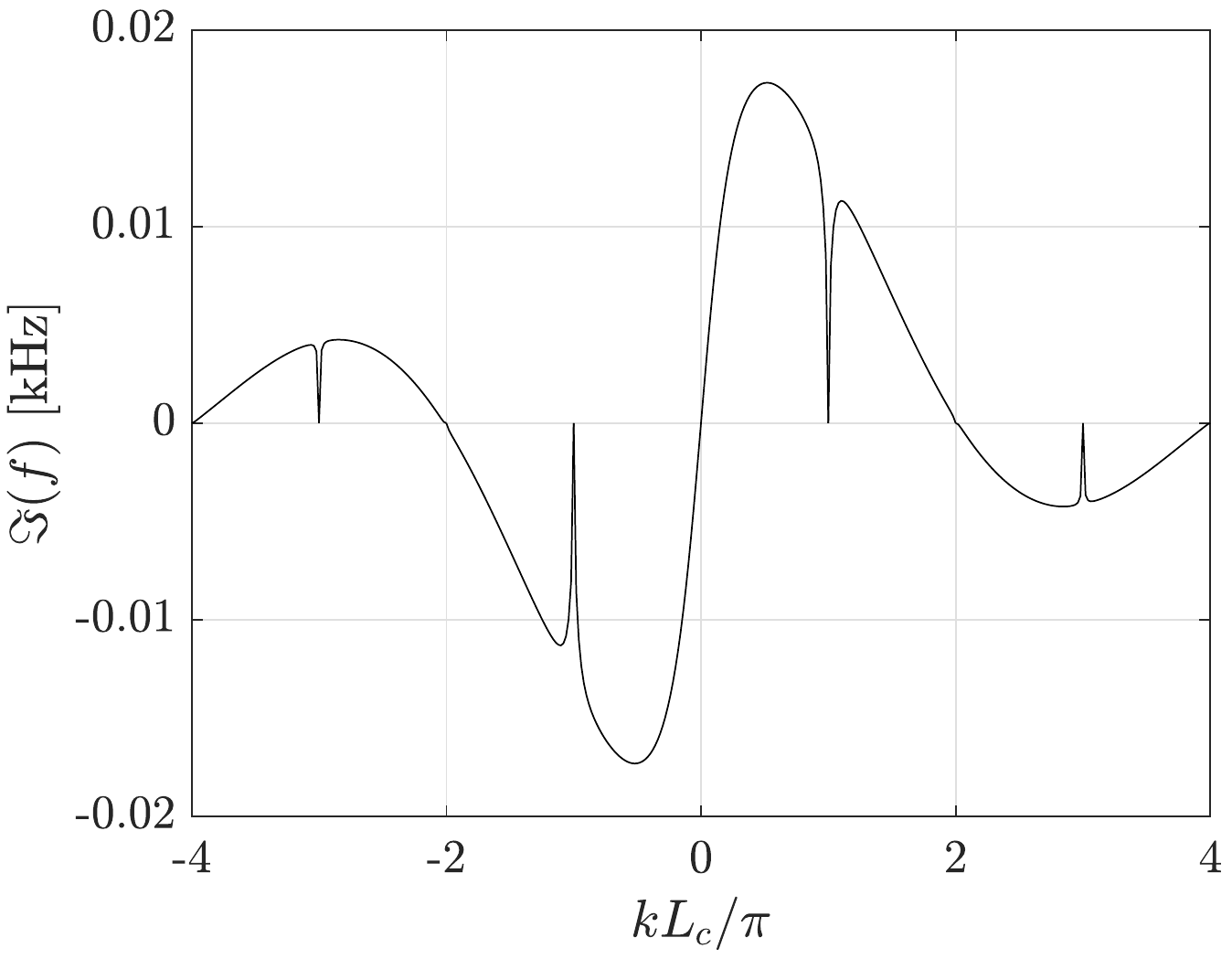}\label{dispersionb}}
\caption{ Dispersion relation depicted as the (a) real frequency against wavenumber and (b) imaginary frequency against wavenumber.}
\label{fig:dispersion}
\end{figure}

 Figure~\ref{fig:stableresponse} shows that setting $\gamma_I = -0.0015$, the transient response  is indeed stable regarding the tone-burst excitation at the middle of the metastructure, with a central frequency of $250 Hz$. Non-reciprocity is achieved due to the NHSE, as illustrated by the concentration of energy, resulting in the increase of pressure at $x_0$. This response could also be predicted by the imaginary part of the dispersion diagram \cite{braghini2021non} presented in Fig.~\ref{dispersionb}, which is not symmetric with respect to the wavenumber axis. This implies attenuation for positive wavenumbers (forward traveling waves) and amplification for negative wavenumbers (backward traveling waves). 
 
\begin{figure}[H]
\centering
\subfigure[]{\includegraphics[width=0.495\textwidth]{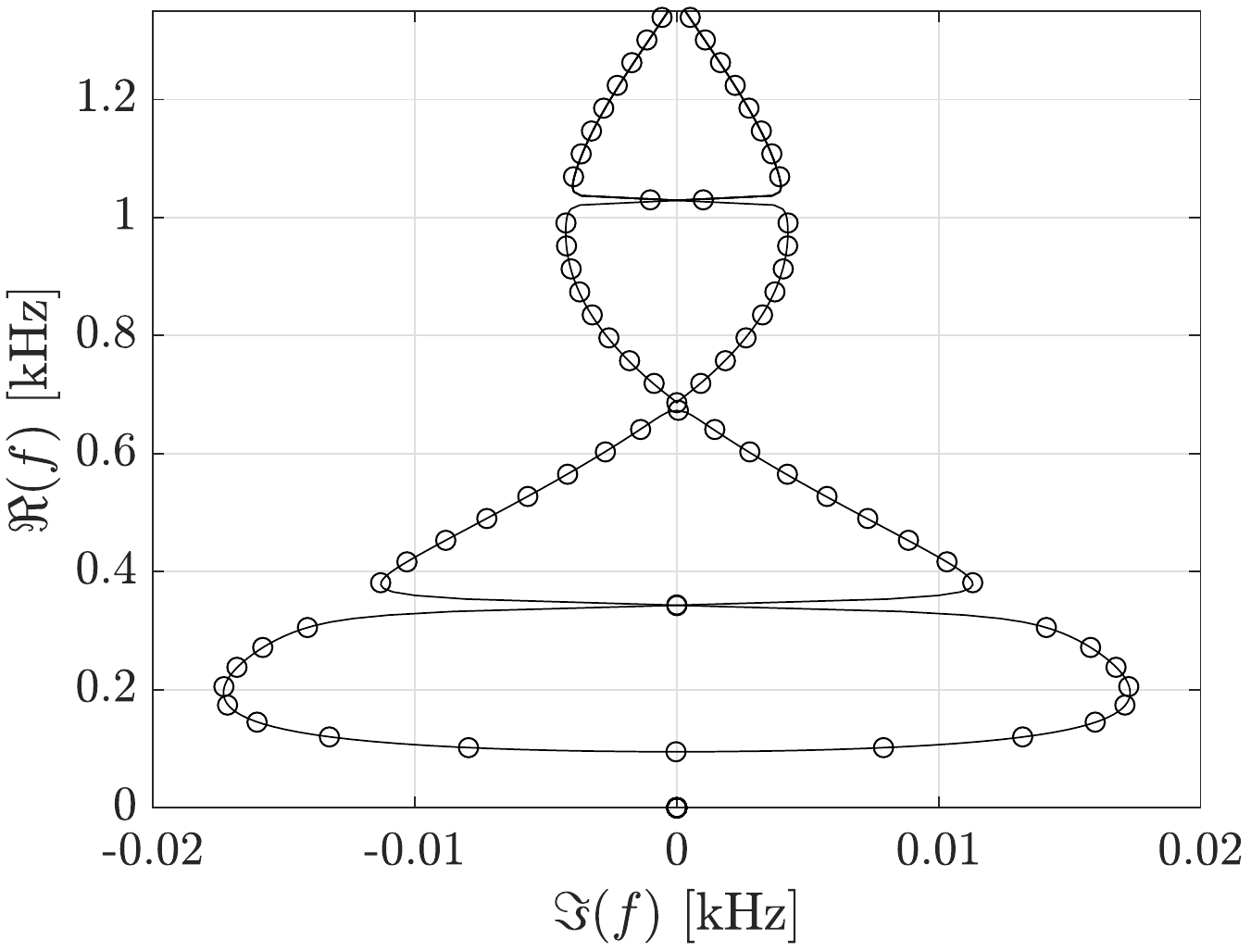}\label{dispersion2a}}
\subfigure[]{\includegraphics[width=0.495\textwidth]{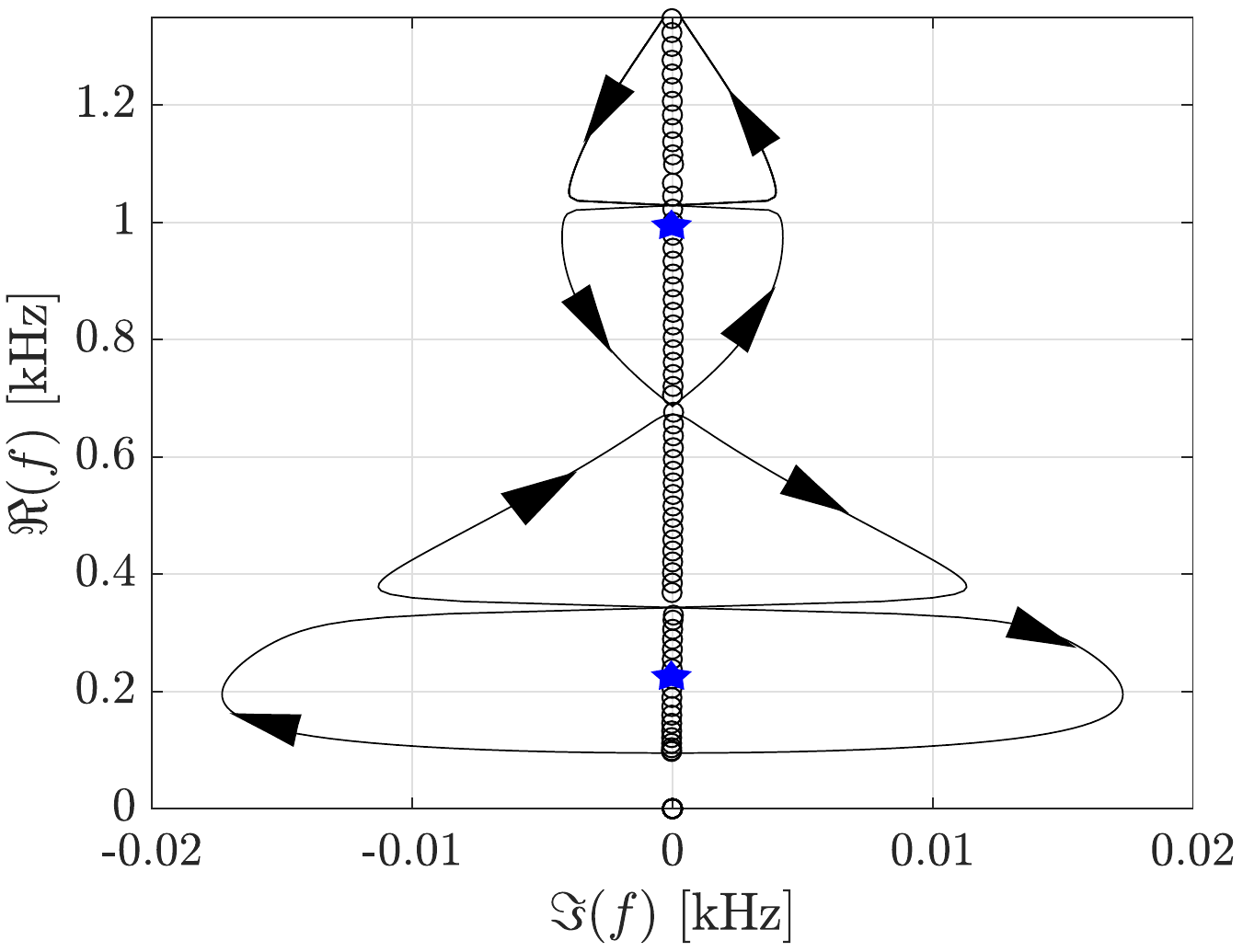}\label{dispersion2b}}
\caption{Complex frequency plane for the system with the feedback defined in Eq.~(\ref{eq:pidfb}), which is local ($a=0$) and integral type ($\gamma_D = \gamma_P = 0$), with gain $\gamma_I = -0.0015$. Dispersion relations (solid curves) and eigenmodes of the structure (circles) are compared under (a) PBC and (b) OBC. Black arrows are used to indicate the sign of the winding number and blue stars highlight particular skin modes.} 
\label{fig:bulkboundary}
\end{figure}

In Fig.~\ref{fig:bulkboundary}, the bulk-boundary relation for both kinds of boundary conditions is again confirmed, as shown in Fig.~\ref{fig:PBCxOBC}. In Fig.~\ref{dispersion2b}, the eigenfrequencies lie on the real axis, as it happens for Hermitian systems, confirming the sensitivity of the system with respect to the  boundary conditions. The arrows indicate negative winding number for the first and second bands, and positive winding number for the third and fourth bands.

\begin{figure}[H]
\centering
\subfigure[]{\includegraphics[width=0.495\textwidth]{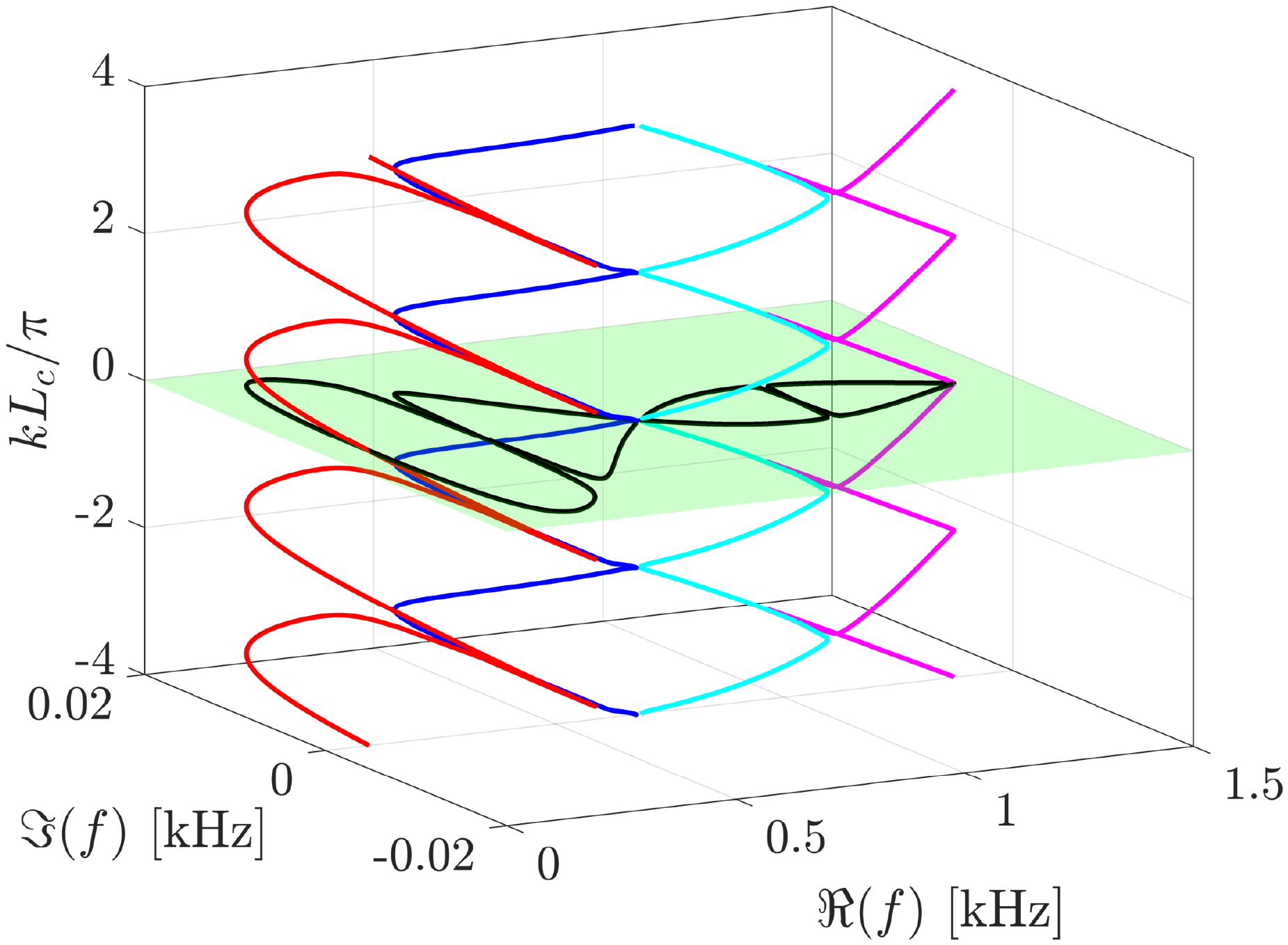}\label{3D1}}
\subfigure[]{\includegraphics[width=0.495\textwidth]{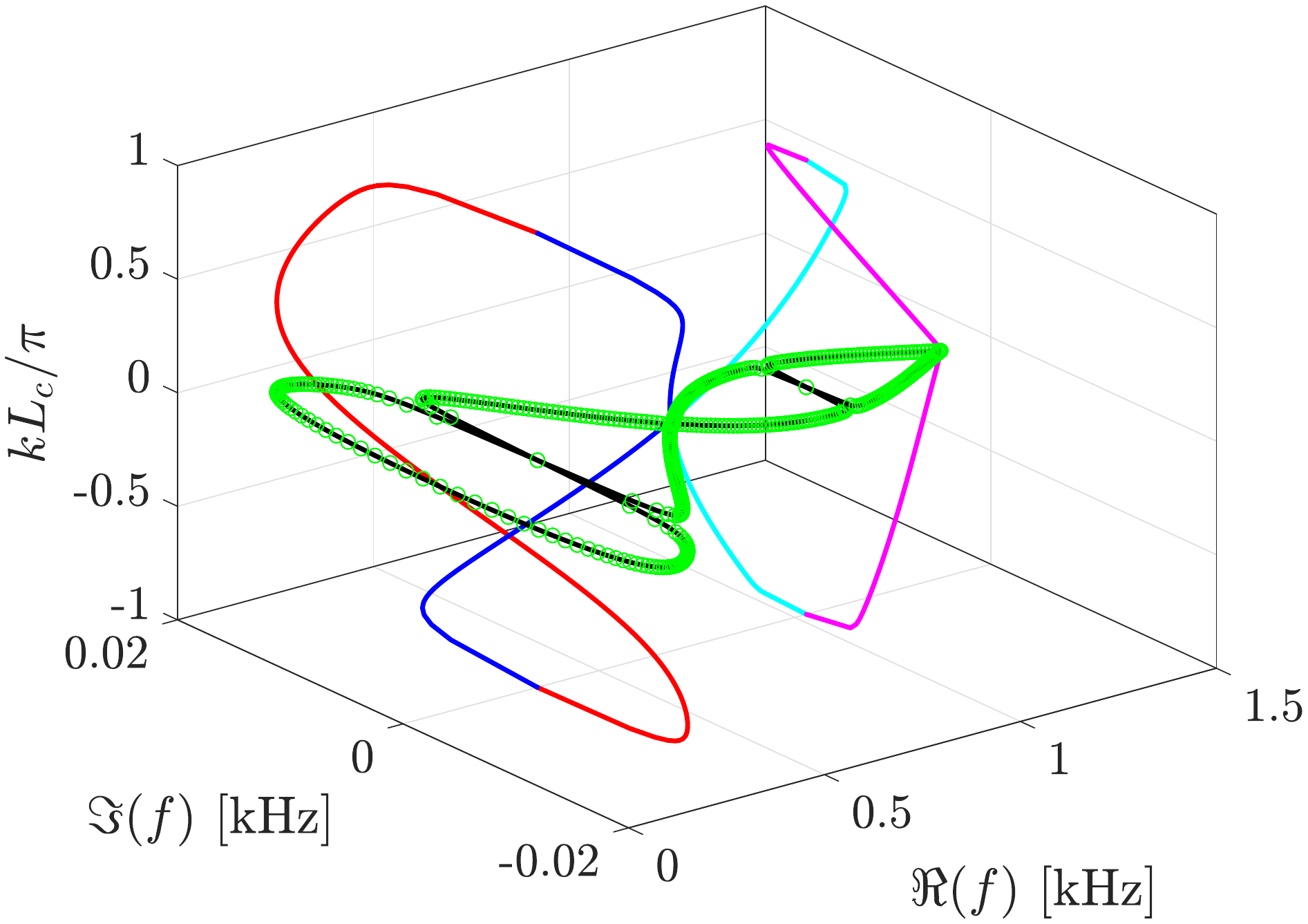}\label{3D2}}
\caption{3D plot of complex frequencies versus real wavenumber, spotting the projection on the complex plane, as shown in Fig.~\ref{fig:bulkboundary}. (a) Dispersion relation (solid colored lines, each color corresponding to a different band) and its projection on the complex frequency plane, in shaded green, by solid black lines. (b) Dispersion restricted to $BZ_1$ and eigenmodes of the structure (green circles) under PBC.}
\label{3D}
\end{figure}

The values of the winding number can be taken from the geometry of the curves drawn by the bands in the reciprocal space (Fig.~\ref{3D}), evolving from negative to positive wavenumber values through the first Brillouin Zone $ BZ_1 = [-\pi/L_c, \pi/L_c) $. The projection of these bands on the complex plane ($k=0$) together with the modes under PBC, as shown in Fig.~\ref{dispersion2a}, is depicted again with the eigenmodes highlighted as green circles. For the sake of visualization, each band is depicted in a different color. Note that the bands coincide at the edges of the Brillouin zones, i.e, whenever $k = n \pi$, $\forall n \in  \mathbb{N}$. As can be seen, both the first and second bands have winding number $\nu = -1$ (one clockwise rotation), whereas the third and forth bands present $\nu = 1$ (one counter-clockwise rotation), as indicated by the arrows in Fig.~\ref{dispersion2b}. Thus, one should expect localized modes at $x=0$ and $x=9m$ (the boundaries of the metastructure) for the first pair (1BB and 2BB) and the second pair (3BB and 4BB), respectively \cite{gong2018topological}.       

\begin{figure}[H]
\centering
\subfigure[]{\includegraphics[width=0.495\textwidth]{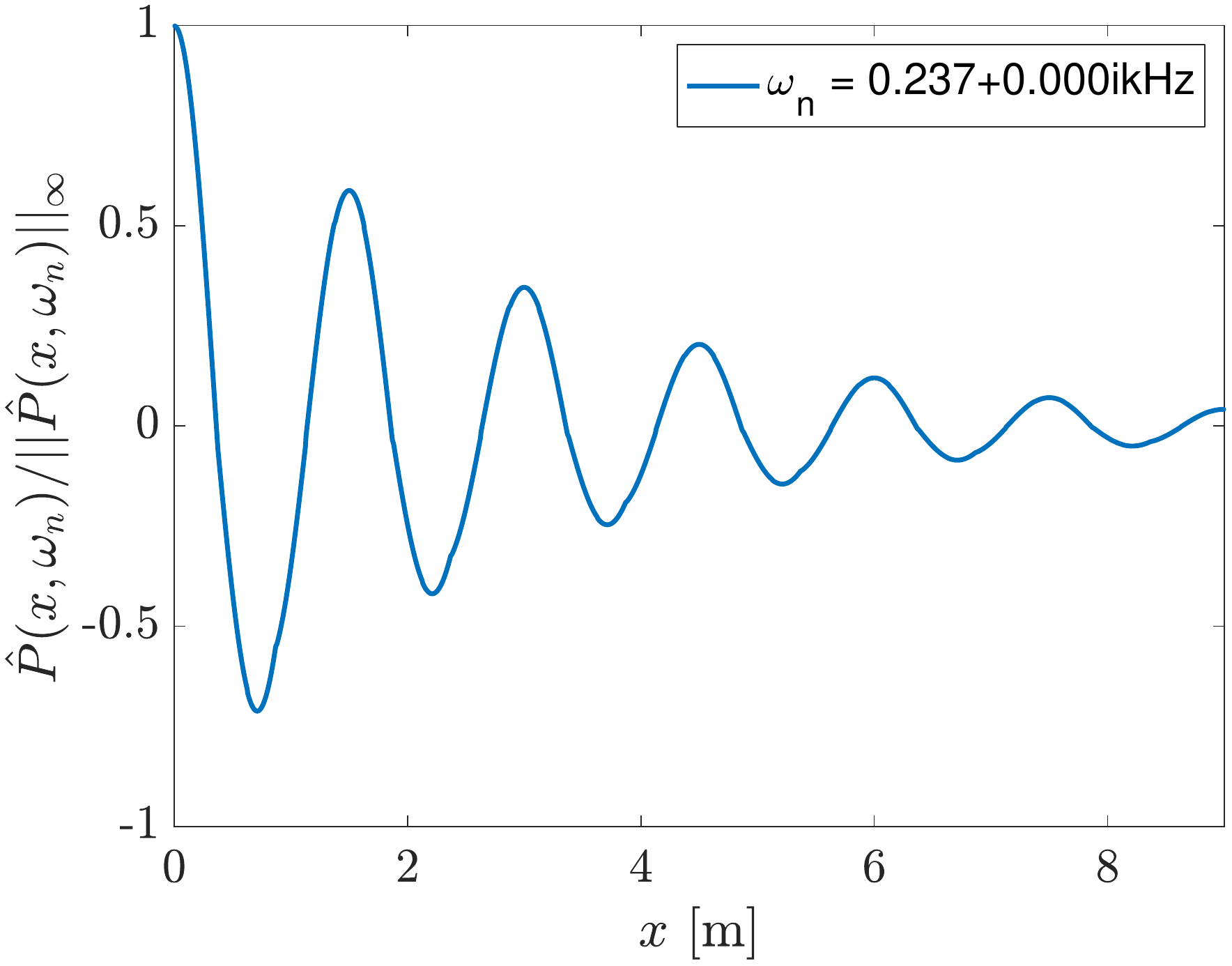}\label{skinmode1}}
\subfigure[]{\includegraphics[width=0.495\textwidth]{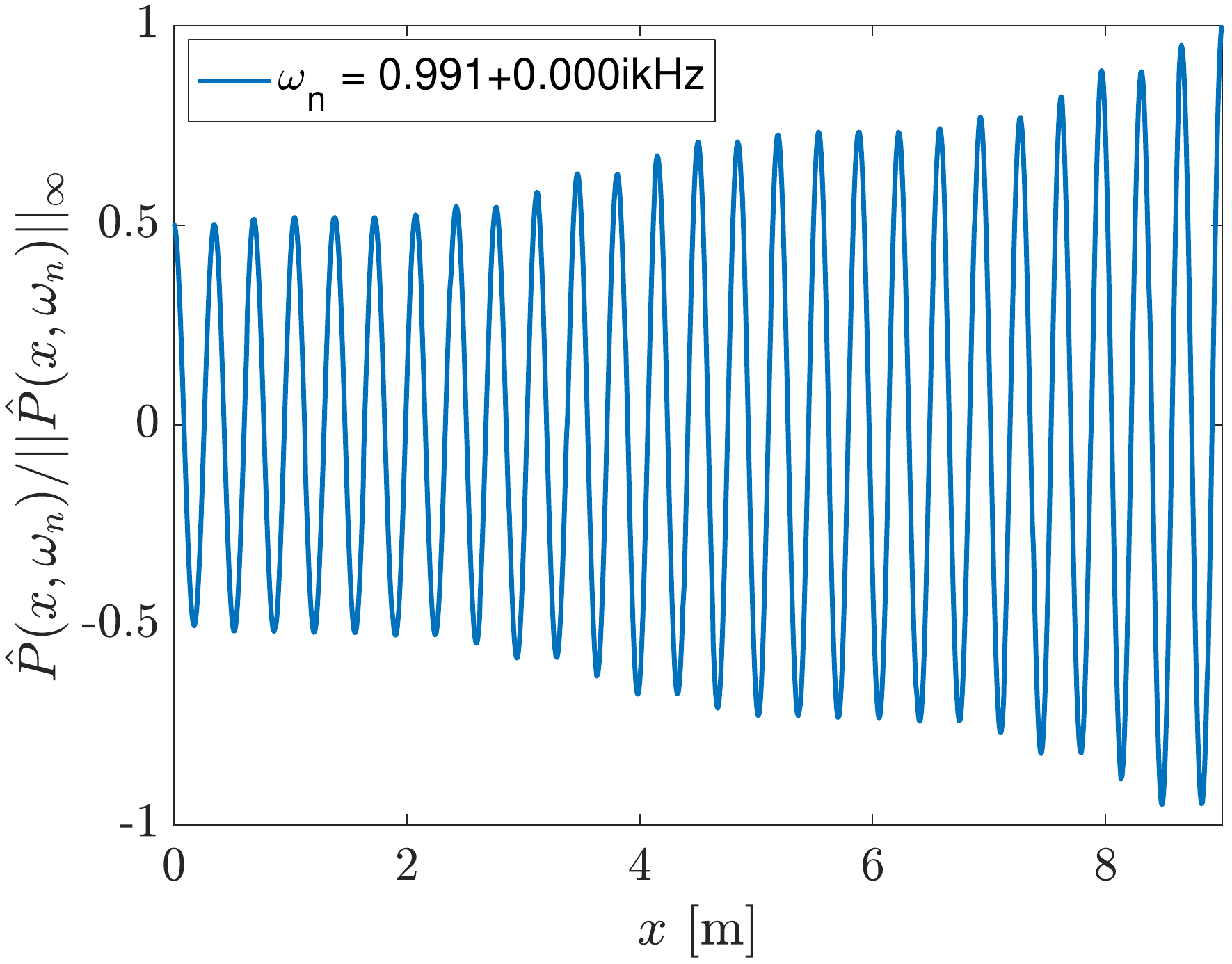}\label{skinmode2}}
\caption{Normalized skin modes corresponding to the purely real eigenfrequencies highlighted by blue stars in Fig.~\ref{dispersion2b}. (a) Localized mode at $x=0$ from 1BB, which has negative winding number and (b) localized mode at $x=9 m$ from 3BB, which has positive winding number.}
\label{fig:skinmodes}
\end{figure}

Fig.~\ref{fig:skinmodes} depicts the eigenmodes corresponding to the eigenfrequencies of the metastructure under OBC highlighted in Fig.~\ref{dispersion2b}, confirming the localization predicted by the topology of the bands.

\begin{figure}[H]
\centering
\subfigure[]{\includegraphics[width=0.495\textwidth]{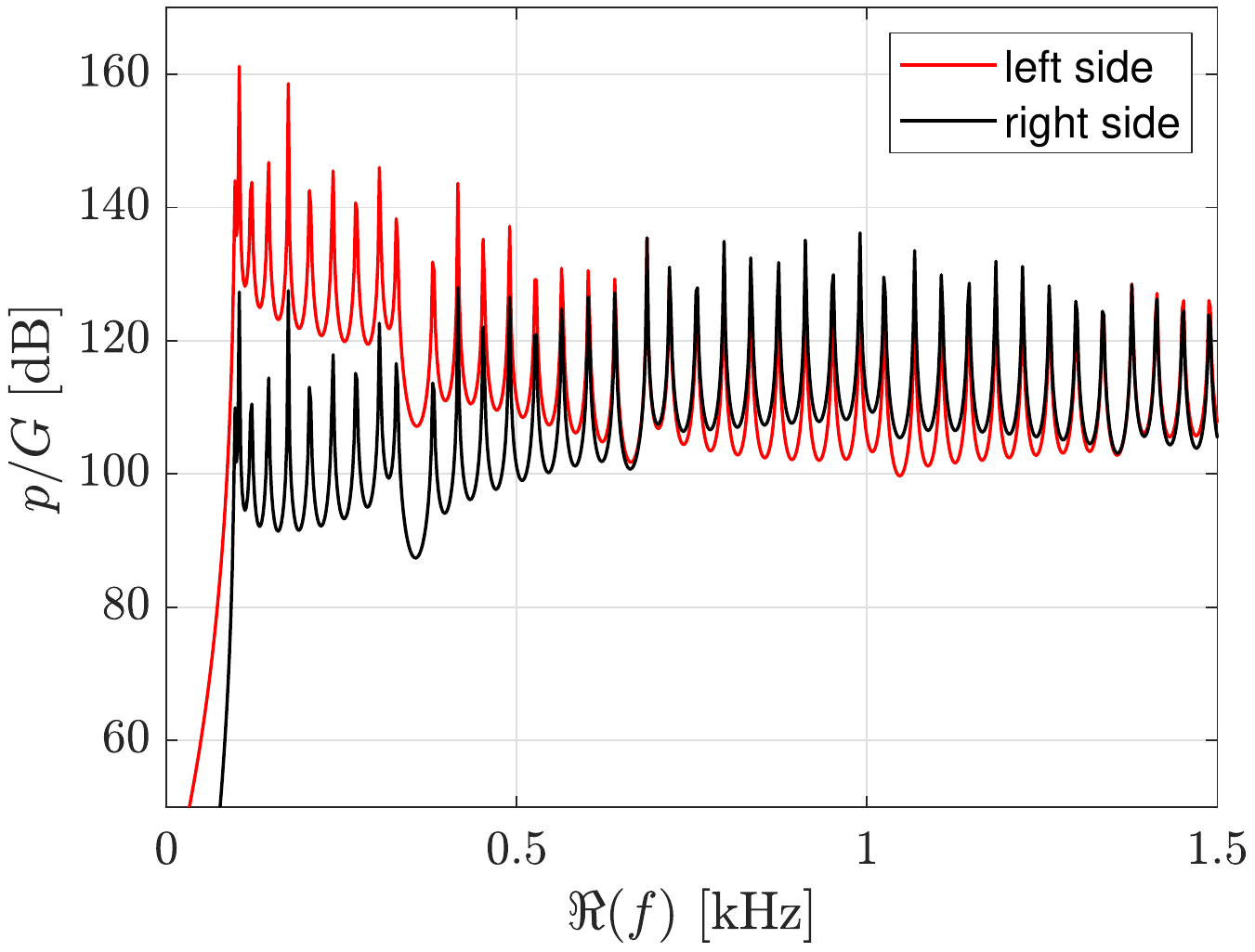}\label{FRFa}}
\subfigure[]{\includegraphics[width=0.495\textwidth]{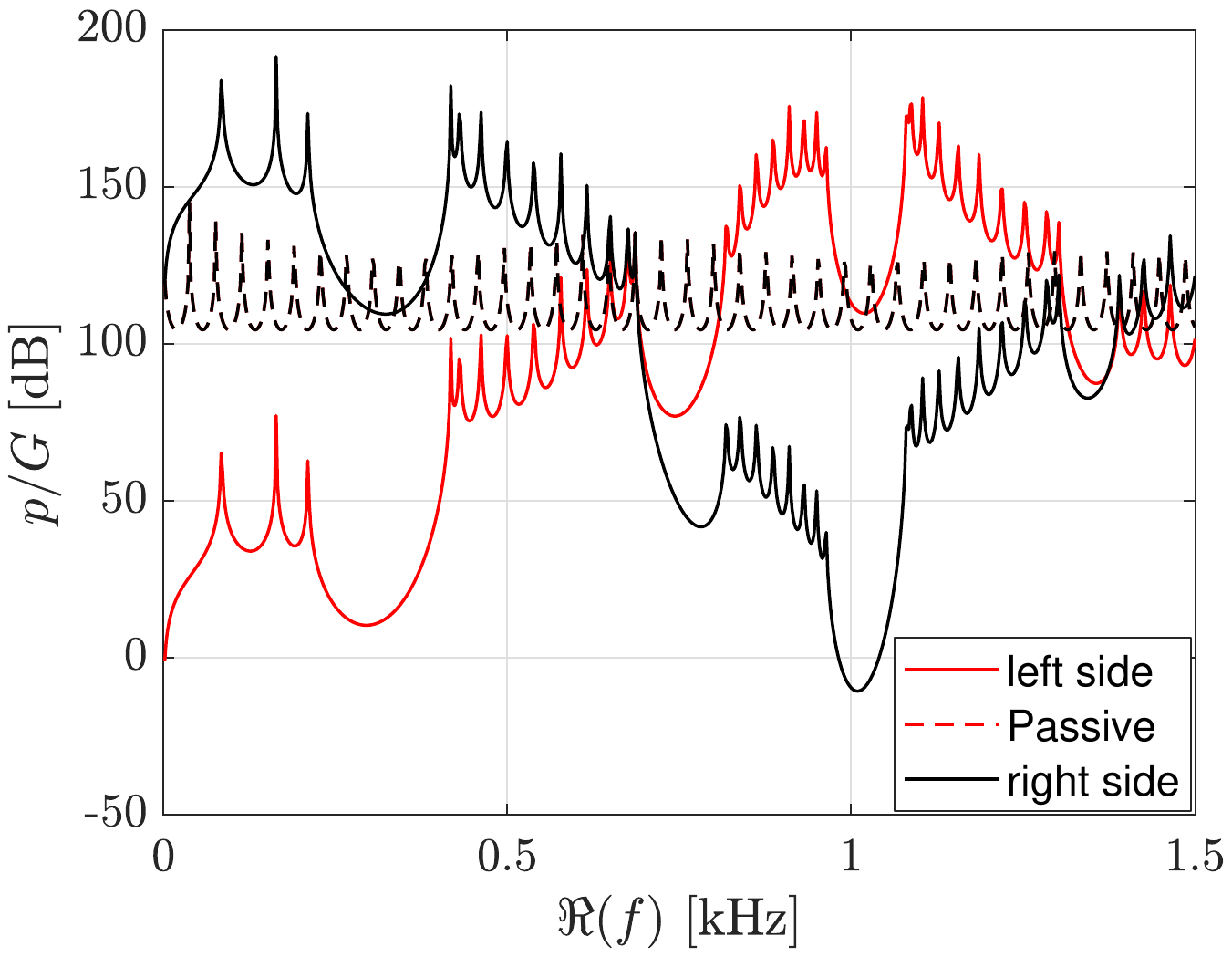}\label{FRFb}}
\caption{FRF for integral feedback under OBC (a) $k_i = -0.0015$ (b) $k_i = 0.015$. }
\label{fig:FRFi}
\end{figure}

In Fig.~\ref{fig:FRFi}, the analysis of the Frequency Response Function (FRF) for two integral gains was used to depict the difference in energy flow through the structure. It shows the response at each extremity of the structure with a different color: red for the measurement at the left end and black for the measurement at the right end. Both responses were simulated with excitation at the middle of the structure.

One can notice that the energy localization on the left or on the right side of the structure, depending on each BB, as previously observed in \cite{braghini2021non}. It can be inferred that, in this system, energy flows to the right side at lower frequencies, whereas it flows to the left side at higher frequencies. This effect becomes clear in Fig.~\ref{FRFb}, with a higher value of the feedback gain, and, thus, more energy applied to the system. In this Figure, the dashed lines are the FRF corresponding to the passive structure. As expected, the response of the passive structure is equal at both ends, and, thus, this structure does not manifest spatial concentration of vibration.

It is important to highlight that the parametric analysis of Fig.~\ref{fig:parametric} was performed for different feedback laws: purely proportional, derivative, integral, and combinations of those. Still, only the integral case showed a range of gains for which the structure is certified to be stable, according to the conservative, yet sufficient, definition used here. To get around this issue, band-pass filters were designed and added to the feedback loop. Nonetheless, any realizable filter (non-ideal) showed a negative effect on the topology of the reciprocal space, dramatically reducing the non-reciprocity and localization of edge states.

\subsection{Long-range (non-local)  coupling}
\label{longerange}

\begin{figure}[H]
\centering
\subfigure[]{\includegraphics[width=0.495\textwidth]{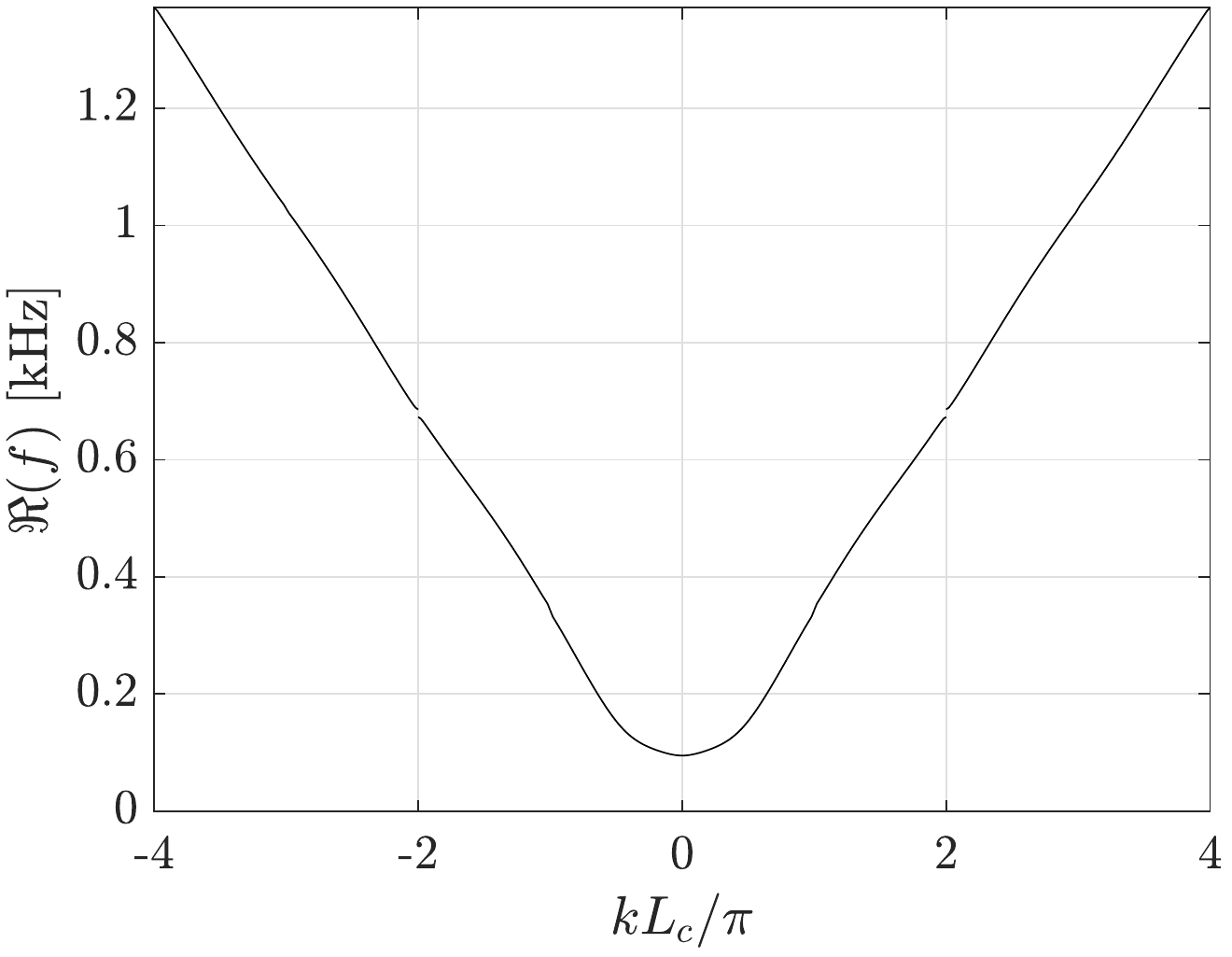}\label{dispersion1a=1}}
\subfigure[]{\includegraphics[width=0.495\textwidth]{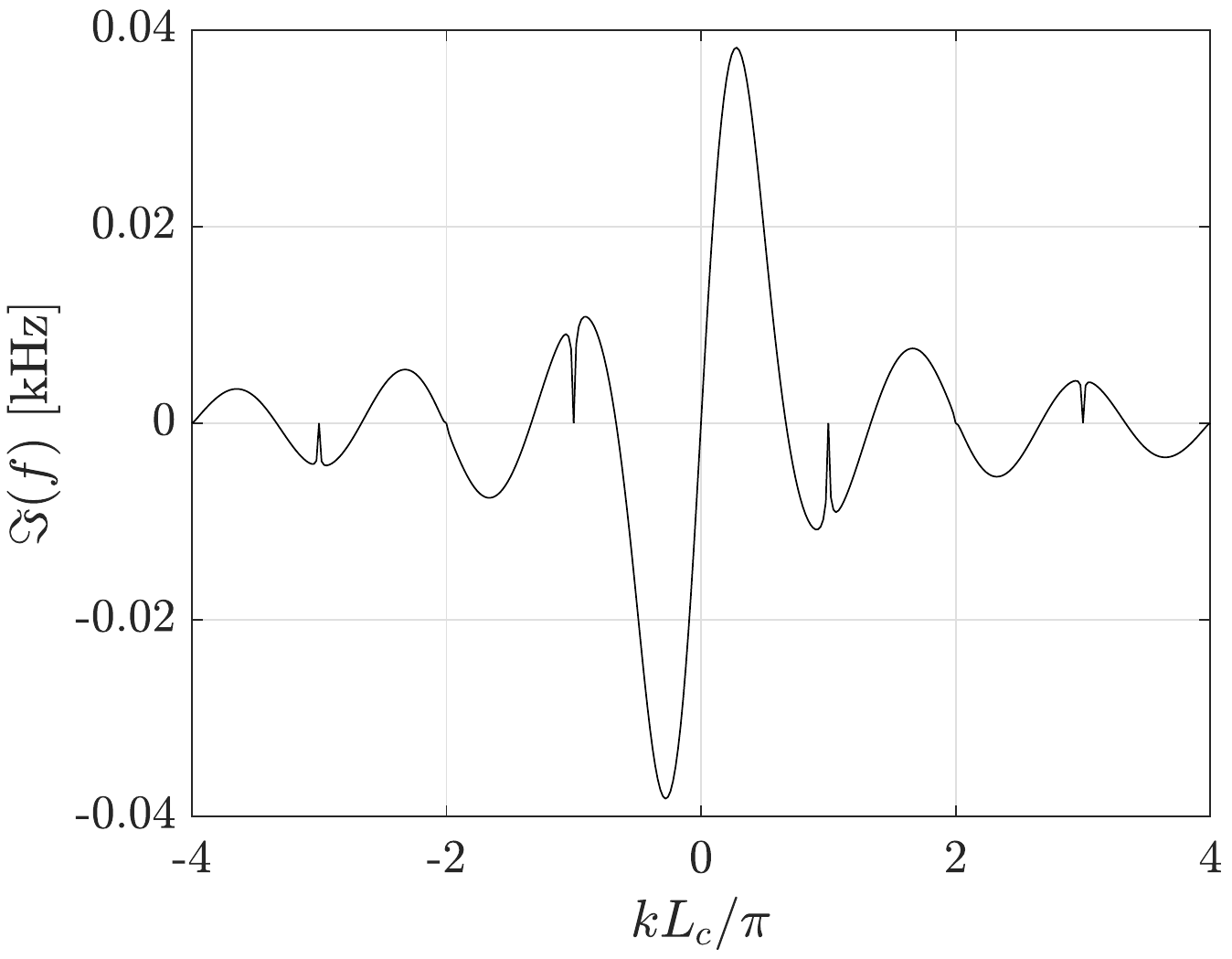}\label{dispersion2a=1}}
\caption{First four BB of the dispersion relation for non-local integral feedback with $k_i = -0.0015$ and $a=1$ (a) real frequency against wavenumber (b) imaginary frequency against wavenumber. }
\label{fig:dispersion_nonlocal1}
\end{figure}

\begin{figure}[H]
\centering
\subfigure[]{\includegraphics[width=0.495\textwidth]{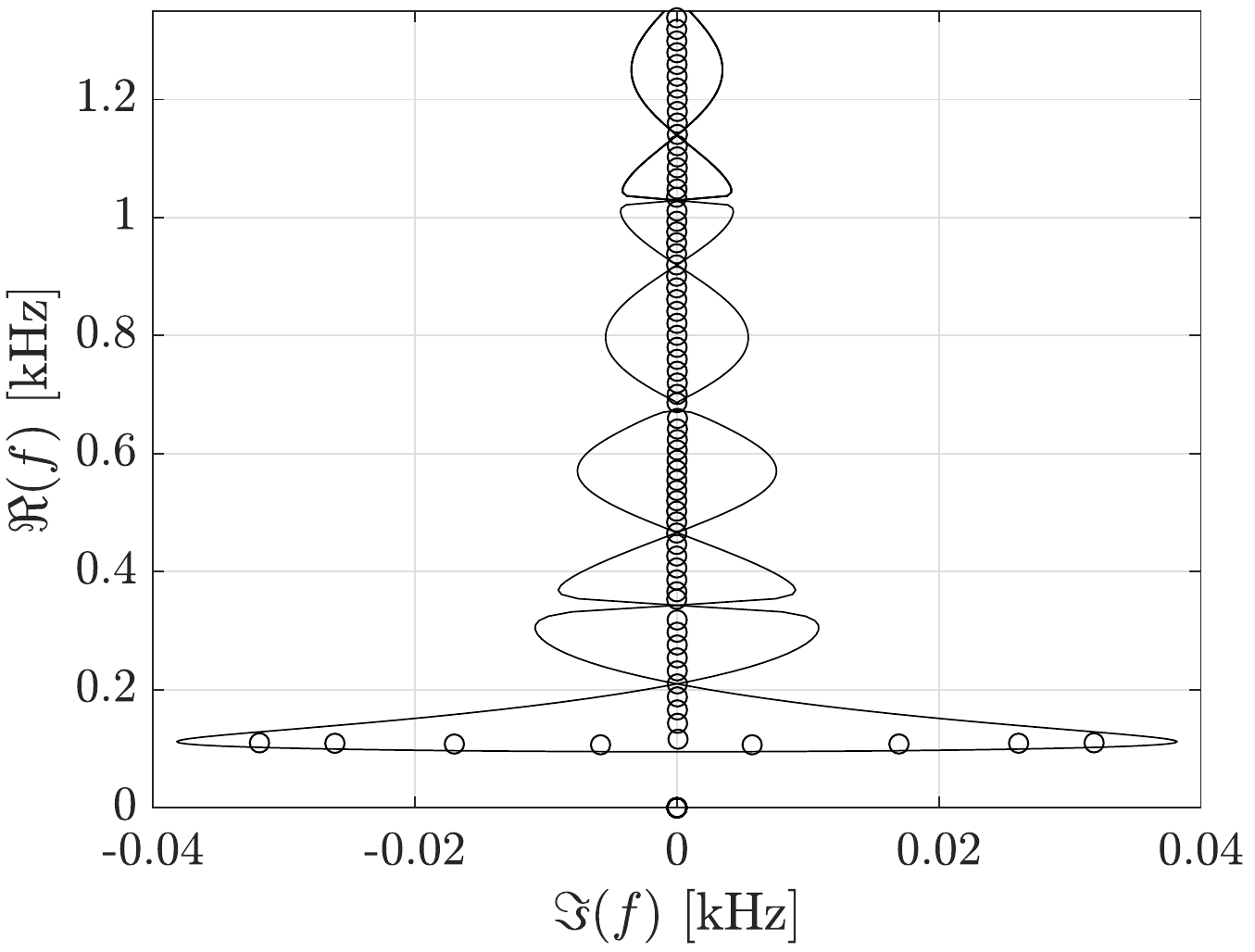}\label{bulkboundarya=1}}
\subfigure[]{\includegraphics[width=0.495\textwidth]{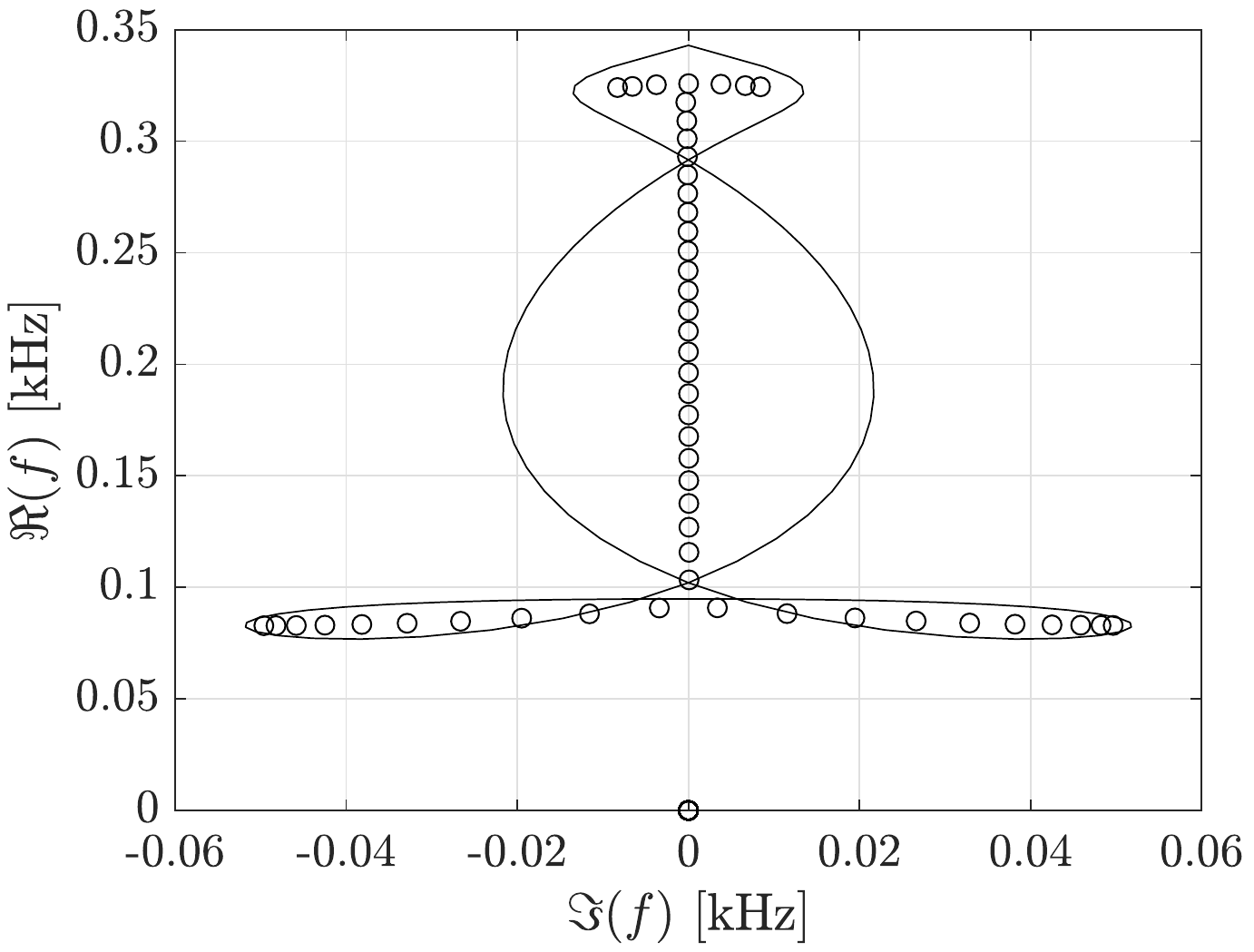}\label{bulkboundara=2}}
\caption{Bulk-boundary correspondence for non-local integral feedback with $k_i = -0.0015$ and (a) $a=1$(b) $a = 2$. }
\label{fig:bulkboundary_nonlocal}
\end{figure}

By applying non-local feedback interactions we can emulate long-range, non-reciprocal coupling in the metamaterial \cite{wang2021generating}. For instance, setting the range $a=1$, meaning a distance of one cell between the sensor and the actuator, and the same configuration defined in section \ref{stable}, the resulting dispersion for four BB is displayed in Fig.~\ref{fig:dispersion_nonlocal1}. Fig.~ \ref{fig:bulkboundary_nonlocal} shows the eigenmodes of a metastructure under OBC (free-free boundary conditions).

\begin{figure}[H]
\centering
\subfigure[]{\includegraphics[width=0.495\textwidth]{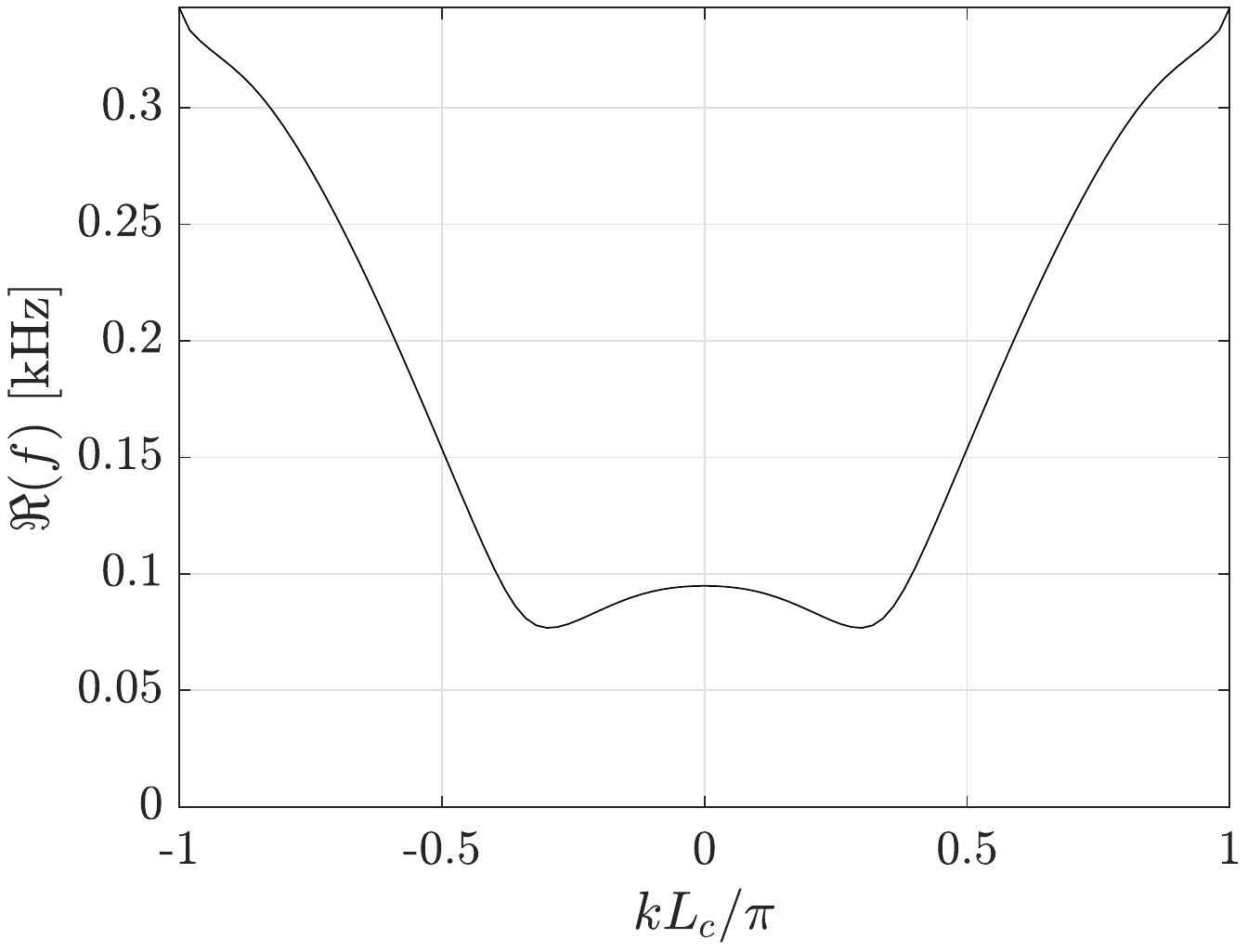}\label{dispersion1a=2}}
\subfigure[]{\includegraphics[width=0.495\textwidth]{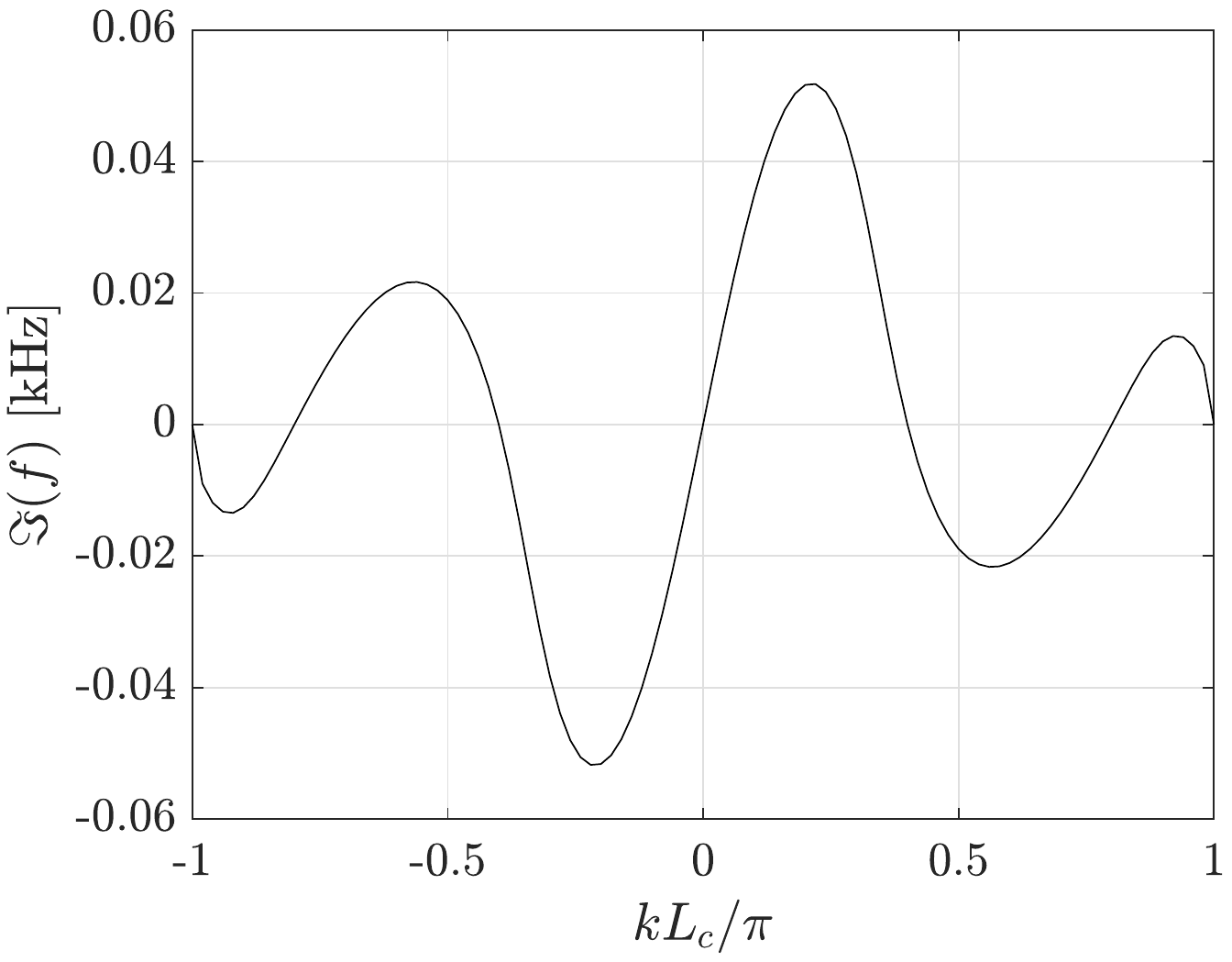}\label{dispersion2a=2}}
\caption{First BB of the dispersion relation for non-local integral feedback with $k_i = -0.0015$ and $a=2$ (a) real frequency against wavenumber (b) imaginary frequency against wavenumber.}
\label{fig:dispersion_nonlocal2}
\end{figure}

By using this strategy, each individual band splits into multiple regions with alternating directions of wave amplification and attenuation. The degree of this splitting is proportional to the locality parameter $a$ \cite{braghini2021non}. To exemplify, Fig.~\ref{fig:dispersion_nonlocal2} shows only the first band split in four regions by setting $a=2$.

To confirm again the odd bulk-boundary correspondence of NH systems, Fig.~\ref{fig:bulkboundary_nonlocal} shows the eigenmodes of the structures from previous non-local feedback designs over the corresponding dispersion relations. Four bands are depicted on Fig.~\ref{bulkboundarya=1} and just the first band on Fig.~\ref{bulkboundara=2}.

\subsection{Comparing methods}

In this section, different numerical methods are compared to corroborate previous numerical results. Fig.~\ref{fig:pwe} compares SEM and PWE methods of computing the dispersion relations, as described in the Appendix.

\begin{figure}[H]
\centering
\subfigure[]{\includegraphics[width=0.495\textwidth]{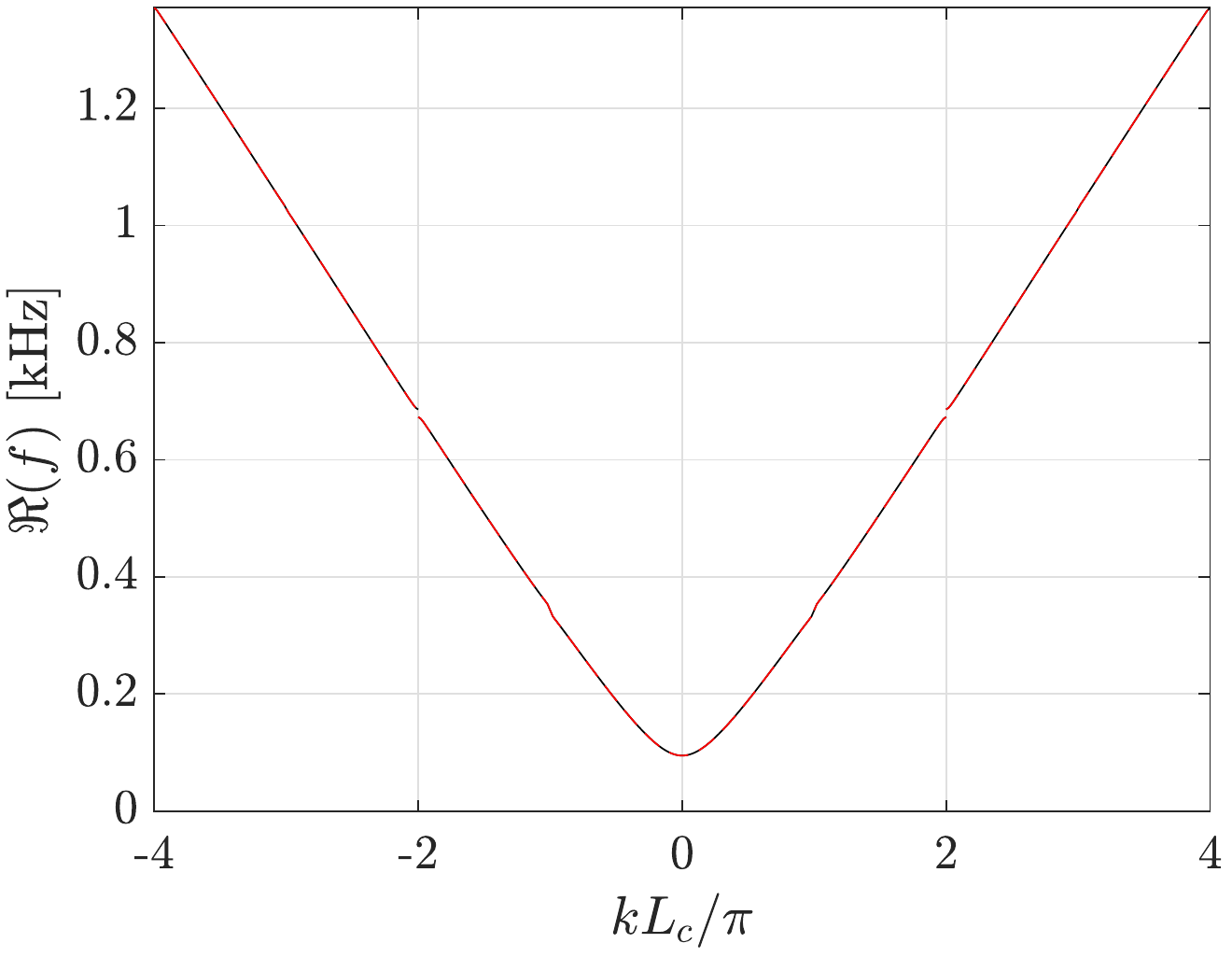}\label{pwe1}}
\subfigure[]{\includegraphics[width=0.495\textwidth]{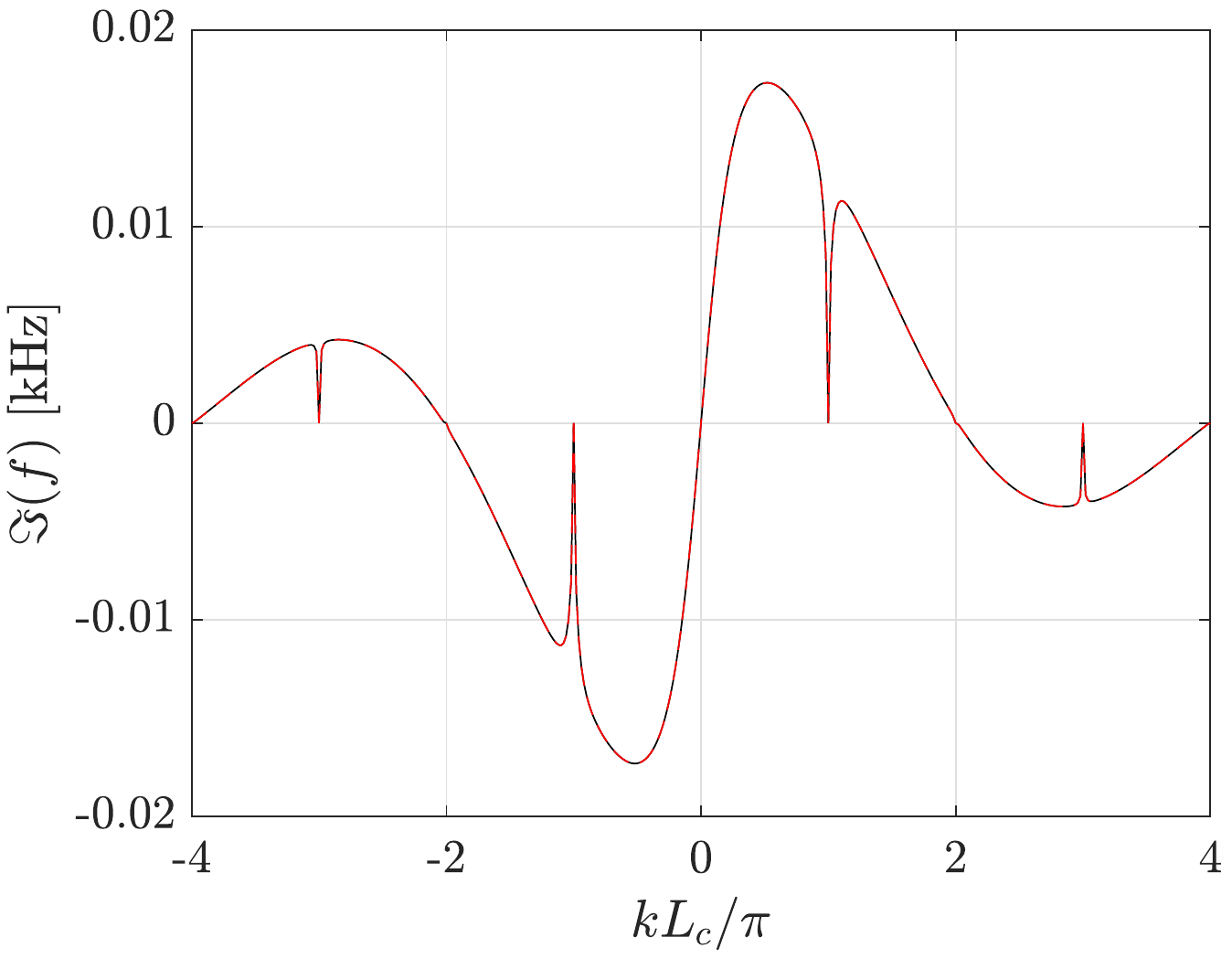}\label{pwe2}}
\caption{Compared dispersion diagrams obtained with PWE (dashed red lines) and SEM (solid black lines).}
\label{fig:pwe}
\end{figure}

Next, a more complex model based on a 3D solid finite element mesh is used to compute the non-Hermitian properties of the acoustic waveguides built in previous sections.

\begin{figure}[H]
\centering
\includegraphics[width=\textwidth]{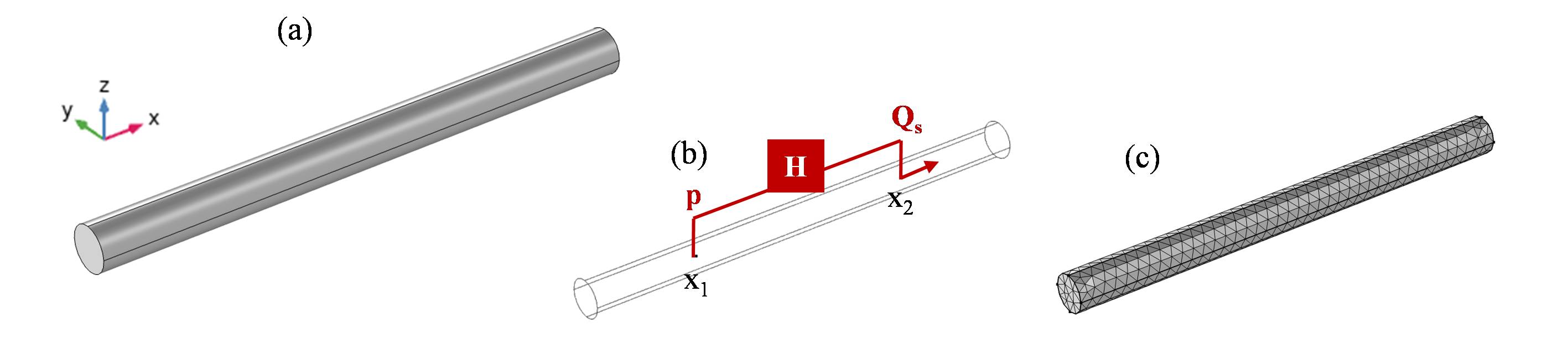}
\caption{3D finite element model in COMSOL Multiphysics\textsuperscript{\textregistered} for an acoustic duct cell with local feedback (a), position of the sensor - $x_1$ - and actuator - $x_2$ (b), and FE mesh respecting 8 elements per wavelength (c). This unit cell has the same geometry of the 1D model in Fig.~\ref{cell}, i.e. $a = 0.5$m, $x1 = 0.25 a$, $x_2 = 0.75 a$ and $d = 0.04$m. The acoustic duct is filled with air at ambient conditions, $\rho_0 \approx 1.225$ kg/m$^3$ and $c \approx 343$ m/s, with proportional damping in the sound speed of $n_c = 0.01$.}
	\label{fig:3DFE_1}
\end{figure}

\begin{figure}[H]
\centering
\includegraphics[width=\textwidth]{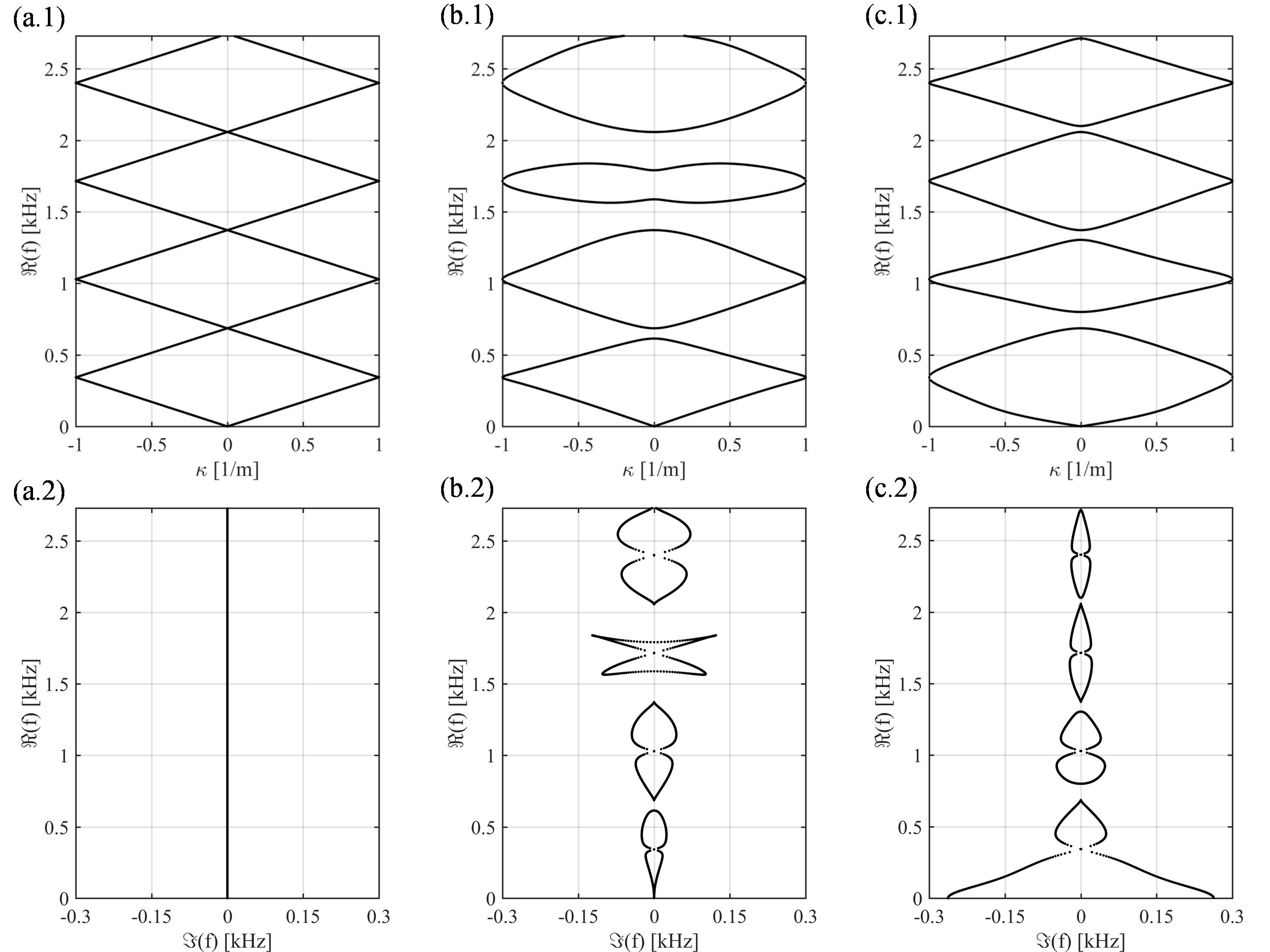}
\caption{Spectral properties with periodic boundary conditions: (a) no feedback, (b) derivative feedback with $\gamma_{D} = 5.0e-10$, and (c) integral feedback with $\gamma_{I} = 1.5e-2$.}
\label{fig:3DFE_2}
\end{figure}

Figure \ref{fig:3DFE_2} shows the dispersion relation for both derivative (Figure \ref{fig:3DFE_2}(b)) and integral (Figure \ref{fig:3DFE_2}(c)) feedback actions in comparison with the Hermitian counterpart (no feedback) computed with this high-fidelity numerical model. Figure \ref{fig:3DFE_2}(b) confirms the one-dimensional results of Fig.~\ref{fig:PBCxOBC}, and hence, validates the dispersion properties by using the spectral models. 

\begin{figure}[H]
\centering
\includegraphics[scale=0.75]{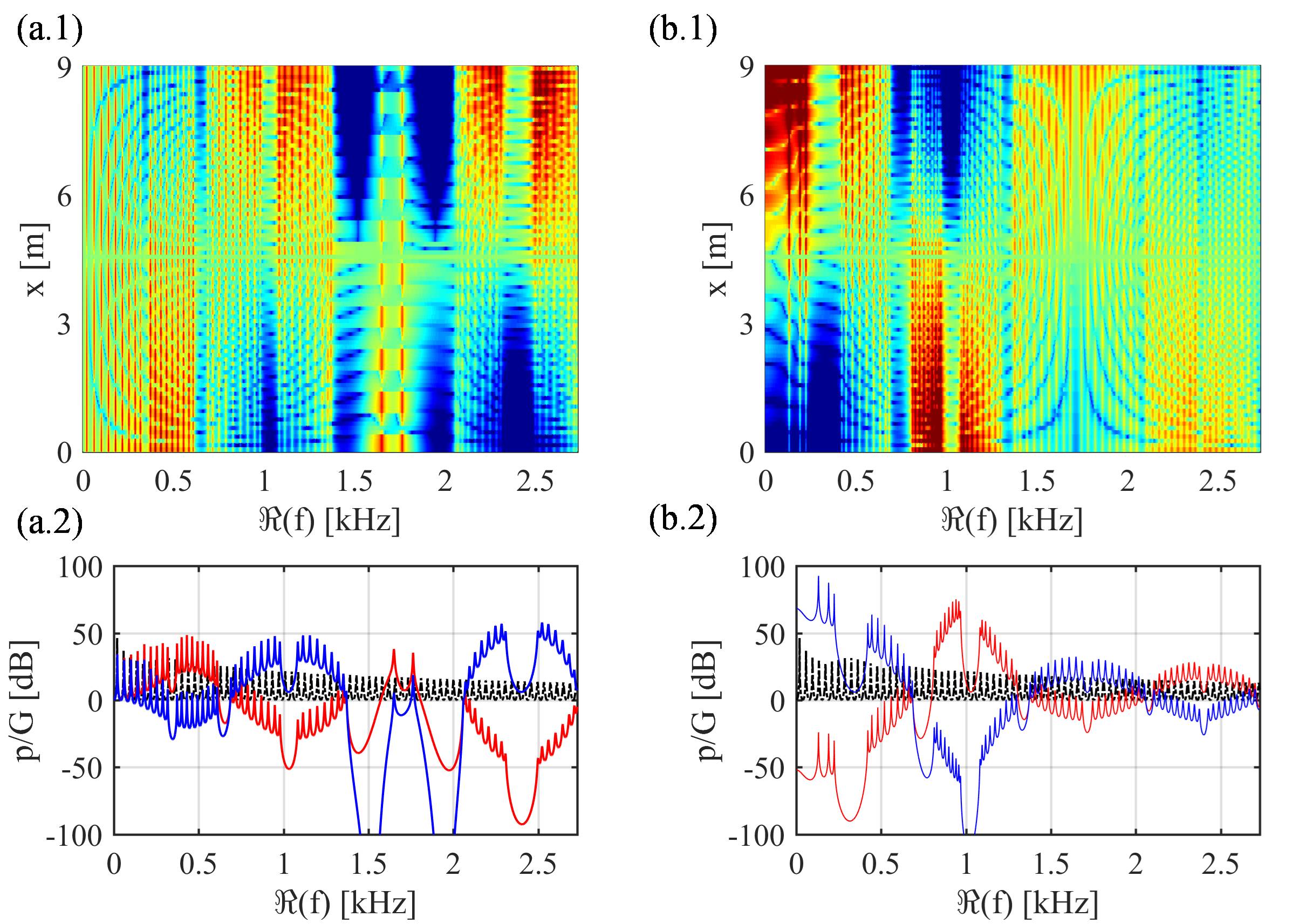}
\caption{Frequency response with excitation at the center (i.e., $x = 4.5$ m): (a) derivative feedback with $\gamma_{D} = 5.0e-10$ and (b) integral feedback with $\gamma_{I} = 1.5e-2$. Left end (red), right end (blue), no feedback case (black).}
\label{fig:3DFE_3}
\end{figure}

Figure \ref{fig:3DFE_3} presents the harmonic response by exciting the acoustic duct at its center (i.e., at x = 4.5 m). Again, as predicted by the spectral models, some frequency zones are related to attenuation or amplification, which appear in an alternate pattern for both  

\begin{figure}[H]
\centering
\includegraphics[scale=0.50]{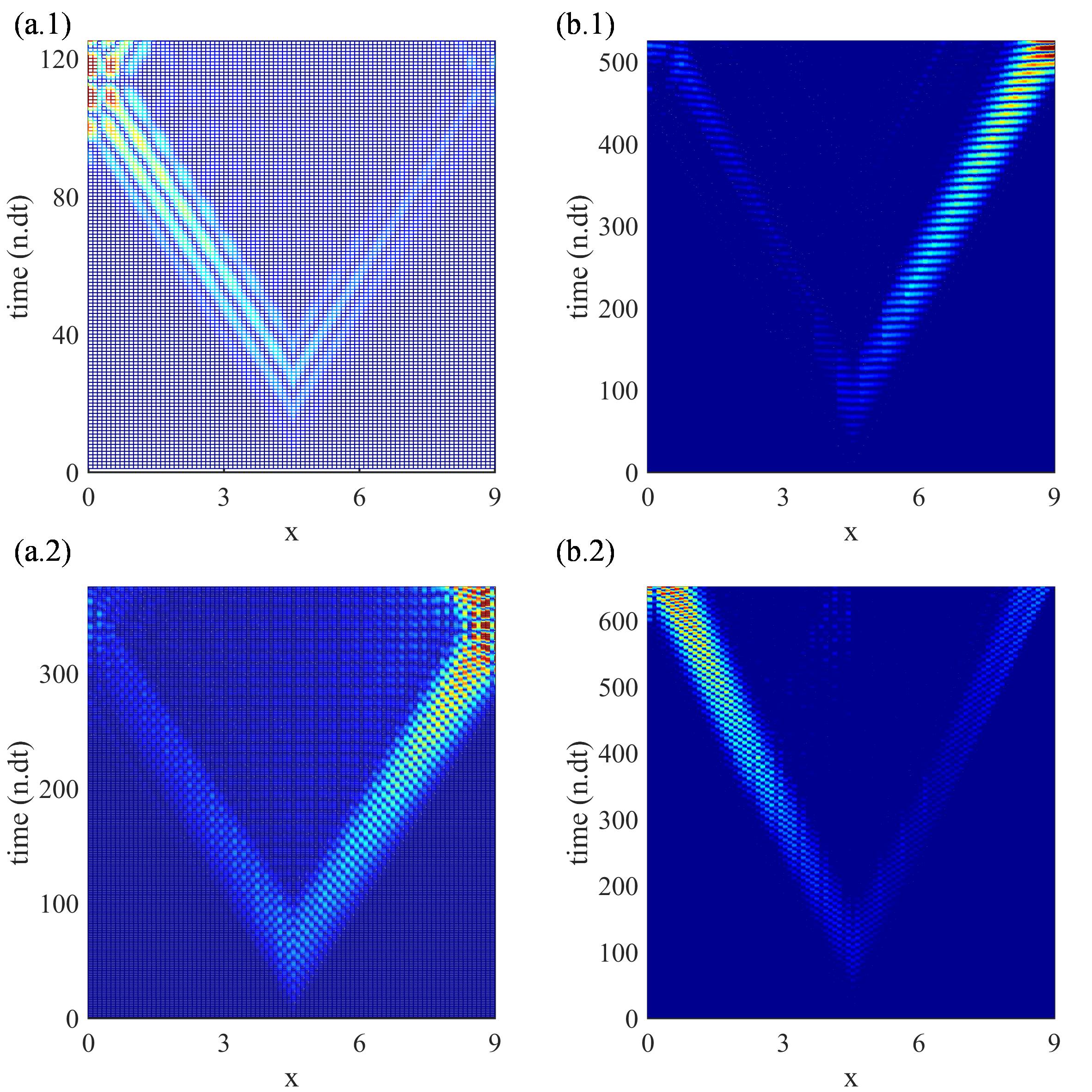}
\caption{Transient time response with excitation at the center (i.e., $x = 4.5$ m): (a) derivative feedback with $\gamma_{D} = 5.0e-10$ and (b) integral feedback with $\gamma_{I} = 1.5e-2$. Left end (red), right end (blue), no feedback case (black).}
\label{fig:3DFE_4}
\end{figure}

\section{Conclusions}
\label{sec: concl}

This work expands previous results published on \cite{braghini2021non} to different feedback laws and highlights that the stability of metastructures built based on the proposed unit cell with feedback should be better investigated, as this is essential for practical applications. Further investigations should be conducted in this regard by combining numerical simulations with experiments.
The numerical results presented here, confirmed by high-fidelity FE simulations, can guide the experimental observations, as well as practical implementations in mechanical engineering applications such as sensing, energy harvesting, structural health monitoring, and control of filaments and membranes in biological systems. 

\section*{Acknowledgments}
The authors acknowledge the financial support provided by the São Paulo Research Foundation (FAPESP) through grant Nos. \#2018/15894-0, \#2018/18774-6, \#2021/14611-8, and \#2021/05140-1. Matheus I. N. Rosa gratefully acknowledges the support from the National Science Foundation (NSF) through the EFRI 1741685 grant and from the Army Research office through grant W911NF-18-1-0036.

\section{Author's contributions:}
Danilo Braghini carried out the methodology, software and investigation using FEM and SEM and writing of the original draft.
Vinicius de Lima carried out the methodology, software and investigation using PWE.
Danilo Beli carried out the investigations using high-fidelity simulation on the software COMSOL\textsuperscript{\textregistered} and performed the review of the manuscript.
MINR conceived the investigation and conduced preliminary simulation results.
JRFA provided the computational resources, acquisition of funding, conceived and designed the study, performed the review of the manuscript, and overall supervision.  All authors read and
approved the manuscript.




\bibliographystyle{unsrtnat}

\newpage
\begin{appendix}
\section{ Linear acoustic system with mass flow source}
\label{A2: PDE}

In this section, we derive the forced linear wave equation, which models the dynamics of the unit cell of the active acoustic metamaterial. This equation is used to derive the numerical models of the following sections. These methods aim one of the following objectives: 1. search for a solution of the $n$-coupled wave equations, which model the dynamics of the metastructure with $n$-cells under defined initial conditions, boundary conditions, and exogenous excitation (FEM); 2. Give the dispersion relation of the periodic metamaterial (SEM and PWE).

\subsection{State equation and entropy transport equation}

The acoustic wave traveling through a quiescent fluid can be assumed as an adiabatic process. The state equation of a perfect gas in such conditions can be stated as follows

\begin{equation}\label{adibaticstate}
P(\mathbf{x},t)=P_0\left(\frac{\rho(\mathbf{x},t)}{\rho_0}\right)^{C_p},
\end{equation}

where the scalar fields $P, \rho : D \subset \mathbb{R}^3 \times [0,T] \to \mathbb{R}$ representing general time varying pressure and mass density, $P_0, \rho_0 \in \mathbb{R}$ the pressure and density of the fluid in its mean state and $C_p \in \mathbb{R}_{>0}$ being the heat capacity ratio. 
The dependency on space and time variables will be omitted in the equations henceforth for the sake of clarity. With the assumption of small acoustic disturbances, one can expand Eq.~(\ref{adibaticstate}) in Taylor series and neglect the derivatives of second order and higher. Thus, recalling also that any thermodynamic variable can be written as a function of two other independent thermodynamic variables, one gets the expression for the pressure as an affine function of density for densities close enough to $\rho_0$

\begin{equation}\label{adibaticstate_taylor}
P(\rho, \eta)=P_0+ \left. \frac{\partial P (\rho,\eta)}{\partial\rho} \right |_{\rho=\rho_0} (\rho-\rho_0).
\end{equation}

Defining the Bulk modulus $B$ as 
\begin{equation}\label{bulkmodulus}
B(\rho,\eta)=\rho_0 \left( \frac{\partial P}{\partial\rho} \right)_{\eta},
\end{equation}
 
 where $(\cdot)_{\eta}$ indicates that entropy $\eta$ remains constant while taking the partial derivative. Omitting also the dependency on thermodynamic variables, define pressure and density disturbances $p'$ and $\rho'$, respectively as
\begin{equation}\label{relativepressure}
p'=P-P_0,
\end{equation}
\begin{equation}\label{relativedensity}
\rho'=\rho-\rho_0.
\end{equation}

Now, assuming the additional hypothesis of inviscid fluid flow, one can conclude that the entropy transport equation is

\begin{equation}
\frac{D \eta}{Dt} = 0,
\end{equation}

meaning that the material time derivative of entropy equals zero on the system. Thus, the only variable is $\rho$, $B(\rho,\eta) =B(\rho)$, and Eq.~(\ref{adibaticstate_taylor}) can be rewritten as the following linear relation between acoustic variables
\begin{equation}
\label{pressureandcondensation}
p'=\frac{B(\rho_0)}{\rho_0} \rho'.
\end{equation}

\subsection{Continuity}

Consider an infinitesimal volume $ dV= dx dy dz$. The volume is fixed in space, and the fluid flows through it. The rate of mass flowing along each direction with velocity field $\mathbf{u} : D \subset \mathbb{R}^3 \times [0,T] \to \mathbb{R}^3 $ such that $\mathbf{u}(\mathbf{x},t) = (u_x(\mathbf{x},t)), u_y(\mathbf{x},t)), u_z(\mathbf{x},t)))$ is given by three scalar equations:

\begin{equation}\label{massflowx}
\rho u_x dy dz-\left(\rho u_x dy dz+\frac{\partial (\rho u_x)}{\partial x} dx dy dz\right)=
-\frac{\partial (\rho u_x)}{\partial x} dV,
\end{equation}
\begin{equation}\label{massflowy}
\rho u_y dy dz-\left(\rho u_y dy dz+\frac{\partial (\rho u_y)}{\partial y} dx dy dz\right)=
-\frac{\partial (\rho u_y)}{\partial y} dV,
\end{equation}
\begin{equation}\label{massflowz}
\rho u_z dy dz-\left(\rho u_z dy dz+\frac{\partial (\rho u_z)}{\partial z} dx dy dz\right)=
-\frac{\partial (\rho u_z)}{\partial z} dV.
\end{equation}

Combining these equations, the flux of mass through the boundaries of the volume is

\begin{equation}\label{totalmassflow}
F = -\left(\frac{\partial (\rho u_x)}{\partial x} + \frac{\partial (\rho u_y)}{\partial y} + \frac{\partial (\rho u_z)}{\partial z} \right)
=-\textrm{div}(\rho \mathbf{u}). 
\end{equation}


If there is a source injecting mass with a rate per unit of volume $Q : D \subset \mathbb{R}^3 \times [0, T] \to \mathbb{R}$, the principle of balance of mass results in the transport equation of mass,

\begin{equation}\label{timederivativecondensation}
\frac{\partial \rho}{\partial t} = Q - \textrm{div}(\rho \mathbf{u}),
\end{equation}

which can be rewritten using the identity: $\textrm{div}(\rho \mathbf{u}) = \mathbf{\nabla \rho} \cdot \mathbf{u} + \rho  \textrm{div} (\mathbf{u})$ and expanding the variables in terms of the acoustic disturbances as

\begin{equation}\label{timederivativecondensation2}
\frac{\partial (\rho_0+\rho')}{\partial t} = Q - \mathbf{\nabla  (\rho_0+\rho')} \cdot (\mathbf{u}_0+\mathbf{u}') + (\rho_0+\rho')  \textrm{div} (\mathbf{u}_0+\mathbf{u}'),
\end{equation}

having in mind that, since the fluid movement is a consequence of pressure disturbances, $\mathbf{u}$ can also be written as $\mathbf{u}(\mathbf{x},t) = \mathbf{u}_0 + \mathbf{u}'(\mathbf{x},t)$. $\mathbf{u}_0 = 0$, since the fluid is quiescent, and using again the hypothesis of small disturbances one gets

\begin{equation}\label{timederivativecondensation_}
\frac{\partial \rho'}{\partial t}  + \rho_0  \textrm{ div} (\mathbf{u}') = Q,
\end{equation}

which can be rewritten with the relation between pressure and density given by Eq.~(\ref{pressureandcondensation}):

\begin{equation}\label{timederivativecondensation__}
\frac{\rho_0}{B(\rho_0)}\frac{\partial p'}{\partial t}+ \rho_0 \textrm{ div}( \mathbf{u})=Q.
\end{equation}

\begin{remark}
\label{remark 1}
Note that one can define the mass rate $\dot{m}$ entering the domain $D$ as the volume integral of the field $Q$, i.e,

\begin{equation}
\label{mdotdefinition}
    \dot{m} = \int_D Q dV .
\end{equation}

Now, taking the concentrated feedback described in \ref{A1: methods}, $Q(\mathbf{x},t) = \bar{Q}(t) \delta(\mathbf{x}-\mathbf{x_0})$. Considering also the 1D system in Fig.~\ref{cell}, $dV(x) = A(x) dx$ and  $\int_D Q(\mathbf{x},t) dV(x) = \int_0^{L_c}  \bar{Q}(t) \delta(x-x_2)A(x) dx = \bar{Q}(t) \int_0^{L_c}A(x) \delta(x-x_2) dx$. Recalling the properties of Dirac's delta function yields

\begin{equation}
\label{mdotQrelation}
    \dot{m} =  \bar{Q}(t) A(x_2).
\end{equation}

Now, since $\bar{Q}(t) = H(P(\mathbf{x_1},t))$ and $P$ is given by Eq.~(\ref{relativepressure}), one can also write $\bar{Q}(t) = H(p'(\mathbf{x_1},t))$. Consequently, $\bar{Q}, \dot{m} << 1$. Thus, one can define the volume velocity $G$ as 

\begin{equation}
\label{Gdefinition}
    G = \frac{\dot{m}}{\rho_0},
\end{equation}

with $G << 1$. Therefore, $\dot{m} = G \rho_0 \approx G \rho$  under small disturbances. Substituting Eq.~(\ref{mdotQrelation}), one gets 

\begin{equation}\label{QnGrelationlinear}
    \bar{Q} = \frac{\rho_0 G}{A(x_2)},
\end{equation}

which is the relation between the excitation used in PWE, $\bar{Q}(t)$ (see on Eq.~(\ref{pwe})) and volume velocity $G(t)$, used in the other methods.

\end{remark}
	
\subsection{Euler's equation}

The difference of force caused by the pressure on each opposite side of the element in each direction is given by three scalar equations:

\begin{equation}\label{equilibriumpressurex}
P dy dz-\left(P+\frac{\partial P}{\partial x} dx\right) dy dz=-\frac{\partial P}{\partial x} dV,
\end{equation}
\begin{equation}\label{equilibriumpressurey}
P dx dz-\left(P+\frac{\partial P}{\partial y} dy\right) dx dz=-\frac{\partial P}{\partial y} dV,
\end{equation}
\begin{equation}\label{equilibriumpressurez}
P dx dz-\left(P+\frac{\partial P}{\partial z} dz\right) dx dy=-\frac{\partial P}{\partial z} dV.
\end{equation}

Combining these equations results in

\begin{equation}\label{equilibriumpressuregradient}
d\mathbf{f}=-\mathbf{\nabla P} ,
\end{equation}

where $\mathbf{f}$ is the force per unit of volume. The resultant force acting on the element $dV$ with mass $dm=\rho  dV$ is

\begin{equation}\label{equilibriumvolumemassacceleration}
d\mathbf{f} dV=\mathbf{a} dm=\mathbf{a}\rho  dV=\rho\left(\frac{\partial\mathbf{u}}{\partial t}+(\nabla \mathbf{u})\bullet \mathbf{u}\right) dV .
\end{equation}

From Eqs.(~\ref{equilibriumpressuregradient}) and (\ref{equilibriumvolumemassacceleration}), it follows that:

\begin{equation}\label{eulerequation}
-\mathbf{\nabla} p=\rho\left(\frac{\partial\mathbf{u}}{\partial t}+(\nabla\mathbf{u})\bullet \mathbf{u}\right).
\end{equation}

Finally, expanding all the acoustic variables as small acoustic disturbances in a quiescent fluid, one gets the linearized \textit{Euler} equation: 

\begin{equation}\label{eulerlinear}
-\mathbf{\nabla} p'=\rho_0\frac{\partial\mathbf{u}'}{\partial t}.
\end{equation}

\subsection{Linear wave equation with acoustic source}

The time derivative of Eq.~(\ref{timederivativecondensation__}), multiplied by $A(x)$ is

\begin{equation}\label{secondtimederivativecondensation_linear}
\frac{\rho_0 A(x)}{B(\rho_0)}\frac{\partial^2 p'}{\partial t^2}+ \textrm{div} \left(A(x) \rho_0\frac{\partial\mathbf{u'}}{\partial t}\right)=A(x)\frac{\partial Q}{\partial t},
\end{equation}

Applying the divergent operator to Eq.~(\ref{eulerlinear}) multiplied by $A(x)$ we get the following equation

\begin{equation}\label{gradienteulerlinear}
-\textrm{div}( A(x)  \nabla p')= \textrm{div}\left( A(x)\rho_0 \frac{\partial\mathbf{u'}}{\partial t}\right).
\end{equation}

 Substituting the result in Eq.~(\ref{secondtimederivativecondensation_linear}), we finally derive the linear wave equation with a source of mass as follows:

\begin{equation}\label{waveequationsource_linear}
A(x)\frac{1}{c^2}\frac{\partial^2 p'}{\partial t^2}-\textrm{div}(A(x)\nabla p')= A(x)\frac{\partial Q}{\partial t},
\end{equation}

where $c=\sqrt{B(\rho_0)/\rho_0}$ is the speed of sound in the fluid. In particular, for 1-D systems, 

\begin{equation}\label{waveequationsource_linear2}
\frac{\partial}{\partial x} \left( A(x) \frac{\partial p'}{\partial x}\right) -\frac{A(x)}{c^2}\frac{\partial^2 p'}{\partial t^2}= -A(x)\frac{\partial Q}{\partial t}.
\end{equation}

Multiplying the previous equation by $c^2$ and taking the concentrated source of mass described on remark \ref{remark 1} with uniform circular area $A(x) = A$ of diameter $d$ (Fig.~\ref{cell}) yields the following wave equation governing the dynamics of each unit cell

\begin{equation}\label{waveequationsource_linear_volumevelocity}
c^2\frac{\partial^2 p'}{\partial^2 x} - \frac{\partial^2 p'}{\partial t^2}=-\frac{4B}{\pi d^2}\frac{d G}{d t} \delta(x - x_2) \hspace{1 cm} | \hspace{1 cm} x \in \Omega = (x_0,x_3),
\end{equation}

 with solutions $p' : \overline{\Omega} \times [0,T] \to \mathbb{R}$ for the appropriate boundary and initial conditions.
 
\section{ Numerical methods}
\label{A1: methods}
\subsection{Finite Element Method (FEM)}

Consider the system of $nd$ ordinary differential equations, where $nd  \in \mathbb{N}$ is the number of degrees of freedom (DOF) considered in the FEM model representing the metastructure built with the metamaterial whose unit cell is displayed in Fig.~\ref{cell} by the periodic arrangement of $nc \in \mathbb{N}$ unit cells

\begin{equation}
    \mathbf{M}_s \ddot{\bxi}(t) + \mathbf{C}_s \dot{\bxi}(t) + \mathbf{K}_s \mathbf{\bxi}(t) = \mathbf{f}(t) + \mathbf{u}_a(t),
    \label{eq:fem}
\end{equation}

where ($\mathbf{M}_s$,  $\mathbf{C}_s$,   $\mathbf{K}_s$) are matrices $\in \mathbb{R}^{nd \times nd}$ assembled by FEM for linear acoustics, which can be found, for instance, in~\cite{atalla2015finite}. $\xi : [0,T] \to \mathbb{R}^{nd}$ is a vector-valued time signal, and $\bxi(t)$ is the vector of which the entries are the physical variables, acoustic pressures in this case, at every node of the mesh (in this case there is one DOF per node). The vector-valued signal $f$ represents an external perturbation, whereas $u_a$ represents the applied feedback effort. The matrix $\mathbf{C}_s$ can be built to model structural (hysteretic) damping or viscous (\textit{Rayleigh}) damping. Using Eq.~(\ref{waveequationsource_linear_volumevelocity}), the discretization given at Eq.(\ref{eq:fem}), as described on \cite{atalla2015finite} (pg. 74, remark 4), gives excitation of the form $-\frac{B}{A}\frac{d G}{d t}$ on some corresponding node. 

\subsubsection{Physical model}

First, consider a simple feedback law of a proportional-integral-derivative (PID) type, defined for applied volume velocity on the $n$-th cell $G_{fb}^n : [0,T] \to \mathbb{R}$ with respect to the concentrated measured pressure $p_s^{n-a} : [0,T] \to \mathbb{R}$ of the corresponding sensor located $a \in \mathbb{Z}$ cells distant. Thus, the feedback law is defined, with proportional, integral, and derivative gains $\gamma_P, \gamma_I, \gamma_D \in \mathbb{R}$ as follows 

\begin{equation}
    G_{fb}^n(t) = \gamma_P p_s^{n-a}(t) + \gamma_I \int_0^tp_s^{n-a}(t) dt + \gamma_D \frac{d p_s^{n-a}(t)}{dt} \hspace{1 cm}\forall n \in \mathbb{Z} \mid a < n \leq nc.
    \label{eq:pidfb}
\end{equation}

But on Eq.~(\ref{eq:fem}), the applied feedback effort is given in volume acceleration \cite{atalla2015finite}, so we need to define the feedback effort signal $u : [0,T] \to \mathbb{R}^{n_c}$ such that $\mathbf{u}(t) =  -\frac{B}{A}\frac{d \mathbf{G}_v}{d t}$, being $\mathbf{G}_v$ a vector with all the concentrated efforts on actuators along the structure. Then, $\mathbf{u}_a(t) = \mathbf{T}\mathbf{u}(t)$ is the image vector of $u$ at any instant of time transformed to $\mathbb{R}^{n_d}$ by completing the non-actuated  DOFs with zeros, using the following matrix coordinates for $\mathbf{T}$

\begin{equation}
    T_{ij} = \begin{cases}
    1  \textrm{ if $ i = n_2+(j+a-1)n_e$ } \forall j \in \mathbb{Z} \mid 1\leq j \leq n_c-a;\\
    0 \textrm{ otherwise}.
    \end{cases}
\end{equation}

where $n_e$ is the number of finite elements per unit cell. Thus, the feedback law in Eq.~(\ref{eq:pidfb}) in matrix form and units of volume acceleration is: 

\begin{equation}
    \mathbf{u}_a(t) =  \mathbf{\Gamma}_P \dot{\bxi}(t)+ \mathbf{\Gamma}_I \bxi(t) + \mathbf{\Gamma}_D \ddot{\bxi}(t),
\end{equation}

with the matrices $\Gamma$ defined with a generalized feedback gain $\gamma \in \mathbb{R}$ as coordinates in the following manner- $n_2$ is the DOF where the actuator signal is applied, $x_2$ on Fig.~\ref{cell}, in the first unit cell of the arrangement 

\begin{equation}
    \Gamma_{i,j} = \begin{cases}
    -\frac{B}{A}\gamma \textrm{ if $ i = (k+a) n_2$ and $j = k n_1$}, \forall k \in \mathbb{Z}\mid 1\leq k \leq n_c-a;\\
    0 \textrm{ otherwise}.
    \end{cases}
\end{equation}

Substitution of the above law on (\ref{eq:fem}) leads to
\begin{equation}
    \left( \mathbf{M}_s - \mathbf{\Gamma}_D \right)  \ddot{\bxi}(t) +  \left( \mathbf{C}_s - \mathbf{\Gamma}_P  \right) \dot{\bxi}(t) +  \left( \mathbf{K}_s - \mathbf{\Gamma}_I\right) \bxi(t) = \mathbf{f},
\end{equation}

which can be numerically integrated to give the solution $\bxi(t)$.

\subsubsection{State-space model}

By using state-space realizations of (\ref{eq:fem}), it is possible to decouple the passive system and the feedback law. As we show here, this allows an algorithmic procedure to build a numerical model of the metastructure with a generalized feedback law, periodically applied or not, with next-neighborhood or even long-range, non-reciprocal coupling (local or non-local feedback). 

First, the state vector $\mathbf{x}(t) = (\mathbf{\xi}(t) , \dot{\mathbf{\xi}}(t))$ is defined. Then, (\ref{eq:fem}) can be rewritten in matrix form, with $w : [0,T] \to \mathbb{R}$ denoting the external input signal, which is a real-valued function of time, as follows

\begin{equation}
    \dot{\mathbf{x}}(t) = 
    \begin{bmatrix}
     \mathbf{0} & \mathbf{I}\\
     -\mathbf{M}_s^{-1} \mathbf{K}_s &  -\mathbf{M}_s^{-1} \mathbf{C}_s
    \end{bmatrix}
    \mathbf{x}(t)+
    \begin{bmatrix}
     \mathbf{0}\\
       \mathbf{M}_s^{-1}\mathbf{F}
    \end{bmatrix}
    w(t)+
    \begin{bmatrix}
     \mathbf{0} \\
      \mathbf{M}_s^{-1}\mathbf{T} 
    \end{bmatrix}
    \mathbf{u}(t),
    \label{dynamiceq}
\end{equation}

such that $\mathbf{F} \in \mathbb{R}^{n_d}$ defines the point in the structure where the concentrated external load $w(t)$ acts by the following matrix coordinates: 

\begin{equation}
    F_{i} = \begin{cases}
    1  \textrm{ if $ i$ corresponds to the excitation dof};\\
    0 \textrm{ otherwise}.
    \end{cases}
\end{equation}

On the other hand, for the output signal $z : [0,T] \to \mathbb{R}^{n_d}$ to give the physical variables, the following equation is defined

\begin{equation}
    \mathbf{z}(t) = 
    \begin{bmatrix}
     \mathbf{I} & \mathbf{0}
    \end{bmatrix}
    \mathbf{x}(t).
    \label{outputeq}
\end{equation}

For the measured states given by the sensors, $y: [0,T] \to \mathbb{R}^{n_c-a} $, the equation is: 

\begin{equation}
    \mathbf{y}(t) = 
    \begin{bmatrix}
     \mathbf{Y} & \mathbf{0}
    \end{bmatrix}
    \mathbf{x}(t),
    \label{outputeq2}
\end{equation}

such that $\mathbf{Y}$ selects the sensor DOFs as outputs, with the matrix coordinates below. $n_1$ is the DOF where the sensor is placed, represented by $x_1$ on Fig.~\ref{cell}, in the first unit cell of the arrangement. Note that, in the arbitrary non-local feedback case, we have $n_c-a$ sensors and actuators, since the first $a$ cells of the finite arrangement are endowed with sensors but no actuator, whereas the last $a$ cells are endowed only with actuators.

\begin{equation}
    Y_{ij} = 
    \begin{cases}
    1  \textrm{ if $ j = n_1+(i-1)n_e$ } \forall i \in \mathbb{Z} \textrm{ such that } 1\leq i \leq n_c-a;\\
    0 \textrm{ otherwise}.
    \end{cases}
\end{equation}

Thus, we can define the passive (open-loop) linear, time-invariant, system $S$ with inputs $w$ and $u$ and outputs $z$ and $y$ as expressed in \ref{S}. Particularly, $\mathbf{D}_{11}$, $\mathbf{D}_{12}$, $\mathbf{D}_{21}$ and $\mathbf{D}_{22}$ are all null matrices.

\begin{equation}
    \begin{cases}
    \dot{\mathbf{x}}(t) = \mathbf{A} \mathbf{x}(t) + \mathbf{B}_1w(t) + \mathbf{B}_2 \mathbf{u}(t)\\
    \mathbf{z}(t) = \mathbf{C}_1 \mathbf{x}(t) + \mathbf{D}_{11}w(t) + \mathbf{D}_{12} \mathbf{u}(t),\\
    \mathbf{y}(t) = \mathbf{C}_2 \mathbf{x}(t) + \mathbf{D}_{21}w(t) + + \mathbf{D}_{22} \mathbf{u}(t).\\    
    \end{cases}
    \label{S}
\end{equation}

The feedback law can be expressed in the state-space formulation for the system with $n_c-a$ inputs (measured signals) and $n_c-a$ outputs (actuation efforts) and any given transfer function $H(s)$. In particular, if the system is periodic, the same transfer function acts on every unit cell, and each cell has at most one sensor and one actuator, meaning that the state-space representation $H^(1)(s)$, with the measured signal $y_{i-a}(t) \in \mathbb{R}$ and the feedback effort $u_i(t) \in \mathbb{R}$, applied to the $i$-th cell, $\forall a+1 \leq i < n_c$, can be expressed generally as the $n_{xi}$-order linear system below, whose state vector is $\mathbf{x}_i(t) \in \mathbb{R}^{n_{xi}}$

\begin{equation}
    \begin{cases}
    \dot{\mathbf{x}}_i(t) = \mathbf{A}_c^{(1)} \mathbf{x}_i(t) + \mathbf{B}_c^{(1)} y_i(t),\\
    u_i(t) = \mathbf{C}_c^{(1)} \mathbf{x}_i(t) + \mathbf{D}_c^{(1)}y_i(t).\\  \end{cases}
    \label{Hi}
\end{equation}

Now, define $H^{(k-1)}$ as the $k-1$ system of a sequence of systems augmented by adding one feedback relation at each term of the sequence, such that the vectors $\mathbf{x}_i^{(k-1)}(t) = (\mathbf{x}_i(t),\mathbf{x}_i(t),...)$, $\mathbf{y}_i^{(k-1)}(t) = (y_i(t),y_i(t),...)$ and $\mathbf{u}_i^{(k-1)}(t) = (u_i(t),u_i(t),...)$ were augmented with $\mathbf{x}_i(t)$, $u_i(t)$ and $y_i(t)$, respectively, $k-1$ times. Then, since the feedback laws are independent from each other, the $(k-1) \cdot n_{xi}$-order system $H^{(k-1)}$ has, by construction, block diagonal matrices ($\mathbf{A}_c^{(k-1)}$,$\mathbf{B}_c^{(k-1)}$,$\mathbf{C}_c^{(k-1)}$,$\mathbf{D}_c^{(k-1)}$), in the following way

\begin{eqnarray}
    \mathbf{A}_c^{(k-1)}=\begin{bmatrix}
    \mathbf{A}_c^{(1)} & \cdots &\mathbf{0}\\
    \vdots &   \ddots & \vdots\\
     \mathbf{0} & \cdots &\mathbf{A}_c^{(1)}\\
    \end{bmatrix},
    \mathbf{B}_c^{(k-1)}=\begin{bmatrix}
    \mathbf{B}_c^{(1)} & \cdots &\mathbf{0}\\
    \vdots &   \ddots & \vdots\\
     \mathbf{0} & \cdots &\mathbf{B}_c^{(1)}\\
    \end{bmatrix},\\
        \mathbf{C}_c^{(k-1)}=\begin{bmatrix}
    \mathbf{C}_c^{(1)} & \cdots &\mathbf{0}\\
    \vdots &   \ddots & \vdots\\
     \mathbf{0} & \cdots &\mathbf{C}_c^{(1)}\\
    \end{bmatrix},
        \mathbf{D}_c^{(k-1)}=\begin{bmatrix}
    \mathbf{D}_c^{(1)} & \cdots &\mathbf{0}\\
    \vdots &   \ddots & \vdots\\
     \mathbf{0} & \cdots &\mathbf{D}_c^{(1)}\\
    \end{bmatrix}
\end{eqnarray}

Thus, the next term of this sequence, $H^{(k)}$, is defined with the vectors $\mathbf{x}_i^{(k)}(t) = (\mathbf{x}_i^{(k-1)}(t),\mathbf{x}_i(t))$, $\mathbf{y}_i^{(k)}(t) = (\mathbf{y}_i^{(k-1)}(t),y_i(t))$ and $\mathbf{u}_i^{(k)}(t) = (\mathbf{u}_i^{(k-1)}(t),u_i(t))$. This system is clearly a $k \cdot n_{xi}$-order linear system represented by block diagonal matrices ($\mathbf{A}_c^{(k)}$,$\mathbf{B}_c^{(k)}$,$\mathbf{C}_c^{(k)}$,$\mathbf{D}_c^{(k)}$) augmented from the previously defined matrices, in block form, as follows:

\begin{eqnarray}
    \mathbf{A}_c^{(k)}=\begin{bmatrix}
    \mathbf{A}_c^{(k-1)} &\mathbf{0}\\
     \mathbf{0} & \mathbf{A}_c^{(1)}\\
    \end{bmatrix}
    , \mathbf{B}_c^{(k)}=\begin{bmatrix}
    \mathbf{B}_c^{(k-1)} &\mathbf{0}\\
     \mathbf{0} & \mathbf{B}_c^{(1)}\\
    \end{bmatrix},\\
    \mathbf{C}_c^{(k)}=\begin{bmatrix}
    \mathbf{C}_c^{(k-1)} &\mathbf{0}\\
     \mathbf{0} & \mathbf{C}_c^{(1)}\\
    \end{bmatrix},
     \mathbf{D}_c^{(k)}=\begin{bmatrix}
    \mathbf{D}_c^{(k-1)} &\mathbf{0}\\
     \mathbf{0} & \mathbf{D}_c^{(1)}\\
    \end{bmatrix}
    \label{Hk}
\end{eqnarray}

Thus, by induction, we conclude that, by defining $H = H^{(n_c-a)}$ we have the state-space representation of the feedback interactions as a $(n_c-a) \cdot n_{xi}$-order linear system with block diagonal matrices, as follows:

\begin{equation}
    \begin{cases}
    \dot{\mathbf{x}}_c(t) = \mathbf{A}_c \mathbf{x}_c(t) + \mathbf{B}_c \mathbf{y}(t),\\
    \mathbf{u}(t) = \mathbf{C}_c \mathbf{x}_c(t) + \mathbf{D}_c\mathbf{y}(t).\\  \end{cases}
    \label{H}
\end{equation}

Finally, the interconnection $(S,H)$ (closed-loop) can be written in terms of the previously defined state-space matrices, as

\begin{equation}
    \begin{cases}
    \dot{\mathbf{x}}_a(t) = \mathbf{A}_{cl} \mathbf{x}_a(t) + \mathbf{B}_{cl} w(t), \\
    \mathbf{z}(t) = \mathbf{C}_{cl} \mathbf{x}_a(t) + \mathbf{D}_{cl}w(t), \\
    \end{cases}
    \label{SH}
\end{equation}

with the augmented state $\mathbf{x}_a(t) = (\mathbf{x}(t),\mathbf{x}_c(t))$

\begin{figure}[H]
    \centering
    \includegraphics[width=10cm]{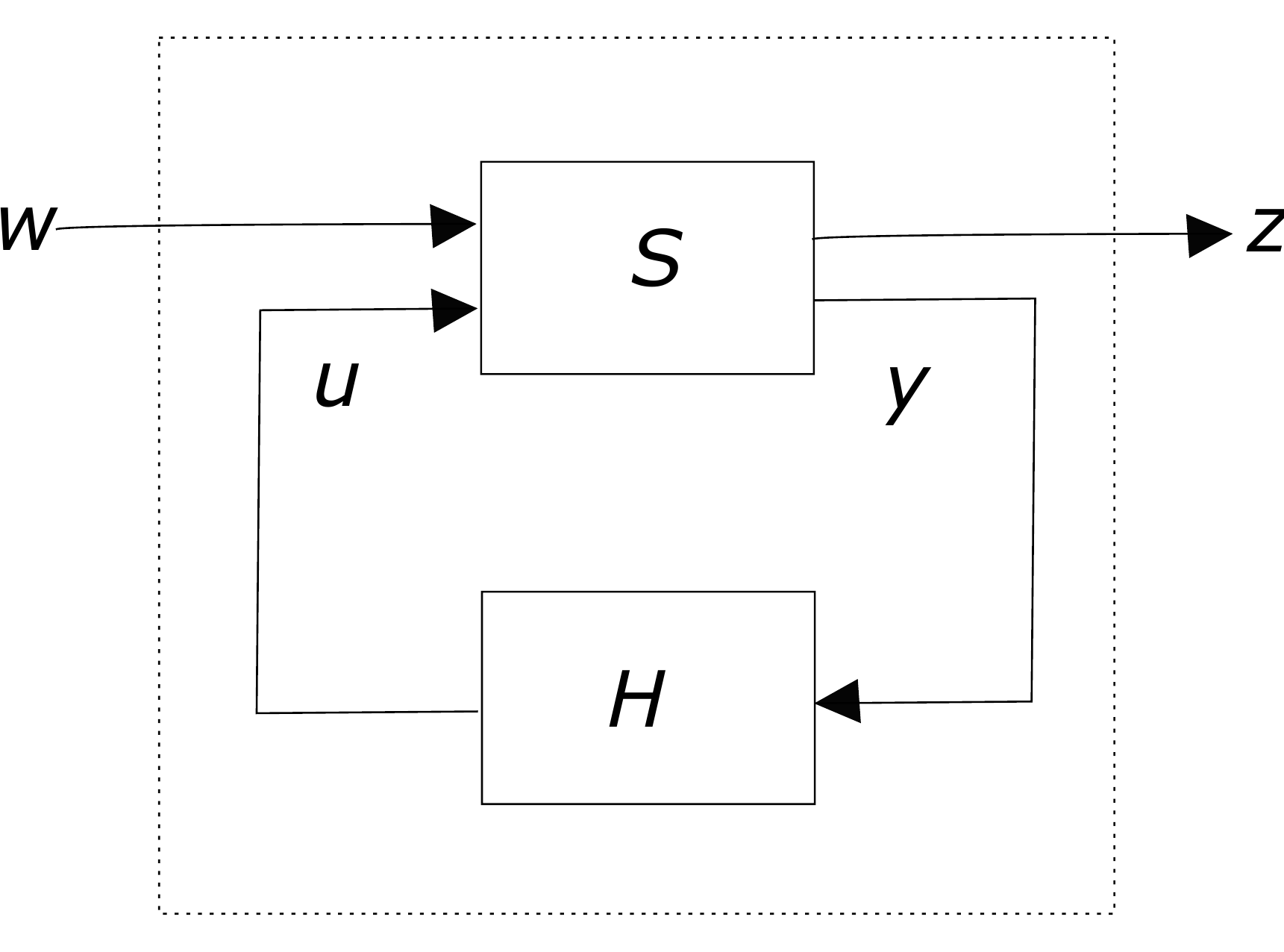}
    \caption{$(S,H)$ interconnection scheme with corresponding signals $w$, $z$, $u$ and $y$.}
    \label{fig:iterconnection}
\end{figure}

It can be shown that the closed-loop matrices are written, in terms of the previously defined matrices, as

\begin{eqnarray}
    \mathbf{A}_{cl} = \begin{bmatrix}
    \mathbf{A}+\mathbf{B}_2\mathbf{D}_{c}\mathbf{C}_{2} & \mathbf{B}_{2} \mathbf{C}_{c}\\
    \mathbf{B}_{c} \mathbf{C}_{2} & \mathbf{A}_{c}
    \end{bmatrix},
    \mathbf{B}_{cl} = \begin{bmatrix}
    \mathbf{B}_1+\mathbf{B}_2\mathbf{D}_{c}\mathbf{D}_{21}\\
    \mathbf{B}_{c} \mathbf{D}_{21} 
    \end{bmatrix},\\
    \mathbf{C}_{cl} = \begin{bmatrix}
    \mathbf{C}_{1}+\mathbf{D}_{21}\mathbf{D}_{c}\mathbf{C}_{2} & \mathbf{D}_{21}\mathbf{C}_{c}
    \end{bmatrix},
    \mathbf{D}_{cl} = \begin{bmatrix}
    \mathbf{D}_{11}+\mathbf{D}_{12}\mathbf{D}_{c}\mathbf{D}_{21}    
    \end{bmatrix}
\end{eqnarray}

Thus, the closed loop is a linear, time-invariant system that gives $z$ as output for any external load $w$. The internal stability of this system can also be analyzed by computing the eigenvalues of $\mathbf{A}_{cl}$. This is a conservative way to analyze the behavior of $z$ for any given load $w$, as internal stability is a sufficient, but not a necessary condition for input-output stability. 

\begin{remark}
\label{remark 2}
 Define the signals in the frequency domain as given by the \textit{Laplace} transform, for instance, indicated by $\hat{(\cdot)}$. Then, $H(s) = \hat{u}(s)_i/ \hat{y}(s)_{i-a}$. Particularly for the acoustic system, $y_{i-a}=p_s$ is the measured pressure, and $u_i$ is the volume acceleration. Thus, the feedback law in terms of volume velocity, as previously defined on (\ref{eq:pidfb}), is actually given more generically by the transfer function $H_v(s) = \frac{H(s)}{s} $.
\end{remark}

\begin{remark}
\label{remark 3}
Note that, in this model, since we separate the passive ($S$) from the active part ($H$) of the structure, the coupling between measurements and applied feedback signals can be generically selected by the state-space model of $S$ through the matrices involved in the relation between $y$ and $u$ on \ref{S} and \ref{H}, following the design requirements.

For instance, in the lumped models used in \cite{rosa2020dynamics}, the feedback law was defined on the $i-th$ lumped element as $ F_i = k (\xi_{i-a} - \xi_{i-a-1}) $ with the same range $a$ defined here, but with $ a+1\leq i \leq n_d$. Thus, $n_c = n_d-a-1$ and we just need to replace $C_{2ss}$ by the composition of $Y^l$, a bi-diagonal matrix of dimension $n_c x n_c+1$ (defined on the following) with a zero matrix on the same way as in \ref{outputeq}.

\begin{equation}
    Y^l_{ij} = \begin{cases}
    -1  \textrm{ if $ j = i$ } \textrm{ , }  1  \textrm{ if $ j = i+1$ } \forall i \in \mathbb{Z} \textrm{ such that } 1\leq i \leq n_c;\\
    0 \textrm{ otherwise}.
    \end{cases}
\end{equation}

\end{remark}

\subsection{Spectral Element Method (SEM)}
 
In SEM, the goal is to find the analytical dynamic stiffness matrix $\mathbf{D_c}(\omega)$ for all frequencies $\omega \in \mathbb{C}$. With this matrix, one can easily find the transfer matrix as explained in plenty of references, such as \cite{hussein2014dynamics}. The eigenvalues of the transfer matrix have well-known structure given by \textit{Bloch}'s theorem and, thus, the problem of finding $\omega(k)$, i.e., the dispersion relation of an infinite waveguide as the complex frequency for imposed real values of wavenumber $k$, can be solved. In this work, all the dispersion relations were computed based on this approach.

Consider, again, the unit cell given by Fig.~\ref{cell}, with the concentrated applied feedback in volume velocity. One spectral element can be assigned for each continuous segment of the domain. This implies three spectral elements for this cell, one for each segment connecting the nodes $x_0$, $x_1$, $x_2$ and $x_3$.

The SEM formulation for a (homogeneous) one-dimensional linear acoustic duct is given by Eq.(\ref{SEMac}) as a function of the frequency \cite{rosa2017numerical}. $L_i$ represents the length of the element and $k(\omega) = \frac{\omega}{c}$ the local wavenumber related to the constant phase velocity $c$ of acoustic waves traveling through the $i-th$ homogeneous segment of the periodic system (metastructure) under analysis. 


\begin{equation}
    \mathbf{K_i}(\omega) =\frac{A}{\rho c j (1-e^{-2jk(\omega)})}
    \begin{bmatrix}
    e^{-2 j k(\omega) L_i}+1  &  -2e^{-j k(\omega) L_i}\\
     -2e^{-j k(\omega)L_i}           &  e^{-2 j k(\omega) L_i}+1 
    \end{bmatrix}.
     \label{SEMac}
\end{equation}
\vspace{1 pt}

The dynamic stiffness matrix $\mathbf{D}_c(\omega)$ relates volume velocities and pressures in the frequency domain (via Fourier transform) at the nodes of the element.

\begin{equation}
    \begin{bmatrix}
    \hat{G}_0(\omega)\\
    \hat{G}_1(\omega)\\
    \hat{G}_2(\omega)\\
    \hat{G}_3(\omega)\\
    \end{bmatrix}
    = \mathbf{D}_c(\omega)
    \begin{bmatrix}
    \hat{p}_0(\omega)\\
    \hat{p}_1(\omega)\\
    \hat{p}_2(\omega)\\
    \hat{p}_3(\omega)\\
    \end{bmatrix}.
\end{equation}
\vspace{1 pt}

Consider a harmonic solution in both space and time. In this case $P(x,t) = \tilde{P}(k,\omega) exp[j(\omega t - k x)] = \hat{P}(x,\omega) e^{j\omega t}$ for each $\omega$ and $k$. In linear systems, a general solution will be obtained by the superposition of harmonic solutions for every $\omega$ and $k$. Thus, one can deduce that, if the system is periodic in space, the relation between $\hat{P}(x_s-aL_c,\omega)$ and the pressure $\hat{P}(x_s,\omega)$ is expressed as $\hat{P}(x_s-aL_c,\omega) = e^{ j k a L_c} \hat{P}(x_s,\omega)$, i.e. a phase delay proportional to the distance $a L_c$ on the complex-valued function $\hat{P}(x_s,\omega)$.

Adding the previously defined feedback input as $v = \begin{bmatrix}0\\ 0\\ \hat{G}_{fb}(j\omega)\\ 0 \end{bmatrix} = \begin{bmatrix}
        0 & 0 & 0 & 0\\
        0 & 0 & 0 & 0\\
        0 & H_{v}(j \omega) e^{j k a L_c} & 0 & 0\\
        0 & 0 & 0 & 0 
\end{bmatrix} \begin{bmatrix}
        \hat{p}_0(\omega)\\
    \hat{p}_1(\omega)\\
    \hat{p}_2(\omega)\\
    \hat{p}_3(\omega)\\
\end{bmatrix}$ to the left hand side of the equation, such that $H_{v}$ is the transfer function of pressure by volume velocity corresponding to the feedback law being considered.

Thus, $v$ can be added to the dynamic stiffness matrix in the following way

\begin{equation}
    \mathbf{D}_c(\omega) = 
    \begin{bmatrix}
    \mathbf{K}_1(1,1) & \mathbf{K}_1(1,2) & 0 & 0\\
    \mathbf{K}_1(2,1)    & \mathbf{D}_1  &   \mathbf{K}_2(1,2) & 0\\
    0 & \mathbf{C} & \mathbf{D}_2 & \mathbf{K}_3(1,2)\\
    0 & 0 & \mathbf{K}_3(2,1) & \mathbf{K}_3(2,2)
    \end{bmatrix},
\end{equation}
\vspace{1 pt}

with the auxiliary variables defined as

\begin{equation}
    \begin{cases}
    \mathbf{D}_1 = \mathbf{K}_1(2,2) + \mathbf{K}_2(1,1),\\
    \mathbf{D}_2 = \mathbf{K}_2(2,2) + \mathbf{K}_3(1,1),\\
    \mathbf{C} = \mathbf{K}_2(2,1) - H_{v}(j \omega) e^{j k a L_c}.
    \end{cases}
\end{equation}
\vspace{1 pt}

\subsection{Plane wave expansion (PWE)}
\label{pwe}

Applying the Fourier transform to variable $t$ in Eq.~(\ref{waveequationsource_linear}) yields
\begin{equation}\label{waveequationsource_linear_varea_frequency}
\frac{\partial}{\partial x}\left[A\frac{\partial \hat{p} (\omega)}{\partial x}\right]+\frac{\omega^2}{c^2}A\hat{p}(\omega)+j\omega A\hat{Q}(\omega)=0.
\end{equation}

Hereafter, the frequency dependency of signals will be omitted. Thus, $\hat{(\cdot)} = f(\mathbf{x},\omega)$ denotes a variable transformed to the frequency domain via \textit{Fourier} transform.

The \textit{Bloch-Floquet} theorem \cite{bloch1929quantenmechanik} for wave propagation in the longitudinal direction of a periodic system is

\begin{equation}\label{blochfloquetcondition}
\hat{p}=p_ke^{-jkx}.
\end{equation}

Expanding the periodic function $p_k$ as a Fourier series yields

\begin{equation}\label{blochfloquettheorem_series}
\hat{p}=e^{-jkx}\sum_{m=-\infty}^{+\infty}\bar{P}_k(g)e^{-jgx}=\sum_{m=-\infty}^{+\infty}\bar{P}_k(g)e^{-j(k+g)x},
\end{equation}

with $g=2\pi m/L_c$ $\forall$ $m \in \mathbb{Z}$. The section area also can be expanded as Fourier series

\begin{equation}\label{area_series}
A=\sum_{\bar{m}=-\infty}^{+\infty}\bar{A}(\bar{g})e^{-j\bar{g}x},
\end{equation}

with $\bar{g}=2\pi \bar{m}/L_c$ $\forall$ $m \in \mathbb{Z}$.

Applying Eqs.~(\ref{blochfloquettheorem_series})-(\ref{area_series}) on each term of Eq.~(\ref{waveequationsource_linear_varea_frequency}):

\begin{equation}\label{waveequationsource_linear_varea_frequency_1}
\frac{\partial}{\partial x}\left[A\frac{\partial \hat{p}}{\partial x}\right]= -
\sum_{\bar{m}=-\infty}^{+\infty}\sum_{m=-\infty}^{+\infty}(k+g)(k+g+\bar{g})\bar{A}(\bar{g})\bar{P}_k(g)e^{-j(k+g+\bar{g})x},
\end{equation}

\begin{equation}\label{waveequationsource_linear_varea_frequency_2}
A\frac{\omega^2}{c^2}\hat{p}=\frac{\omega^2}{c^2}
\sum_{\bar{m}=-\infty}^{+\infty}\sum_{m=-\infty}^{+\infty}\bar{A}(\bar{g})\bar{P}_k(g)e^{-j(k+g+\bar{g})x}.
\end{equation}

Recalling the relation on Eq.~(\ref{QnGrelationlinear}), the non-local feedback law in the frequency domain, with gain $\gamma$ and $\hat{Q}(\omega)$ denoting the \textit{Fourier} transform of $\bar{Q}(t)$, becomes

\begin{equation}\label{feedbacklaw}
\hat{Q}=\sum_{n=-\infty}^{+\infty}\frac{\rho_0}{A(x_2)} \gamma H(j\omega)\hat{p}(x_1+(n-a)L_c)\delta(x-(x_2+nL_c)),
\end{equation}

since the spatial domain was expanded to infinity, where $x_2$ is the point of excitation, $x_1$ is the point of measurement and $H(j\omega)$ depends on the type of gain. For instance, 
\begin{equation}\label{typeofH}
H=1\,\text{(proportional)},
H=j\omega\,\text{(derivative)},
H=\frac{1}{j\omega}\,\text{(integrative)}
\end{equation}

The third therm in Eq.~(\ref{waveequationsource_linear_varea_frequency}) is

\begin{equation}\label{waveequationsource_linear_varea_frequency_3}
j\omega A \hat{Q}=j\omega \sum_{\bar{m}=-\infty}^{+\infty}\bar{A}(\bar{g})e^{-j\bar{g}x}
\sum_{n=-\infty}^{+\infty}\frac{\rho_0}{A(x_2)}\gamma H(j \omega)\hat{p}(x_1+(n-a)L_c)\delta(x-(x_2+nL_c)).
\end{equation}

From \textit{Bloch} wave condition
\begin{equation}\label{blochwavecondition}
\hat{p}(x_1+(n-a)L_c)=\hat{p}(x_1)e^{-jk(n-a)L_c},
\end{equation}

Eq.~\ref{waveequationsource_linear_varea_frequency_3} becomes

\begin{equation}\label{waveequationsource_linear_varea_frequency_31}
j\omega A \hat{Q}=j\omega\frac{\rho_0}{A(x_2)}\gamma H(j \omega)\hat{p}(x_1)\sum_{\bar{m}=-\infty}^{+\infty}\bar{A}(\bar{g})e^{-j\bar{g}x}
\sum_{n=-\infty}^{+\infty}e^{-jk(n-a)L}\delta(x-(x_2+nL_c)).
\end{equation}

Since $a\in\mathbb{Z}$ and is constant,

\begin{equation}\label{diraccomb1}
\sum_{n=-\infty}^{+\infty}e^{-jk(n-a)L_c}\delta(x-(x_2+nL_c))=
e^{jkaL_c}\sum_{n=-\infty}^{+\infty}e^{-jknL_c}\delta(x-(x_2+nL_c)).
\end{equation}

The summation on the right side of Eq.~(\ref{diraccomb1}) holds only if $x=x_2+nL_c$. This is equivalent of sampling the $f(x)=e^{jkx}$

\begin{equation}\label{diraccomb2}
e^{jkaL_c}\sum_{n=-\infty}^{+\infty}e^{-jknL_c}\delta(x-(x_2+nL_c))=
e^{jkaL_c}e^{-jk(x-x_2)}\sum_{n=-\infty}^{+\infty}\delta(x-(x_2+nL_c)).
\end{equation}

The Fourier series of the series of Dirac distribution is

\begin{equation}\label{diraccomb3}
\sum_{n=-\infty}^{+\infty}\delta(x-(x_2+nL_c))=\frac{1}{L_c}\sum_{m=-\infty}^{+\infty}e^{-jg(x-x_2)}.
\end{equation}

Applying Eqs.~(\ref{diraccomb1})-(\ref{diraccomb3}) results in
\begin{equation}\label{waveequationsource_linear_varea_frequency_32}
j\omega AQ=j\frac{\omega \gamma \rho_0  H(j \omega)}{L_c A(x_2)}\hat{p}(x_1)e^{jk(aL_c+x_2)}\sum_{\bar{m}=-\infty}^{+\infty}
\sum_{m=-\infty}^{+\infty}\bar{A}(\bar{g})e^{-jg(x-x_2)}e^{-j(k+\bar{g})x}.
\end{equation}

Also, applying the change of variable $\tilde{g}=\bar{g}+g$ in Eqs. (\ref{waveequationsource_linear_varea_frequency_1}), (\ref{waveequationsource_linear_varea_frequency_2}) and (\ref{waveequationsource_linear_varea_frequency_32}),

\begin{equation}\label{waveequationsource_linear_varea_frequency_11}
\frac{\partial}{\partial x}\left[A\frac{\partial \hat{p}}{\partial x}\right]=
-\sum_{\tilde{m}=-\infty}^{+\infty}\sum_{m=-\infty}^{+\infty}(k+g)(k+\tilde{g})\bar{A}(\tilde{g}-g)\bar{P}_k(g)e^{-j(k+\tilde{g})x},
\end{equation}

\begin{equation}\label{waveequationsource_linear_varea_frequency_21}
A\frac{\omega^2}{c^2}\hat{p}=\frac{\omega^2}{c^2}
\sum_{\tilde{m}=-\infty}^{+\infty}\sum_{m=-\infty}^{+\infty}\bar{A}(\tilde{g}-g)\bar{P}_k(g)e^{-j(k+\tilde{g})x},
\end{equation}

\begin{equation}\label{waveequationsource_linear_varea_frequency_33}
j\omega A \hat{Q}=\alpha H(j\omega) \hat{p}(x_1)e^{jk(aL_c+x_2)}\sum_{\tilde{m}=-\infty}^{+\infty}
\sum_{m=-\infty}^{+\infty}\bar{A}(\tilde{g}-g)e^{jgx_2}e^{-j(k+\tilde{g})x},
\end{equation}

with $\alpha = j\frac{\omega \gamma \rho_0}{L_c A(x_2)}$. Next, orthogonality of the exponential function is used in the following manner: multiplying Eqs. (\ref{waveequationsource_linear_varea_frequency_11})- (\ref{waveequationsource_linear_varea_frequency_33}) by $e^{j\bar{g}x}$, dividing by the unit cell length $L_c$ and integrating over the unit cell after factoring $e^{-jkx}$ ($e^{-jkx}\ne0$). Finally, substituting on Eq.(\ref{waveequationsource_linear_varea_frequency}) yields

\begin{multline}\label{waveequationsource_linear_varea_frequency_integral}
-\sum_{\tilde{m}=-\infty}^{+\infty}\sum_{m=-\infty}^{+\infty}(k+g)(k+\tilde{g})\bar{A}(\tilde{g}-g)\bar{P}_k(g)\frac{1}{L_c}\int_{-L_c/2}^{L_c/2}e^{-j(\tilde{g}-\bar{g})x}dx+\\
\frac{\omega^2}{c^2}
\sum_{\tilde{m}=-\infty}^{+\infty}\sum_{m=-\infty}^{+\infty}\bar{A}(\tilde{g}-g)\bar{P}_k(g)\frac{1}{L_c}\int_{-L_c/2}^{L_c/2}e^{-j(\tilde{g}-\bar{g})x}dx+\\
\alpha H(j\omega) \hat{p}(x_1) e^{jk(aL_c+x_2)}\sum_{\tilde{m}=-\infty}^{+\infty}\sum_{m=-\infty}^{+\infty}\bar{A}(\tilde{g}-g)e^{jgx_2}
\frac{1}{L_c}\int_{-L_c/2}^{L_c/2}e^{-j(\tilde{g}-\bar{g})x}dx=0.
\end{multline}

Recalling that, due to the orthogonality previously mentioned, the \textit{Kronecker} delta equals

\begin{equation}\label{kroenecker}
\delta_{\tilde{g}\bar{g}}=\frac{1}{L_c}\int_{-L_c/2}^{L_c/2}e^{j(\tilde{g}-\bar{g})x}dx,
\end{equation}

and it is nonzero only when $\tilde{g}=\bar{g}$. Expanding $\hat{p}(x_1)$, with $ \hat{g} = 2 \pi q / L_c$, $\forall q \in \mathbb{Z}$ results in

\begin{equation}\label{expansionpx2}
\hat{p}(x_1)=\sum_{q=-\infty}^{+\infty}\bar{P}_k(\hat{g})e^{-j(k+\hat{g})x_1}.
\end{equation}

The series of Eqs.~(\ref{expansionpx2}) and (\ref{expansionpx2}) can be truncated limiting the indexes to $m,\bar{m},q=[-M,M]$. Thus, Eq.~(\ref{waveequationsource_linear_varea_frequency_integral}) becomes

\begin{multline}\label{waveequationsource_linear_varea_frequency_final}
-\sum_{m=-M}^{M}(k+g)(k+\bar{g})\bar{A}(\bar{g}-g)\bar{P}_k(g)+\\
\frac{\omega^2}{c^2}
\sum_{m=-M}^{M}\bar{A}(\bar{g}-g)\bar{P}_k(g)+\\
\alpha H(j\omega) e^{jk(aL+x_2)}\sum_{m=-M}^{M}\bar{A}(\bar{g}-g)e^{jgx_2}\mathbf{U}_{M}^T\mathbf{P}_{M}=0,
\end{multline}

wherein $\mathbf{U}_{M},\mathbf{P}_{M} \in \mathbb{R}^{2M+1 x 1}$ are column matrices containing the terms $e^{-j(k+\hat{g})x_1}$ and coefficients $\bar{P}_k(\hat{g})$, respectively, corresponding to the truncation of Eq.~(\ref{expansionpx2}). Superscript $T$ indicates the transpose of a matrix.

Eq.~(\ref{waveequationsource_linear_varea_frequency_final}) is valid $\forall \hat{m}$. Thus, in matrix form, results in the following eigenvalue problem

\begin{equation}\label{pwematrixequation}
(\mathbf{K}-\omega^2\mathbf{M}-j\omega H(j \omega) \mathbf{C})\mathbf{P}_{M}=0,
\end{equation}

where $\mathbf{\bar{A}}, \mathbf{K}, \mathbf{M} \in \mathbb{R}^{2M+1 x 2M+1}$ are matrices such that its elements satisfy: $\bar{A}_{ij} = \bar{A}(\bar{g}_i - g_j)$ and  $K_{ij} = (k+g_j)(k+\bar{g}_i) \bar{A}_{ij}$ and $M_{ij} =\frac{1}{c^2} \bar{A}_{ij}$, where $g_i = 2 \pi i / L_c$ is the subscript notation used for $g$ and $\bar{g}$. Matrix  $\mathbf{C} \in \mathbb{R}^{2M+1 x 2M+1}$ is defined as follows

\begin{align*}
\mathbf{C}=\alpha e^{jk(aL+x_2)}\mathbf{E}\mathbf{U}_{M}^T,\\
\mathbf{E}=\begin{bmatrix}
\sum_{p=1}^{2M+1}D_{1,p}\\\vdots\\\sum_{p=1}^{2M+1}D_{2M+1,p}
\end{bmatrix},\\
\mathbf{D}=\mathbf{\bar{B}}\circ\mathbf{\bar{A}},\\
\mathbf{\bar{B}}=\begin{bmatrix}
e^{jx_2g_{-M}}&\dots& e^{jx_2g_{M}}
\end{bmatrix} \otimes \mathbf{1},\\ 
\end{align*}

where the symbols $\circ$ represents the \textit{Hadamard} product, $\otimes$ the \textit{Kronecker} product and $\mathbf{1} \in \mathbb{R}^{2M+1 x 1}$ is a column matrix wherein all entries equals 1.

\end{appendix}


\begin{thebibliography}{10}

\bibitem{xiong2018does}
Y.~Xiong, ``Why does bulk boundary correspondence fail in some non-hermitian
  topological models,'' {\em Journal of Physics Communications}, vol.~2, no.~3,
  p.~035043, 2018.

\bibitem{koch2020bulk}
R.~Koch and J.~C. Budich, ``Bulk-boundary correspondence in non-hermitian
  systems: stability analysis for generalized boundary conditions,'' {\em The
  European Physical Journal D}, vol.~74, no.~4, pp.~1--10, 2020.

\bibitem{kunst2018biorthogonal}
F.~K. Kunst, E.~Edvardsson, J.~C. Budich, and E.~J. Bergholtz, ``Biorthogonal
  bulk-boundary correspondence in non-hermitian systems,'' {\em Physical review
  letters}, vol.~121, no.~2, p.~026808, 2018.

\bibitem{yao2018edge}
S.~Yao and Z.~Wang, ``Edge states and topological invariants of non-hermitian
  systems,'' {\em Physical review letters}, vol.~121, no.~8, p.~086803, 2018.

\bibitem{yao2018non}
S.~Yao, F.~Song, and Z.~Wang, ``Non-hermitian chern bands,'' {\em Physical
  review letters}, vol.~121, no.~13, p.~136802, 2018.

\bibitem{kawabata2019symmetry}
K.~Kawabata, K.~Shiozaki, M.~Ueda, and M.~Sato, ``Symmetry and topology in
  non-hermitian physics,'' {\em Physical Review X}, vol.~9, no.~4, p.~041015,
  2019.

\bibitem{okuma2020topological}
N.~Okuma, K.~Kawabata, K.~Shiozaki, and M.~Sato, ``Topological origin of
  non-hermitian skin effects,'' {\em Physical review letters}, vol.~124, no.~8,
  p.~086801, 2020.

\bibitem{bergholtz2021exceptional}
E.~J. Bergholtz, J.~C. Budich, and F.~K. Kunst, ``Exceptional topology of
  non-hermitian systems,'' {\em Reviews of Modern Physics}, vol.~93, no.~1,
  p.~015005, 2021.

\bibitem{edvardsson2022sensitivity}
E.~Edvardsson and E.~Ardonne, ``Sensitivity of non-hermitian systems,'' {\em
  Physical Review B}, vol.~106, no.~11, p.~115107, 2022.

\bibitem{wang2018topological}
S.~Wang, G.~Ma, and C.~T. Chan, ``Topological transport of sound mediated by
  spin-redirection geometric phase,'' {\em Science advances}, vol.~4, no.~2,
  p.~eaaq1475, 2018.

\bibitem{ma2019topological}
G.~Ma, M.~Xiao, and C.~T. Chan, ``Topological phases in acoustic and mechanical
  systems,'' {\em Nature Reviews Physics}, vol.~1, no.~4, pp.~281--294, 2019.

\bibitem{huber2016topological}
S.~D. Huber, ``Topological mechanics,'' {\em Nature Physics}, vol.~12, no.~7,
  pp.~621--623, 2016.

\bibitem{delplace2020geometry}
P.~Delplace and A.~Venaille, ``From the geometry of foucault pendulum to the
  topology of planetary waves,'' {\em Comptes Rendus. Physique}, vol.~21,
  no.~2, pp.~165--175, 2020.

\bibitem{zhong2021nontrivial}
J.~Zhong, K.~Wang, Y.~Park, V.~Asadchy, C.~C. Wojcik, A.~Dutt, and S.~Fan,
  ``Nontrivial point-gap topology and non-hermitian skin effect in photonic
  crystals,'' {\em Physical Review B}, vol.~104, no.~12, p.~125416, 2021.

\bibitem{okugawa2020second}
R.~Okugawa, R.~Takahashi, and K.~Yokomizo, ``Second-order topological
  non-hermitian skin effects,'' {\em Physical Review B}, vol.~102, no.~24,
  p.~241202, 2020.

\bibitem{kawabata2020higher}
K.~Kawabata, M.~Sato, and K.~Shiozaki, ``Higher-order non-hermitian skin
  effect,'' {\em Physical Review B}, vol.~102, no.~20, p.~205118, 2020.

\bibitem{hofmann2020reciprocal}
T.~Hofmann, T.~Helbig, F.~Schindler, N.~Salgo, M.~Brzezi{\'n}ska, M.~Greiter,
  T.~Kiessling, D.~Wolf, A.~Vollhardt, A.~Kaba{\v{s}}i, {\em et~al.},
  ``Reciprocal skin effect and its realization in a topolectrical circuit,''
  {\em Physical Review Research}, vol.~2, no.~2, p.~023265, 2020.

\bibitem{nassar2020nonreciprocity}
H.~Nassar, B.~Yousefzadeh, R.~Fleury, M.~Ruzzene, A.~Al{\`u}, C.~Daraio, A.~N.
  Norris, G.~Huang, and M.~R. Haberman, ``Nonreciprocity in acoustic and
  elastic materials,'' {\em Nature Reviews Materials}, vol.~5, no.~9,
  pp.~667--685, 2020.

\bibitem{scheibner2020odd}
C.~Scheibner, A.~Souslov, D.~Banerjee, P.~Sur{\'o}wka, W.~T. Irvine, and
  V.~Vitelli, ``Odd elasticity,'' {\em Nature Physics}, vol.~16, no.~4,
  pp.~475--480, 2020.

\bibitem{chen2021realization}
Y.~Chen, X.~Li, C.~Scheibner, V.~Vitelli, and G.~Huang, ``Realization of active
  metamaterials with odd micropolar elasticity,'' {\em Nature communications},
  vol.~12, no.~1, pp.~1--12, 2021.

\bibitem{zhang2020correspondence}
K.~Zhang, Z.~Yang, and C.~Fang, ``Correspondence between winding numbers and
  skin modes in non-hermitian systems,'' {\em Physical Review Letters},
  vol.~125, no.~12, p.~126402, 2020.

\bibitem{helbig2020generalized}
T.~Helbig, T.~Hofmann, S.~Imhof, M.~Abdelghany, T.~Kiessling, L.~Molenkamp,
  C.~Lee, A.~Szameit, M.~Greiter, and R.~Thomale, ``Generalized bulk--boundary
  correspondence in non-hermitian topolectrical circuits,'' {\em Nature
  Physics}, vol.~16, no.~7, pp.~747--750, 2020.

\bibitem{ghatak2020observation}
A.~Ghatak, M.~Brandenbourger, J.~van Wezel, and C.~Coulais, ``Observation of
  non-hermitian topology and its bulk--edge correspondence in an active
  mechanical metamaterial,'' {\em Proceedings of the National Academy of
  Sciences}, vol.~117, no.~47, pp.~29561--29568, 2020.

\bibitem{rosa2020dynamics}
M.~I. Rosa and M.~Ruzzene, ``Dynamics and topology of non-hermitian elastic
  lattices with non-local feedback control interactions,'' {\em New Journal of
  Physics}, vol.~22, no.~5, p.~053004, 2020.

\bibitem{braghini2021non}
D.~Braghini, L.~G. Villani, M.~I. Rosa, and J.~R. de~F~Arruda, ``Non-hermitian
  elastic waveguides with piezoelectric feedback actuation: non-reciprocal
  bands and skin modes,'' {\em Journal of Physics D: Applied Physics}, vol.~54,
  no.~28, p.~285302, 2021.

\bibitem{zhang2021acoustic}
L.~Zhang, Y.~Yang, Y.~Ge, Y.-j. Guan, Q.~Chen, Q.~Yan, F.~Chen, R.~Xi, Y.~Li,
  D.~Jia, {\em et~al.}, ``Acoustic non-hermitian skin effect from twisted
  winding topology,'' {\em arXiv preprint arXiv:2104.08844}, 2021.

\bibitem{longhi2021non}
S.~Longhi, ``Non-hermitian skin effect beyond the tight-binding models,'' {\em
  Physical Review B}, vol.~104, no.~12, p.~125109, 2021.

\bibitem{doi:10.1063/5.0097530}
Y.~Jin, W.~Zhong, R.~Cai, X.~Zhuang, Y.~Pennec, and B.~Djafari-Rouhani,
  ``Non-hermitian skin effect in a phononic beam based on piezoelectric
  feedback control,'' {\em Applied Physics Letters}, vol.~121, no.~2,
  p.~022202, 2022.

\bibitem{wang2021generating}
K.~Wang, A.~Dutt, K.~Y. Yang, C.~C. Wojcik, J.~Vu{\v{c}}kovi{\'c}, and S.~Fan,
  ``Generating arbitrary topological windings of a non-hermitian band,'' {\em
  Science}, vol.~371, no.~6535, pp.~1240--1245, 2021.

\bibitem{PhysRevApplied.18.014067}
R.~Cai, Y.~Jin, Y.~Li, T.~Rabczuk, Y.~Pennec, B.~Djafari-Rouhani, and
  X.~Zhuang, ``Exceptional points and skin modes in non-hermitian metabeams,''
  {\em Phys. Rev. Applied}, vol.~18, p.~014067, Jul 2022.

\bibitem{hussein2006dispersive}
M.~I. Hussein, G.~M. Hulbert, and R.~A. Scott, ``Dispersive elastodynamics of
  1d banded materials and structures: analysis,'' {\em Journal of sound and
  vibration}, vol.~289, no.~4-5, pp.~779--806, 2006.

\bibitem{khalil2015nonlinear}
H.~K. Khalil, {\em Nonlinear control}, vol.~406.
\newblock Pearson New York, 2015.

\bibitem{gong2018topological}
Z.~Gong, Y.~Ashida, K.~Kawabata, K.~Takasan, S.~Higashikawa, and M.~Ueda,
  ``Topological phases of non-hermitian systems,'' {\em Physical Review X},
  vol.~8, no.~3, p.~031079, 2018.

\bibitem{atalla2015finite}
N.~Atalla and F.~Sgard, {\em Finite element and boundary methods in structural
  acoustics and vibration}.
\newblock CRC Press, 2015.

\bibitem{hussein2014dynamics}
M.~I. Hussein, M.~J. Leamy, and M.~Ruzzene, ``Dynamics of phononic materials
  and structures: Historical origins, recent progress, and future outlook,''
  {\em Applied Mechanics Reviews}, vol.~66, no.~4, 2014.

\bibitem{rosa2017numerical}
M.~Rosa, V.~Lima, J.~Santos, and J.~Arruda, ``Numerical and experimental
  investigation of interface modes in periodic acoustic waveguides,'' {\em
  Proceedings of ICEDYN, Ericeira, Portugal}, 2017.

\bibitem{bloch1929quantenmechanik}
F.~Bloch, ``{\"U}ber die quantenmechanik der elektronen in kristallgittern,''
  {\em Zeitschrift f{\"u}r physik}, vol.~52, no.~7, pp.~555--600, 1929.

\end{thebibliography}
\end{document}